\documentclass[11pt,a4paper,english,twoside]{article}

\usepackage{a4wide}
\usepackage{amssymb, amsmath, amsthm}
\usepackage{graphicx}
\usepackage[mathscr]{euscript}
\usepackage{subcaption}
\usepackage[all]{xy}
\usepackage[pdftex,svgnames]{xcolor}
\usepackage[pdftex,colorlinks=true,
pdfstartview=FitV,
pdfnewwindow=true,
linktoc = page,
linkcolor= blue,
citecolor= red,
urlcolor= blue,
hyperindex=true,
hyperfigures=false]{hyperref}
\hypersetup{linktocpage}
\usepackage{dsfont}
\usepackage{empheq}
\usepackage{cite}
\usepackage{float}
\usepackage{cancel}
\usepackage{relsize}
\usepackage{soul}
\usepackage{enumerate}
\usepackage{enumitem}
\usepackage{hhline}
\usepackage{multirow}
\usepackage{xspace}
\usepackage{bbm}
\usepackage{pdfpages}
\usepackage{setspace}
\usepackage{comment}
\usepackage{booktabs}

\usepackage[normalem]{ulem} 

\newcommand{\nn}{\nonumber}

\newcommand*\widefbox[1]{\fbox{\hspace{1em}#1\hspace{1em}}}

\def\beq{\begin{equation}}
\def\eeq{\end{equation}}
\def\bea{\begin{eqnarray}}
\def\eea{\end{eqnarray}}

\def\del {\partial}
\def\d {{\rm d}}
\def\w {\wedge}
\def\tr {{\rm tr}}
\def\Tr {{\rm Tr}}

\begin{document}
\numberwithin{equation}{section}

\begin{titlepage}
\begin{center}

\phantom{DRAFT}

\vspace{1.4 cm}

{\LARGE \bf{Axionic Quintessence\\\vspace{0.4 cm} in Heterotic String Theory }}\\

\vspace{2.0 cm} {\Large David Andriot${}^1$, Margherita Putti${}^{2,3}$, L\'eo Veissi\`ere${}^1$}\\

\vspace{0.8 cm} ${}^1$ {\small\slshape Laboratoire d'Annecy de Physique Th\'eorique (LAPTh),\\[0.15cm]
CNRS, USMB, UMR 5108,\\[0.08cm]
9 Chemin de Bellevue, 74940 Annecy, France}\\

\vspace{0.4 cm} ${}^2$ {\small\slshape
Van Swinderen Institute for Particle Physics and Gravity,\\[0.15cm]
University of Groningen, Nijenborgh 3, 9747 AG Groningen, The Netherlands}\\

\vspace{0.4 cm} ${}^3$ {\small\slshape Dutch Institute for Emergent Phenomena, The Netherlands}\\

\vspace{0.9cm} {\upshape\ttfamily andriot@lapth.cnrs.fr, m.putti@rug.nl, veissiere@lapth.cnrs.fr}\\

\vspace{2.cm}
{\bf Abstract}
\vspace{0.1cm}
\end{center}
\begin{quotation}
We investigate axionic quintessence in heterotic string compactifications. Starting from the ten-dimensional theory, we derive the four-dimensional axion effective action, including gauge couplings and non-perturbative potentials, and identify the curvature terms that can correct it. We examine the geometric requirements for ultralight axions generated by exponentially suppressed worldsheet instantons, finding significant restrictions through an exploratory scan of toric Calabi–Yau hypersurfaces. In an illustrative K3-fibred setup, the minimal benchmark also encounters a tension with the positivity of the one-loop corrected gauge kinetic functions. Aligned worldsheet-instanton contributions provide a way to alleviate this tension and yield an ultralight axion with a phenomenologically relevant decay constant. Our analysis also identifies lower-action instantons whose contributions must be controlled to preserve the ultralight axion. We then study the resulting class of quintessence potentials and qualitatively compare their late-time cosmological solutions with observationally preferred ranges.
\end{quotation}

\end{titlepage}

\tableofcontents

\section{Introduction}
Axions are among the best-motivated light scalar degrees of freedom beyond the Standard Model. Originally introduced in connection with the strong CP problem, they have since found applications across particle physics and cosmology, including dark matter, dark radiation, cosmic birefringence and dark energy; see e.g.~\cite{Marsh:2015xka,Ai:2026ovp} and references therein. They are particularly natural from the viewpoint of string theory, where axion-like fields arise generically from the dimensional reduction of higher-dimensional form fields~\cite{Svrcek:2006yi}. String compactifications therefore typically contain several axions spanning a broad range of masses and decay constants, giving rise to the so-called axiverse and a variety of possible cosmological signatures~\cite{Arvanitaki:2009fg}.

This picture has been studied extensively in explicit string compactifications, particularly in type IIB~\cite{Cicoli:2012sz,Demirtas:2018akl,Demirtas:2021gsq, Gendler:2023kjt}. Axions also arise naturally in heterotic compactifications, descending from the ten-dimensional NSNS two-form either as the universal, model-independent axion dual to the four-dimensional two-form, or as model-dependent axions associated with the internal cohomology of the compactification manifold. Recent work has begun to systematically characterize the heterotic axiverse, including its mass spectrum, gauge couplings and phenomenological implications~\cite{Agrawal:2024ejr,Leedom:2025mlr,Benabou:2026jtv,Benabou:2026lcu}. In particular, explicit studies have shown how the structure of heterotic compactifications constrains axion masses and couplings, providing a useful starting point for cosmological applications.

A separate challenge is to embed these axionic sectors into a controlled cosmological background in which the remaining moduli are stabilized. Progress on heterotic moduli stabilization has involved NSNS three-form flux, gauge bundles, worldsheet instantons, gaugino condensation and related non-perturbative effects~\cite{deCarlos:1992kox,Gukov:2003cy,Gaillard:2007jr,Serone:2007sv,Anderson:2010mh,Dundee:2010sb,Parameswaran:2010ec,Cicoli:2013rwa,Leedom:2022zdm}, together with perturbative $(\alpha')$, string-loop and non-perturbative corrections to the K\"ahler potential~\cite{Shenker:1990uf,Candelas:1990rm,Barreiro:1997rp,Becker:2002nn,Anguelova:2010ed}. These developments suggest new possibilities for constructing controlled heterotic cosmological models.

In this work, we explore a different cosmological role for such a light string axion: we ask whether one axionic degree of freedom arising in a heterotic compactification can act as a quintessence field and account for the dark energy driving the late-time acceleration of the Universe.

Dark energy is currently the main constituent in our universe. It is responsible for the acceleration of its expansion. Such a behaviour cannot be achieved through (non-relativistic) matter or (relativistic) radiation, which makes the nature of dark energy difficult to understand. A first possibility is to have a constant dark energy density, also known as a cosmological constant $\Lambda$. Over the past decades, this option has led to the well-tested standard model of cosmology, a.k.a.~$\Lambda$CDM. An alternative is to have a dynamical dark energy, whose energy density changes with time. One realization is provided by quintessence models, where scalar field(s) roll(s) along a scalar potential $V$, the latter contributing to the varying energy density together with the field(s) kinetic energy. Quintessence models have been revived, in recent years both for theoretical and observational motivations. Whether such models offer a better description of the late-time Universe than  $\Lambda$CDM is currently under investigation (see e.g.~\cite{Giare:2026oti}).

On the theoretical side, string theory naturally provides 4-dimensional (4d) effective theories with scalar fields and a scalar potential. Getting solutions with rolling fields is therefore ubiquitous, even though this is not sufficient to guarantee an adequate quintessence solution. The difficulties in getting a well-controlled de Sitter solution, i.e.~a positive extremum of the scalar potential corresponding to a cosmological constant, in a model derived from string theory, have led to investigating the alternative offered by quintessence. Recent reviews on attempts of getting these various forms of dark energy from string theory include \cite{Cicoli:2023opf, Andriot:2026lac, Anchordoqui:2026hys}. 

On the observational side, the late time evolution of our universe has recently been probed to unprecedented precision by DES \cite{DES:2024jxu} and DESI \cite{DESI:2024mwx, DESI:2025zgx}. Those have provided hints of a dynamical dark energy. The latest analysis \cite{DESI:2026lnd} gives a discrepancy of $3.0 - 3.5 \,\sigma$ in favor of dynamical dark energy with respect to the $\Lambda$CDM model. While these observations are still on-going, and new experiments are collecting new data, this motivates us to have a new look at the quintessence options.\\

In this work, we focus on 4d quintessence models which can be derived from string theory. Compared to a plethora of possible dynamical dark energy models, studied in the cosmology literature, our approach is motivated by having a quantum gravity and UV completeness for our model. We focus on {\sl axionic quintessence}, where the rolling scalar is a canonically normalized axion $\phi$. We consider two representative potentials:
\bea
\text{Axionic:}&&\quad V_{{\rm Ax}} = \Lambda^4 \, \left( 1 +\cos \left(\frac{\phi}{f} \right) \right) \ , \label{VAxintro}\\
\text{Hilltop-like:}&&\quad V_{{\rm Hill}} = \Lambda^4 \,\cos \left(\frac{\phi}{f} \right) \ . \label{VHillintro}
\eea
Axionic quintessence is attractive because the symmetry structure of axions constrains their interactions with matter and radiation, helping to evade several phenomenological constraints, such as fifth force, variation of fundamental constants, radiative corrections, etc., that one could face for a dynamical quintessence field. Moreover, the potentials~\eqref{VAxintro} and \eqref{VHillintro}\footnote{We will focus on $V(\phi)>0$, so the dynamics on the Hilltop-like potential requires the field to roll close to the maximum. The commonly named ``Hilltop'' potential is $V(\phi) = V_0 \left( 1 - \frac{\kappa^2}{2} \phi^2 \right)$, hence the choice to call our potential ``Hilltop-like''.}, admit thawing quintessence solutions~\cite{Andriot:2024sif}, as they provide cosmological solutions with a growing equation of state parameter $w_{\phi}$ in the late universe. Such solutions have good chances to match the latest observational data; the axionic potential and its parameters were actually directly constrained by DESI in \cite{DESI:2025fii}. Relatedly, the DESI-preferred evolution has been compared with quintessential $\alpha$-attractor models in~\cite{Borys:2026sna}. Although axionic quintessence has been explored in type II string compactifications~\cite{Cicoli:2024yqh,Caraffi:2026wzk,Anchordoqui:2025fgz,Borys:2026sna,Ramos-Hamud:2026wdx} and earlier work considered this possibility in heterotic M-theory~\cite{Choi:1999xn}, its realization in perturbative heterotic string compactifications remains largely unexplored. In this work, we take a step towards addressing this gap.\\

If dark energy is driven by axionic quintessence, its realization in perturbative heterotic compactifications faces demanding constraints: the ultralight axion scale and a phenomenologically relevant effective decay constant must be obtained while maintaining geometric control and consistent gauge couplings. We investigate these requirements together, beginning with a minimal mechanism in which the small potential scale arises from an exponentially suppressed worldsheet-instanton contribution associated with a single large two-cycle volume.

Our analysis establishes three main results. First, an exploratory scan of toric Calabi--Yau hypersurfaces finds no sampled configuration satisfying simultaneously the single-large-cycle requirement, the volume bound and the full geometric-control criteria. In an illustrative K3-fibred effective theory, the minimal benchmark also encounters a tension between the desired decay constant and the positivity of the one-loop corrected gauge kinetic functions. Second, non-trivial worldsheet-instanton charge vectors can alleviate this gauge-sector tension: an aligned effective-theory benchmark yields an ultralight axion with $f\simeq0.4M_p$, positive gauge kinetic functions and the target visible coupling, using small topological coefficients. An explicit two-modulus toric geometry also supports the required aligned axion kinematics within the geometric-control region used in the scan.

Third, the aligned configurations expose a crucial microscopic requirement: competing instantons must not lift the light direction. All 55 candidates retained for an invariant audit in a separate finite alignment search exhibit lower-action classes with nonzero projection on that direction and nonvanishing genus-zero Gopakumar-Vafa invariants. These invariants do not by themselves imply nonzero heterotic superpotential contributions, but they show why the required instanton selection cannot be established from geometry alone. A realization must suppress the dangerous contributions through bundle-dependent zero modes or cancellations while retaining the selected instantons~\cite{BuchbinderDonagiOvrut2002,BeasleyWitten2003,Curio2010}. The aligned effective theory admits axionic quintessence, but its realization requires suppressing competing instanton contributions. A consistent gauge bundle and moduli stabilization preserving the light potential remain to be constructed. 

The paper is organized as follows. In Section \ref{sec:10d4d}, we derive the four-dimensional axion effective theory from ten-dimensional heterotic string theory, including the relevant kinetic terms, gauge couplings and non-perturbative contributions to the scalar potential. We also discuss the parametric size of curvature effects and the assumptions underlying moduli stabilization and the effective single-field description. In Section \ref{sec:CY}, we investigate the geometric requirements of the minimal construction through a scan of toric Calabi–Yau hypersurfaces. We then study an illustrative K3-fibred setup and explore how non-trivial instanton charge vectors can alleviate the constraints encountered in the minimal benchmark. In Section \ref{sec:cosmo}, we analyze the late-time cosmological evolution for the Hilltop-like and Axionic potentials and discuss their phenomenological implications. Finally, Section \ref{sec:ccl} summarizes our results and outlines the remaining challenges and possible directions for future work.

\section{From 10d heterotic string to the 4d cosmological model}\label{sec:10d4d}

Starting from 10-dimensional (10d) heterotic string, we reach in this section a 4d multifield theory of axions. This is achieved thanks to a dimensional reduction on a 6d Calabi-Yau manifold, as described in Sections \ref{sec:pertdimred} and \ref{sec:curv}, together with non-perturbative contributions generating a 4d scalar potential as presented in Section \ref{sec:pot}. This procedure is summarized in Section \ref{sec:sum4d}, where we also explain how this 4d theory can eventually lead to a single field model, to be used for cosmology.

\subsection{Perturbative dimensional reduction}\label{sec:pertdimred}

In this section, we start from perturbative and weakly coupled 10-dimensional (10d) heterotic superstring at order $\alpha'$, and perform a dimensional reduction through a specific compactification ansatz. The resulting 4d theory will describe the physics of axions coupled to gravity. Extra contributions to this 4d theory will be presented in the next subsections.\\

Following conventions of \cite{Andriot:2011iw}, in line with \cite{Bergshoeff:1989de}, we first consider the following target space effective theory of 10d heterotic string. The bosonic action in string frame is given by
\bea
&& {\cal S}_{10d} = \frac{1}{2 \kappa_{10}^2} \int \d^{10}x \sqrt{|g_{10}|} \, e^{-2\phi} \Bigg( {\cal R}_{10} + 4 (\del \phi)^2 - \frac12 |H|^2 \label{S10d} \\
&& \phantom{{\cal S}_{10d} = \frac{1}{2 \kappa_{10}^2} \int \d^{10}x \sqrt{|g_{10}|} \, e^{-2\phi} \Bigg(} + \frac{\alpha'}{4} \left(\tr |{\cal R}_{ie+}|^2 - \tr |{\cal F}_{10}|^2 \right)  + {\cal O}({\alpha'}^2) \Bigg) + {\cal S}_{{\rm GS}} \nn
\eea
where $2 \kappa_{10}^2=(2\pi)^7\, l_s^8$, $\alpha'=l_s^2$, and $|H|^2 =1/6\, H_{MNP} H^{MNP}$ with a 10d spacetime index $M$; our conventions on supergravity $p$-forms can be found in \cite[App.A]{Andriot:2026lac}. The 1-loop contribution ${\cal S}_{{\rm GS}}$ is the topological action necessary to cancel the factorizable anomalies via the Green-Schwarz mechanism; we will come back to it. 

At order $\alpha'$, one has
\beq
\tr |{\cal F}_{10}|^2 = \tr (t_{\mathfrak{a}} t_{\mathfrak{b}}) \, \frac12 {\cal F}^{\mathfrak{a}}_{MN} {\cal F}^{\mathfrak{b}\, MN} \ ,\quad \tr |{\cal R}_{ie+}|^2 = \frac12 {\cal R}_{+\, ABMN}{\cal R}_+^{ABMN} \ .\label{trsquare}
\eeq
The 10d 2-form field strength of the gauge interaction is denoted ${\cal F}_{10} = \frac12 {\cal F}_{MN}\, \d x^M \w \d x^N$, with a color index $\mathfrak{a}$ for the gauge group and $t_{\mathfrak{a}}$ a generator. The trace $\tr$ is the one in the fundamental representation, suited for the gauge group $SO(32)$. Here, we rather pick $E_8 \times E_8$, in the adjoint representation, with trace denoted ${\rm Tr}$. When necessary, one can trade one trace for the other with ${\rm Tr}= 30\, \tr$. 

${\cal R}^A_{+ BMN}$ denotes the Riemann tensor component for a spin connection 1-form $\omega_+^A{}_B$, with the 10d local frame indices $A,B$. The spin connection coefficient, labeled with $+$, is given by $\omega_+^A{}_{BM} = \omega^A{}_{BM} + \frac12 H^A{}_{BM}$, in terms of the Levi-Civita one and the $H$-flux \cite{Bergshoeff:1989de}. It is also convenient to introduce the associated curvature 2-form, and the trace of its (wedge) square, defined as follows
\beq
{\cal R}_+^A{}_B = \frac12 {\cal R}_+^A{}_{BMN}\, \d x^M \w \d x^N \ ,\quad  \tr{\cal R}_{10+}^2 \equiv \tr({\cal R}_{10+} \w {\cal R}_{10+}) = - {\cal R}_+^A{}_B  \w {\cal R}_+^B{}_A \ , \label{R+}
\eeq
where the index $10$ on the Riemann tensor is only used in that 10d trace, for later convenience.

The $H$-flux is defined as
\beq
H = \d B + \frac{\alpha'}{4}  \left( {\rm CS}(\omega_+) -  {\rm CS}({\cal A})  \right) +  {\cal O}({\alpha'}^2) \ , \label{Hflux0}
\eeq
in terms of the $B$-field and the Chern-Simons (CS) terms for the spin and the gauge connections. One has $\d {\rm CS}(\omega_+) = \tr{\cal R}_{10+}^2 $ and $\d {\rm CS}({\cal A}) = \tr{\cal F}_{10}^2$. In other words, the $H$-flux Bianchi identity, in absence of $N\!S_5$-brane, is given by
\beq
\d H = \frac{\alpha'}{4} \left( \tr({\cal R}_{10+} \w {\cal R}_{10+}) - \tr({\cal F}_{10}\w {\cal F}_{10}) \right) +  {\cal O}({\alpha'}^2) \ . \label{BI}
\eeq
\\

We are now going to perform a dimensional reduction to 4d. We consider the 10d spacetime to be a direct product of a 4d (cosmological) spacetime and a 6d compact Calabi-Yau manifold $X$. Our compactification ansatz for the 10d fields will only consider as 4d-dependent fluctuations the $B$-field components (with zero background value): they will eventually correspond to 4d axions. All other fluctuations are assumed frozen or stabilized at given values, thus entering background-valued fields: it is for instance the case of real 6d geometric moduli. That assumption will be further discussed in Section \ref{sec:stab}.

With such an ansatz, the 10d metric splits block-diagonally, each block having its own coordinate dependence, giving a 6d (background) volume denoted as
\beq
\int_X {\rm vol}_6 \equiv \int_X \d^{6}x \sqrt{|g_6|} = {\cal V}\, (2\pi l_s)^6 \ .
\eeq
We also get for the Ricci scalar ${\cal R}_{10}={\cal R}_4 + {\cal R}_6$, where the latter vanishes on $X$. Turning to the 10d gauge field strength, we also take it to split into purely 4d and 6d components, with respective 4d and 6d coordinate dependence as for the metric; the corresponding 2-forms are denoted respectively as ${\cal F}_{10}= {\cal F} + \bar{{\cal F}}$. We introduce similarly the notation of ${\cal R}_+$ and $\bar{{\cal R}}_+$ for purely 4d and purely 6d curvature 2-forms obtained from ${\cal R}_{10+}$; at this stage, the latter could also in principle give mixed (4d and 6d) 2-forms. Finally, we take a constant dilaton: $e^{\phi}=g_s$. 

We now turn to the $H$-flux. We denote the internal $B$-field fluctuation, i.e.~the 6d 2-form with 4d coordinate dependence, as $B_2$. The $H$-flux thus splits as
\beq
H= H|_{4d} + \d B_2 + \frac{\alpha'}{4} (\dots)|_{4,6d} + \frac{\alpha'}{4} (\dots)|_{6d} \ . \label{Hflux}
\eeq
We denote by $H|_{4d}$ the 4d components (possibly including $\alpha'$ contributions). We will verify in the following and in Appendix \ref{ap:curv} that $H|_{4d}$ only depends on 4d coordinates. The forms $\d B_2$ and $(\dots)|_{4,6d}$ have components with mixed 4d and 6d indices, while the last term in \eqref{Hflux} is made of the CS terms in \eqref{Hflux0} with 6d components. The mixed (4d and 6d) term at order $\alpha'$ cannot have any contribution from the gauge fields: this is because each of the 10d $E_8$ splits in a product of 2 orthogonal subgroups (one for 4d and one for 6d), and the adjoint trace between their generators vanishes. There can however be a mixed contribution from ${\rm CS}(\omega_+)$. Overall, we will see that there is no background $H$-flux at order $(\alpha')^0$: there only $B$-field fluctuations contribute.

With this ansatz, a dimensional reduction of the 10d action \eqref{S10d} to 4d first leads, at order $\alpha'$, to the following action
\bea
{\cal S}_{4d} &=& \frac{M_p^2}{2} \int \d^{4}x \sqrt{|g_{4}|} \left( {\cal R}_{4} - \frac{\alpha'}{4} \tr |{\cal F}|^2 - \frac12 |H|_{4d}|^2 + {\cal O}({\alpha'}^2) \right) \label{S4d}\\
&&\!\!\!\! +\, \frac{M_p^2}{2\, {\cal V}\, (2\pi l_s)^6} \int \d^{10}x \sqrt{|g_{10}|} \left( - \frac12 |\d B_2|^2 \right) + {\cal S}_{{\rm GS}} + {\cal S}_{{\rm curv}\, 0} + {\cal S}_{\bar{{\cal F}}}   \ ,\nn
\eea
where the square of the reduced 4d Planck mass is given by
\beq
M_p^2 = \frac{\int_X \d^6 x \sqrt{|g_6|}}{\kappa_{10}^2 g_s^2} =\frac{{\cal V}}{\pi l_s^2 g_s^2} \ .
\eeq
The remaining actions include terms at order $\alpha'$ which appear, at first sight, not relevant to the 4d axion physics to be studied. These curvature terms and 6d gauge terms are given by
\bea
{\cal S}_{{\rm curv}\, 0} &=& \ \ \frac{1}{2 \kappa_{10}^2\, g_s^2}\, \frac{\alpha'}{4} \int \left(\d^{10}x \sqrt{|g_{10}|} \, \tr |{\cal R}_{ie+}|^2 - \d B_2 \w *_{10}\, {\rm CS}(\omega_+)|_{4,6d} \right) \ ,\\ 
{\cal S}_{\bar{{\cal F}}} &=& - \frac{1}{2 \kappa_{10}^2\, g_s^2}\, \frac{\alpha'}{4} \int \d^{10}x \sqrt{|g_{10}|} \, \tr |\bar{{\cal F}}|^2  \ , \label{SF}
\eea
and we will come back to them.\\

For the next steps, we broadly follow \cite{Svrcek:2006yi}. First, one may identify the constant in front of $\tr |{\cal F} |^2$ as $1/(2\, g_{{\rm YM}}^2)$ (recall the contraction $|{\cal F}|^2$ carries the standard $1/2$ factor). One can further introduce $\alpha_{{\rm YM}}$ as follows
\beq
\alpha_{{\rm YM}} = \frac{g_{{\rm YM}}^2}{4\pi} =  \frac{g_s^2}{{\cal V}} \ .
\eeq
As we will discuss around \eqref{S4df2}, the standard model gauge group or a GUT scenario is meant to be realized within the first $E_8$ factor. At tree level, we use $\alpha_{{\rm YM}}\simeq1/25$ as a proxy for the unified visible coupling \cite{Svrcek:2006yi}. Once one-loop threshold corrections are included, this tree-level input must be distinguished from the physical visible and hidden gauge couplings. In addition, the present derivation assumes a perturbative string regime, for which $g_s \leq 1$. The tree-level relation then leads to the preliminary volume bound
\beq
{\cal V} \leq 25 \ , \label{boundVol}
\eeq
which will play an important role in the following\footnote{The relation between the four-dimensional gauge coupling, the string coupling and the compactification volume obtained above holds at tree level. In the presence of one-loop threshold corrections, the physical gauge couplings are instead determined by the real parts of the corresponding corrected gauge kinetic functions. We will use the tree-level volume bound as a preliminary geometric selection criterion, while commenting separately on the additional consistency conditions associated with these corrections.}.

Second, we turn to $H|_{4d}$, for which we first need to discuss the 10d $H$-flux Bianchi identity \eqref{BI}. This identity is an extra constraint that has to be imposed. On the background product connection used to extract the zero-derivative four-dimensional couplings, it decomposes into 4d and 6d components,
\beq
\tr({\cal R}_{10+} \w {\cal R}_{10+}) = \tr({\cal R}_{+} \w {\cal R}_{+}) + \tr(\bar{{\cal R}}_{+} \w \bar{{\cal R}}_{+}) \ .\label{trR+2}
\eeq
Here the notation anticipates the torsionful connection; axion-derivative-dependent mixed components are part of the local curvature sector discussed in Appendix \ref{ap:curv} and are not used in deriving the integrated background tadpole condition.
Focusing on the 6d components of \eqref{BI}, they will appear under integrals over 4-cycles $\Sigma_4$ in $X$. In that case, since the integral of $\d H$ vanishes, one should have
\beq
\int_{\Sigma_4}\tr(\bar{{\cal R}}_+ \w \bar{{\cal R}}_+) = \int_{\Sigma_4} \tr(\bar{{\cal F}}\w \bar{{\cal F}}) \ . \label{tadpole}
\eeq
This equality will be used in the following, and referred to as tadpole cancellation. This condition is satisfied for specific choices on $X$ and on the gauge bundle.

The 4d Bianchi identity should also hold. It can be imposed by introducing a (scalar) Lagrange multiplier $a$ in 4d, adding a term to the action \eqref{S4d}. The complete action for $H|_{4d}$ becomes
\beq
\frac{M_p^2}{2} \int \left( - \frac12 H|_{4d} \w *_4 H|_{4d} \right) + \int \frac{a}{4\pi^2 \alpha'} \left( \d H |_{4d} - \frac{\alpha'}{4} \left( \tr({\cal R}_+ \w {\cal R}_+) - \tr({\cal F}\w {\cal F}) \right)    \right) \ .
\eeq
The normalization of $a$ is chosen for later convenience: it will fix the decay constant $f_a$, and for now, it allows the scalar $a$ to be dimensionless. Integrating out $a$ gives back the previous 4d action, together with the 4d Bianchi identity satisfied. In turn, treating $H|_{4d}$ independently and integrating it out, we obtain the equivalent action for $a$
\beq
\frac{1}{(4\pi^2 \alpha')^2} \frac{2}{M_p^2} \int \left( - \frac12 \d a \w *_4 \d a \right) + \frac{1}{16 \pi^2} \int a  \left( \tr({\cal F}\w {\cal F}) - \tr({\cal R}_+ \w {\cal R}_+)  \right) \ .
\eeq
Doing so, we obtained the relation 
\beq
\d a = 2\pi^2 \alpha' M_p^2\ \, *_4\! H|_{4d} \ . \label{daH4}
\eeq
It should be noted that this procedure only fixes $H|_{4d}$ at order $(\alpha')^0$; this point will be useful in Appendix \ref{ap:curv}. The relation \eqref{daH4} can be rephrased by introducing a 6-form $B_6$ as follows
\bea
&& \d B_6 = *_{10} H|_{4d} = *_4 H|_{4d} \w {\rm vol}_6 \phantom{\Big(}\ ,\\
&& B_6 =\frac{1}{2\pi^2 \alpha' M_p^2}\ a\ {\rm vol}_6 \ ,\quad a = \frac{2\pi}{\alpha_{{\rm YM}}} \times \frac{\int_X B_6}{\int_X {\rm vol}_6}  \ .
\eea
$B_6$ depends both on 4d and 6d coordinates; the latter dependence is however only that of ${\rm vol}_6$. The $2\pi$ factor singled-out in $a$ relates to its periodicity, as will be shown below. $a$ will be referred to as the universal or model-independent axion field. 

Third, we focus on $B_2$. To that end, we introduce a basis of harmonic 2-forms $\beta_i$, where $i=1 \dots h^{1,1}$, with corresponding 2-cycles $\Sigma_i$ in $X$, as follows 
\beq
\int_{\Sigma_j} \beta_i = \delta_{ij} \ ,\quad \quad \frac{1}{(2\pi l_s)^2}\int_X \beta_i \w *_6 \beta_j = \delta_{ij}\, \gamma_i^2  \ .
\eeq
The normalization by $(2\pi l_s)^2$ ensures that both the harmonic forms $\beta_i$ and the coefficients $\gamma_i$ are dimensionless. For convenience, we work in a basis where $\int_X \beta_i \w *_6 \beta_j $ is diagonal. This corresponds to having the (real) K\"ahler metric diagonal, as will be seen e.g.~through \eqref{kinbi}. The 2-form $B_2$ is then expanded on this basis as follows
\beq
\label{eq:B2expansion}
B_2 = \frac{(2\pi l_s)^2}{2\pi} \sum_i b^i(x) \, \beta_i \ ,\quad 
\eeq
where the 4d-dependent coefficients $b^i$ will correspond to model-dependent axions. The $2\pi l_s$ factors ensure that the $b_i$ are dimensionless, while the explicit factor of $2\pi$ has been chosen so that each axion has period $2\pi$. This gives
\beq
\int \d^{10}x \sqrt{|g_{10}|} \left( - \frac12 |\d B_2|^2 \right)  =  \frac{(2\pi l_s)^6}{4 \pi^2} \int \d^{4}x \sqrt{|g_{4}|}\ \left( -\frac12\, \gamma_{i}^2\, \del_{\mu} b^i \del^{\mu} b^i \right)\ .\label{kinbi}
\eeq

Let us introduce the dimensionless volume $v^i$ of each 2-cycle $\Sigma_i$
\beq
\int_{\Sigma_i}  {\rm vol}_{\Sigma_i} = v^i \, (2\pi l_s)^2 \ ,\quad \beta_i =\frac{ {\rm vol}_{\Sigma_i} }{v^i \, (2\pi l_s)^2}  \ .\label{betavol}
\eeq
It allows one to build the K\"ahler moduli $T^i$, together with the axio-dilaton $S$ of 4d ${\cal N}=1$ supergravity, as follows
\beq
T^i = v^i -{\rm i}\, \frac{b^i}{2\pi} \ ,\quad S= \frac{{\cal V}}{g_s^2}-{\rm i}\, \frac{a}{2\pi} \ , \label{moduli}
\eeq
which will play an important role in the 4d theory. With these conventions, we have
\beq
B_2 = - \sum_i \frac{{\rm Im}[T^i]}{{\rm Re}[T^i]} \, {\rm vol}_{\Sigma_i} \ , \quad B_6 = - \frac{{\rm Im}[S]}{{\rm Re}[S]} \, {\rm vol}_6 \ .
\eeq

Thanks to the above three steps, the initial theory \eqref{S4d}, together with the 4d Bianchi identity enforced, provide us with the following 4d action
\bea
{\cal S}_{4d} &=& \int \d^{4}x \sqrt{|g_{4}|} \left( \frac{M_p^2}{2}\, {\cal R}_{4} - \frac{1}{8\pi \,\alpha_{{\rm YM}}}\, \tr |{\cal F}|^2 + {\cal O}({\alpha'}^2) \right) \label{S4df}\\
&&\!\! +\, \int \d^{4}x \sqrt{|g_{4}|} \left( -  \frac12\, f_a^2\, (\del a)^2  \right)  + \frac{1}{16 \pi^2} \int a \ \tr({\cal F}\w {\cal F})  \nn\\
&&\!\! +\, \int \d^{4}x \sqrt{|g_{4}|}\ \left( -\frac12\, f_i^2\, (\del b^i)^2 \right) + {\cal S}_{{\rm GS}} + {\cal S}_{{\rm curv}\, 1} + {\cal S}_{\bar{{\cal F}}} \ ,\nn
\eea
with the following axion decay constants
\bea
f_a^2 &=& \frac{1}{(4\pi^2 \alpha')^2} \frac{2}{M_p^2} = \frac{g_s^2 }{2\pi (2\pi l_s)^2 {\cal V}} \ \ \longleftrightarrow\ \ f_a = \frac{\alpha_{{\rm YM}}}{2\pi} \frac{M_p}{\sqrt{2}} \ , \\
f_i^2 &=&  \frac{\gamma_{i}^2}{ 2 \pi (2\pi l_s)^2 g_s^2} \ \ \longleftrightarrow\ \ f_i = \frac{\alpha_{{\rm YM}}^{\frac13}}{2\pi} \frac{M_p}{\sqrt{2}} \ g_s^{-\frac23} \quad {\rm if} \ \gamma_i^2 \simeq {\cal V}^{\frac13} \ ,
\eea
where the last equality assumes an isotropic compactification, $\gamma_i^2\simeq \mathcal{V}^{1/3}$. In the following sections, we keep the discussion completely general and do not assume isotropy. The curvature terms have been updated towards
\beq
{\cal S}_{{\rm curv}\, 1} = {\cal S}_{{\rm curv}\, 0} - \frac{1}{16 \pi^2} \int a \   \tr({\cal R}_+ \w {\cal R}_+) \ .
\eeq
\\

We now turn to ${\cal S}_{{\rm GS}}$: this action is given by
\bea
&&{\cal S}_{{\rm GS}} = \frac{1}{4 (2\pi)^5 \alpha' 4!} \int B \w X_8 \ ,\\[0.1in]
&& X_8 = \frac{1}{3} \Tr {\cal F}_{10}^4 - \frac{1}{900} \left( \Tr {\cal F}_{10}^2\right)^2 - \frac{1}{30} \Tr {\cal F}_{10}^2\, \tr {\cal R}_{10+}^2 + \frac14 \left( \tr {\cal R}_{10+}^2\right)^2 + \tr {\cal R}_{10+}^4 \ , 
\eea
where one verifies that $X_8$ is dimensionless, and the normalization is taken from \cite{Svrcek:2006yi}. The powers refer to wedge products, as defined with the traces around \eqref{R+}. Standard references express $X_8$ in terms of the geometric Riemann tensor, meaning with respect to the geometric spin connection $\omega$. We extend it here by consistency to the Riemann tensor for $\omega_+$, allowing us to eventually use the $H$-flux Bianchi identity or tadpole cancellation \eqref{tadpole}.

Prior to dimensional reduction, we develop $X_8$ using relations between traces. We label by $1,2$ each $E_8$ factor, with $\Tr = \Tr_1 + \Tr_2$ and similarly for $\tr$. For each $E_8$, one has also the following relations
\beq
\Tr_i = 30 \, \tr_i \ ,\quad \Tr_i {\cal F}_{10}^4 = \frac{1}{100} \left(\Tr_i {\cal F}_{10}^2 \right)^2 \ .
\eeq
This leads to
\bea
&& \hspace{-0.5in} X_8 = 2\left( \tr_1 {\cal F}_{10}^2 \right)^2 + 2 \left( \tr_2 {\cal F}_{10}^2 \right)^2 - 2 \tr_1 {\cal F}_{10}^2 \, \tr_2 {\cal F}_{10}^2  - \left(\tr_1 {\cal F}_{10}^2 +  \tr_2 {\cal F}_{10}^2 \right) \tr {\cal R}_{10+}^2 \\
&& \hspace{-0.5in} \phantom{X_8\ \ } + \frac14 \left( \tr {\cal R}_{10+}^2\right)^2 + \tr {\cal R}_{10+}^4 \ .\nn
\eea
The next step is dimensional reduction. We recall the 4d and 6d split ${\cal F}_{10} = {\cal F} + \bar{{\cal F}}$, and that $\tr_i ({\cal F} \w \bar{{\cal F}})=0$ (see below \eqref{Hflux}). To extract the zero-derivative Green-Schwarz couplings, we evaluate the characteristic forms on the background product connection, for which $\tr {\cal R}_{10}^2=\tr {\cal R}^2+\tr\bar{{\cal R}}^2$. Axion-derivative-dependent torsionful-curvature terms are instead included in ${\cal S}_{{\rm curv}}$ and are discussed separately in Section \ref{sec:curv} and Appendix \ref{ap:curv}. At the background level this gives
\bea
&& X_8 = \tr_1 {\cal F}^2 \, \left(4\tr_1 \bar{{\cal F}}^2 - 2  \tr_2 \bar{{\cal F}}^2 - \tr \bar{{\cal R}}_{+}^2\right)  +  \tr_2 {\cal F}^2 \, \left(4 \tr_2 \bar{{\cal F}}^2  - 2 \tr_1 \bar{{\cal F}}^2 -\tr \bar{{\cal R}}_{+}^2 \right)\\
&& \phantom{X_8}  + \tr {\cal R}_{+}^2  \left(\frac12 \tr \bar{{\cal R}}_{+}^2 - \tr_1 \bar{{\cal F}}^2 - \tr_2 \bar{{\cal F}}^2 \right)  + \tr {\cal R}_{10+}^4 \ . \nn
\eea
For the block-diagonal background connection, $\tr {\cal R}_{10}^4$ has no non-zero eight-form component: its purely external and purely internal pieces vanish dimensionally, while the tangent-bundle trace does not mix the two blocks. Therefore, the background contribution to $X_8$ relevant here is proportional to ${\rm vol}_4$. Since it carries four external and four internal legs, only the purely internal component of $B$, namely $B_2$, contributes to the zero-derivative reduction of ${\cal S}_{{\rm GS}}$. Possible axion-derivative-dependent contributions are part of the unevaluated local curvature sector. With explicit wedge products, we get
\bea
\hspace{-0.3in} {\cal S}_{{\rm GS}}& =& \frac{1}{(4\pi)^4} \sum_i \Bigg[ \int b^i \, \tr_1 ({\cal F}\w {\cal F}) \int_X \beta_i \w \left(\tr_1 (\bar{\cal F} \w \bar{\cal F}) - \frac12 \tr (\bar{{\cal R}}_+ \w \bar{{\cal R}}_+)\right) + 1\leftrightarrow 2 \nn\\
\hspace{-0.3in} && \phantom{ \frac{1}{(4\pi)^4}  \sum_i} \!\!\!\! - \frac{1}{12}  \int b^i \, \tr({\cal R}_+ \w {\cal R}_+) \int_X \beta_i \w \tr (\bar{{\cal R}}_+ \w \bar{{\cal R}}_+)  \Bigg] +  \int \frac{B_2 \w \tr {\cal R}_{10+}^4 }{4 (2\pi)^5 \alpha' 4!}
\eea
where we used the tadpole cancellation \eqref{tadpole}. We introduce the internal coefficients $n_i$, which encode the internal gauge and curvature contribution
\bea
\label{eq:ni}
n_i &=& \ \ \frac{1}{16\pi^2} \int_X \beta_i \w \left(\tr_1 (\bar{\cal F} \w \bar{\cal F}) - \frac12 \tr (\bar{{\cal R}}_+ \w \bar{{\cal R}}_+) \right) \\
&=& - \frac{1}{16\pi^2} \int_X \beta_i \w \left(\tr_2 (\bar{\cal F} \w \bar{\cal F}) - \frac12 \tr (\bar{{\cal R}}_+ \w \bar{{\cal R}}_+) \right) \ ,
\eea
where the equality is again due to tadpole cancellation \eqref{tadpole}.
The action ${\cal S}_{{\rm GS}}$ then gives the following 4d axionic couplings
\beq
{\cal S}_{4d} \supset \frac{1}{16 \pi^2} \int n_i\, b^i \left( \tr_1({\cal F}\w {\cal F}) - \tr_2({\cal F}\w {\cal F})\right) \ , \label{bCS}
\eeq
while the remaining curvature terms should be added to ${\cal S}_{{\rm curv}\, 1}$. Overall, we get the following curvature terms
\begin{empheq}[box=\widefbox]{align}
{\cal S}_{{\rm curv}} =&\  \frac{1}{2 \kappa_{10}^2\, g_s^2}\, \frac{\alpha'}{4} \int \d^{10}x \sqrt{|g_{10}|} \ \tr |{\cal R}_{ie+}|^2  - \d B_2 \w *_{10}\, {\rm CS}(\omega_+)|_{4,6d} \nn\\[0.1in]
&\ - \frac{1}{16 \pi^2} \int \tr({\cal R}_+ \w {\cal R}_+) \left(  a  +   b^i \ \, \frac{1}{12 \times 16 \pi^2}  \int_X \beta_i \w \tr (\bar{{\cal R}}_+ \w \bar{{\cal R}}_+) \right) \label{Scurv} \\[0.1in]
&\ +  \int \frac{B_2 \w \tr {\cal R}_{10+}^4 }{4 (2\pi)^5 \alpha' 4!} \ .\nn
\end{empheq}
We discuss their tensor structure and parametric size in Section \ref{sec:curv} and Appendix \ref{ap:curv}.\\

From \eqref{S4df} and above results, we can finally give the complete 4d action obtained from dimensional reduction of perturbative 10d heterotic string. As a last step, we canonically normalize the axions into $\vartheta_a = f_a \, a$, $\vartheta^i = f_i\, b^i$ (without sum), and the 4d gauge terms (without introducing new symbols) ${\cal F} \rightarrow g_{{\rm YM}}{\cal F} = \sqrt{4\pi \alpha_{{\rm YM}} }{\cal F}$. We eventually obtain
\bea
{\cal S}_{4d} &=& \int \d^{4}x \sqrt{|g_{4}|} \left( \frac{M_p^2}{2} {\cal R}_{4} - \frac{1}{2} \tr_1 |{\cal F}|^2 - \frac{1}{2} \tr_2 |{\cal F}|^2 + {\cal O}({\alpha'}^2) \right) + {\cal S}_{{\rm curv}} + {\cal S}_{\bar{{\cal F}}}   \nn\\
&&\!\! +\, \int \d^{4}x \sqrt{|g_{4}|} \left( -  \frac12\, (\del \vartheta_a)^2   -\frac12\, \sum_i \, (\del \vartheta^i)^2 \right) \label{S4df2}\\
&&\!\! +\, \frac{\alpha_{{\rm YM}}}{4 \pi} \int \left(\frac{\vartheta_a}{f_a}+ n_i\, \frac{\vartheta^i}{f_i} \right) \ \tr_1({\cal F}\w {\cal F}) + \left(\frac{\vartheta_a}{f_a} - n_i\, \frac{\vartheta^i}{f_i} \right) \ \tr_2({\cal F}\w {\cal F}) \ .\nn
\eea
where in the Chern-Simons coupling there is an implicit sum in the repeated indices $i$. The first $E_8$ can be associated to the visible sector, leading to a GUT scenario or eventually a standard model gauge group,\footnote{When considering a non-standard embedding of the standard model into heterotic string, the traces can get modified by a factor $k$, $\tr_i \rightarrow k\, \tr_i$, corresponding to the Kac-Moody level $k$; this leads to defining $\alpha_{{\rm GUT}}= \alpha_{{\rm YM}}/k$ \cite{Svrcek:2006yi}. In the following, we work with gauge generators normalized at level $k=1$. For a more general gauge embedding, the appropriate normalization of the surviving gauge factors should be taken into account.} while the second $E_8$ corresponds to the hidden sector. To match phenomenological notations, one should eventually take for each sector $\tr_i (t_{\mathfrak{a}}t_{\mathfrak{b}})=\tfrac12 \delta_{\mathfrak{a}\mathfrak{b}}$ and further canonically normalize the gauge fields as ${\cal F} \rightarrow \sqrt{2}\, {\cal F}$, such that for one axion in one sector, one gets
\beq
{\cal S}_{4d} \supset \int \d^{4}x \sqrt{|g_{4}|} \left( -\frac{1}{4} {\cal F}_{\mu\nu}^{\mathfrak{a}} {\cal F}^{\mu\nu}_{\mathfrak{a}} - \frac12 \del_{\mu} \vartheta \del^{\mu} \vartheta +\frac{g_{\rm YM}^2}{32 \pi^2}\, \frac{\vartheta}{f}\, {\cal F}_{\mu\nu}^{\mathfrak{a}} \tilde{{\cal F}}^{\mu\nu}_{\mathfrak{a}} \right) \ ,
\eeq
with ${\cal F}=*_4 \tilde{{\cal F}}$ or $\tilde{{\cal F}}= -*_4 {{\cal F}}$, i.e.~$\tilde{{\cal F}}^{\mu\nu}_{\mathfrak{a}} = -\frac{\sqrt{|g_4|}}{2}\, \epsilon^{\mu\nu \rho\sigma}\, {\cal F}_{\rho\sigma\, \mathfrak{a}}$; this uses the convention of \cite{Andriot:2026lac}, standard in mathematics, where $\epsilon_{0123}=1$. We do so for the last line in \eqref{S4df2}, and introduce Chern-Simons couplings $\lambda_v$ (visible) and $\lambda_h$ (hidden), namely
\beq
\hspace{-0.2in} {\cal S}_{4d} \supset  \int \d^{4}x \sqrt{|g_{4}|}\ \Bigg(\!\! \left(  \frac{\lambda_{a,v}}{4 f_a} \,\vartheta_a+ \frac{\lambda_{i,v}}{4 f_i} \, \vartheta^i \right) \big({\cal F}_{\mu\nu}^{\mathfrak{a}} \tilde{{\cal F}}^{\mu\nu}_{\mathfrak{a}} \big)_v + \left( \frac{\lambda_{a,h}}{4 f_a} \, \vartheta_a + \frac{\lambda_{i,h}}{4 f_i} \, \vartheta^i \right) \big({\cal F}_{\mu\nu}^{\mathfrak{a}} \tilde{{\cal F}}^{\mu\nu}_{\mathfrak{a}}\big)_h \Bigg)
\eeq
such that
\beq
\lambda_{a,v} = \lambda_{a,h} = \frac{\alpha_{{\rm YM}}}{2\pi} \ ,\quad \lambda_{i,v} = -  \lambda_{i,h} = \frac{\alpha_{{\rm YM}}}{2\pi} n_i \ .
\eeq
This can be rewritten using the gauge kinetic function $f_{{\rm kin}}$, as we now explain. This quantity can be read from the action prior to gauge canonical normalization, i.e.~here from \eqref{S4df}: it is defined as follows for each sector (see e.g.~\cite{Gukov:2003cy, Blumenhagen:2005ga, Cicoli:2013rwa})
\beq
{\cal S}_{4d} \supset - \frac12 \int \Big({\rm Re}[f_{{\rm kin}}]\ \tr_i ({\cal F} \w *_4 {\cal F}) + {\rm Im}[f_{{\rm kin}}]\ \tr_i ( {\cal F} \w {\cal F}) \Big) \ .
\eeq
This gives
\beq
f_{{\rm kin}} = \frac{1}{4\pi} \left( \frac{{\cal V}}{g_s^2}-{\rm i}\, \frac{a}{2\pi} \right) = \frac{S}{4\pi} \ , \label{fkintree}
\eeq
where $S$ is the 4d axio-dilaton defined in \eqref{moduli}. We then rewrite the above last two couplings as follows
\beq
\lambda_{i,v/h} = \pm\, \frac{n_i}{8\pi^2 \, {\rm Re}[f_{{\rm kin}}] } \ , 
\eeq
for the visible ($+$) or hidden ($-$) sector, where ${\rm Re}[f_{{\rm kin}}]= \frac{1}{g_{{\rm YM}}^2} = \frac{1}{4\pi\, \alpha_{{\rm YM}}} = \frac{1}{4\pi} \frac{{\cal V}}{g_s^2} $.

The expression \eqref{fkintree} of the gauge kinetic function corresponds to the tree-level contribution; it gets corrected at one-loop. In particular, the Green-Schwarz term \eqref{bCS} contributes to the imaginary part, shifting the visible/hidden sector according to $ {\rm Im}[f_{{\rm kin}}] \rightarrow  {\rm Im}[f_{{\rm kin}}] \pm \left( -\frac{1}{8\pi^2} n_i\, b^i \right)$. This correction is naturally incorporated by expressing the gauge kinetic function in terms of the 4d K\"ahler moduli $T^i = v^i -{\rm i}\, \frac{b^i}{2\pi}$ \eqref{moduli}, in agreement with the standard one-loop heterotic threshold structure~\cite{Dixon:1990pc,Blumenhagen:2005ga}; the gauge kinetic function becomes
\beq
f_{{\rm kin}}^{1-loop}= \frac{1}{4\pi} \left( S \pm n_i \, T^i \right) \ .
\eeq
Accordingly, in the level-one normalization used here, the tree-level and physical couplings obey
\beq
\alpha_{\rm YM}^{-1}=s=\frac{{\cal V}}{g_s^2}\,,\qquad
\alpha_v^{-1}=s+n_i v^i\,,\qquad
\alpha_h^{-1}=s-n_i v^i\,.
\label{physicalgaugecouplings}
\eeq
Thus positivity, $s\pm n_iv^i>0$, and matching the physical visible coupling, $\alpha_v^{-1}\simeq25$, are distinct requirements.
This modifies the coupling to the visible ($+$) or hidden ($-$) sector towards
\begin{empheq}[box=\widefbox]{equation}
\lambda_{i,v/h} = \pm\, \frac{g_s^2\, n_i }{2\pi ({\cal V} \pm g_s^2\, n_i\, v^i)} \ .\label{lambdai}
\end{empheq}
Using these results together with \eqref{S4df2}, we rewrite the 4d action obtained through this perturbative dimensional reduction
\begin{empheq}[box=\widefbox]{align}
{\cal S}_{4d} &= \int \d^{4}x \sqrt{|g_{4}|} \left( \frac{M_p^2}{2} {\cal R}_{4} -\frac{1}{4}\left( {\cal F}_{\mu\nu}^{\mathfrak{a}} {\cal F}^{\mu\nu}_{\mathfrak{a}}\right)_v -\frac{1}{4}\left( {\cal F}_{\mu\nu}^{\mathfrak{a}} {\cal F}^{\mu\nu}_{\mathfrak{a}}\right)_h + {\cal O}({\alpha'}^2) \right) \nn\\
& + \int \d^{4}x \sqrt{|g_{4}|} \left( -  \frac12\, (\del \vartheta_a)^2   -\frac12\, \sum_i \, (\del \vartheta^i)^2 \right)  + {\cal S}_{{\rm curv}} + {\cal S}_{\bar{{\cal F}}}  \label{S4df3}\\
&\hspace{-0.3in} + \!\!\!  \int \! \d^{4}x \sqrt{|g_{4}|}\, \Bigg(\!\!\! \left(  \frac{\lambda_{a,v}}{4 f_a} \,\vartheta_a+ \frac{\lambda_{i,v}}{4 f_i} \, \vartheta^i \right)\! \big({\cal F}_{\mu\nu}^{\mathfrak{a}} \tilde{{\cal F}}^{\mu\nu}_{\mathfrak{a}}\big)_v \! +\! \left( \frac{\lambda_{a,h}}{4 f_a} \, \vartheta_a + \frac{\lambda_{i,h}}{4 f_i} \, \vartheta^i \right)\! \big({\cal F}_{\mu\nu}^{\mathfrak{a}} \tilde{{\cal F}}^{\mu\nu}_{\mathfrak{a}}\big)_h \! \Bigg) \nn
\end{empheq}
This action will be completed by a non-perturbative 4d scalar potential discussed in the next subsections. Prior to this, we turn to the curvature terms in ${\cal S}_{{\rm curv}}$, and to ${\cal S}_{\bar{{\cal F}}}$.

\subsection{Evaluation of the curvature terms and impact on axions}\label{sec:curv}

Performing in Section \ref{sec:10d4d} the dimensional reduction of 10d perturbative heterotic string to 4d, we have identified curvature terms gathered in the action ${\cal S}_{{\rm curv}}$ \eqref{Scurv}. These terms are rarely considered in the literature (see Appendix \ref{ap:curv}). One reason is that they correspond to higher derivative contributions. Nevertheless, they appear in the 10d effective action at the same $\alpha'$ or loop level as e.g.~gauge contributions, that are kept. Therefore, it is worthwhile to evaluate their quantitative impact on the 4d effective theory: we do so in this section. 

In Appendix \ref{ap:curv}, we consider various restrictions on the fields entering the curvature terms, often motivated by our ansatz, in particular that we take a 4d FLRW metric and that the axions $\vartheta$ only depend on time. We then compute each term in ${\cal S}_{{\rm curv}}$ as well as in the $H$-flux Bianchi identity \eqref{BI}. We reach a final expression for ${\cal S}_{{\rm curv}}$ in \eqref{Scurvcomputedap}, that we repeat here for convenience (with the dimensionless axions $\hat{\vartheta}= \vartheta/M_p$ and the Hubble parameter ${\cal H}$)
\bea
{\cal S}_{{\rm curv}} \!\!\!\! && = \int \d^4 x \sqrt{|g_4|}\ \frac{M_p^2}{2}\ \Bigg[ \frac32\, l_s^2 \, \big({\cal H}^4 + \big( \dot{{\cal H}} + {\cal H}^2 \big)^2 \big) + \ {\cal O}({\alpha'}^2) \label{Scurvfinal}\\
&& \phantom{= \int \d^4 x \sqrt{|g_4|}\ \frac{M_p^2}{2}\ \Bigg[} - \frac{3}{2} \, l_s^2\, {\cal H}^2 \,  (\del_0 \hat{\vartheta}_a)^2 + 3\, l_s^2\,  {\cal H}^2\ \hat{\vartheta}_a\, \del_0^2 \hat{\vartheta}_a  \nn\\
&& \phantom{= \int \d^4 x \sqrt{|g_4|}\ \frac{M_p^2}{2}\ \Bigg[}  + \frac{3}{8}\, l_s^2 \,  (\del_0 \hat{\vartheta}_a)^4  - \frac34\, l_s^2\, \left(\del_0^2 \hat{\vartheta}_a \right)^2  - 9\ l_s^2\ \hat{\vartheta}_a\, \del_0^2 \hat{\vartheta}_a \, (\del_0 \hat{\vartheta}_a)^2    \nn\\
&& \phantom{= \int \d^4 x \sqrt{|g_4|}\ \frac{M_p^2}{2}\ \Bigg[} + \frac{9}{8} \, l_s^2 \, \sum_i (\del_0 \hat{\vartheta}^i)^4  + \frac{1}{8} \,  l_s^2\, \sum_i \left(\del_0^2 \hat{\vartheta}^i \right)^2  \nn\\
&& \phantom{= \int \d^4 x \sqrt{|g_4|}\ \frac{M_p^2}{2}\ \Bigg[} + \frac18 \ \frac{1}{{\cal V} (2\pi l_s)^6}\int {\rm vol}_6\ \ l_s^2 \ {\cal R}_{mnpq} {\cal R}^{mnpq}  \nn\\
&& \phantom{= \int \d^4 x \sqrt{|g_4|}\ \frac{M_p^2}{2}\ \Bigg[} +\frac12\, l_s^2\ \del_0 \hat{\vartheta}^i \del_0 \hat{\vartheta}^j  \int {\rm vol}_6\ {\cal R}_{mnpq} \ \frac{ \beta_i^{mp} \beta_j^{nq} }{2\, (2\pi l_s)^2\, \gamma_i \gamma_j} \nn\\
&& \phantom{= \int \d^4 x \sqrt{|g_4|}\ \frac{M_p^2}{2}\ \Bigg[} + \frac{1}{2} \, l_s^2\ \del_0 \hat{\vartheta}^i \del_0 \hat{\vartheta}^j\, \int \frac{-\beta_i{}^a{}_b\ \beta_j}{2\, (2\pi l_s)^2\, \gamma_i \gamma_j} \w  *_6\d \omega^b{}_{a} \nn\\
&& \phantom{= \int \d^4 x \sqrt{|g_4|}\ \frac{M_p^2}{2}\ \Bigg[} + \frac{1}{(4\pi)^3}\ \del_0^2 \hat{\vartheta}_a \left( {\cal H}^2   - 3 \, (\del_0 \hat{\vartheta}_a)^2  \right) \, \frac{\sqrt{{\cal V}}}{M_p^2}\ \hat{\vartheta}^i \, \int \frac{\beta_i}{\gamma_i} \w \tr (\bar{{\cal R}}  \w  \bar{{\cal R}} ) \nn\\
&&\hspace{-0.5in} + \frac{1}{(4 \pi)^3}\, \del_0^2 \hat{\vartheta}_a \left( {\cal H}^2  - 3 \, (\del_0 \hat{\vartheta}_a)^2  \right) \, \hat{\vartheta}^i \, \del_0 \hat{\vartheta}^j \del_0 \hat{\vartheta}^k \ \frac{\sqrt{{\cal V}}}{M_p^2} \, {\cal V} (2\pi l_s)^6 \! \int \frac{\beta_i}{\gamma_i} \w \frac{\beta_{j\, r}{}^m \beta_{k\, s}{}^n}{2\, (2\pi l_s)^2\, \gamma_j \gamma_k} \, {\cal R}_{mnpq} \, \d y^{pqrs}  \Bigg] \nn
\eea
The terms containing local contractions of the internal curvature, connection and harmonic forms require the explicit Calabi-Yau metric for a microscopic evaluation. Accordingly, below we use these terms only to estimate their parametric size; harmonicity of the $\beta_i$ does not by itself fix their pointwise contractions or the detailed correction matrix.
Our goal here is to provide an estimate of each term, allowing to assess whether it can be neglected, or whether it may have an impact on 4d physics.\\

Given that we will consider a cosmological setting, we assume that the characteristic timescale is the Hubble time, so that each time derivative introduces one power of $\mathcal H $. We recall the present day value of the Hubble parameter ${\cal H}_0 = 5.9 \cdot 10^{-61} M_p$ \cite{Planck:2018vyg}. As a result, the number $l_s {\cal H}$ must be very small, especially in the context of the recent universe; during inflation, $\mathcal{H}$ is much larger than its present value, so it could be interesting to reconsider such hierarchies. It is then straightforward to see that the first four lines of \eqref{Scurvfinal}, corresponding to purely 4d contributions, are of order ${\cal H}^2 \times l_s^2 {\cal H}^2$. This should be compared to the Einstein-Hilbert contribution, for which ${\cal R}_4 \sim {\cal H}^2$: these terms are negligible.

We are left with the last five lines of \eqref{Scurvfinal} which involve 6d contributions. They depend on the 6d metric, which is not known explicitly for the compactifications considered here. We then propose a rough estimate: we introduce a typical 6d length $L$, and consider each 6d derivative to be of order $1/L$.\footnote{We note that in an anisotropic compactification, the different terms may actually probe different cycle-dependent lengths; for now, we can only propose the simpler estimate with one length $L$.} We also have $\beta_{i\, mn} \sim 1/L^2$, and we recall that $\int_X {\rm vol}_6\, \frac12 \beta_{i\, mn} \beta_j^{mn} = \delta_{ij}\, \gamma_i^2 (2\pi l_s)^2$. For the last two lines, we also recall that $\gamma_i \sim L/l_s$ and ${\cal V} \sim (L/l_s)^6$. We obtain for these five last lines the following estimates
\bea
\hspace{-0.4in} {\cal S}_{{\rm curv}} \supset \!\!\!\! &&  \int \d^4 x \sqrt{|g_4|}\ \frac{M_p^2}{2}\ \Bigg[ \frac18 \ \frac{l_s^2}{L^2} \frac{1}{L^2}  +\frac12\, \frac{l_s^2}{L^2}\ \sum_i (\del_0 \hat{\vartheta}^i)^2  + \frac{1}{2} \, \frac{l_s^2}{L^2}\ \sum_i (\del_0 \hat{\vartheta}^i )^2 \nn\\
\hspace{-0.4in} && \phantom{\supset \!\!\!\! \int \d^4 x \sqrt{|g_4|}\ \frac{M_p^2}{2}\ \Bigg[} + \frac{1}{(4\pi)^3}\ \frac{1}{M_p^2} \ \frac{L^2}{l_s^2}\ \del_0^2 \hat{\vartheta}_a \left( {\cal H}^2 - 3 \, (\del_0 \hat{\vartheta}_a)^2  \right) \sum_i \hat{\vartheta}^i  \nn\\
\hspace{-0.4in} && \phantom{\supset \!\!\!\! \int \d^4 x \sqrt{|g_4|}\ \frac{M_p^2}{2}\ \Bigg[} + \frac{1}{(4 \pi)^3}\ \frac{1}{M_p^2} \ \frac{L^4}{l_s^2}\ \del_0^2 \hat{\vartheta}_a \left( {\cal H}^2  - 3 \, (\del_0 \hat{\vartheta}_a)^2  \right) \, \hat{\vartheta}^i \ \del_0 \hat{\vartheta}^j \del_0 \hat{\vartheta}^k  \Bigg] \label{Scurvestimate}\\
\hspace{-0.4in} \sim &&  \int \d^4 x \sqrt{|g_4|}\ \frac{M_p^2}{2}\ \Bigg[ \frac18 \ \frac{l_s^2}{L^2}\, \frac{1}{L^2}  +\frac12\, \frac{l_s^2}{L^2}\, {\cal H}^2  + \frac{1}{2} \, \frac{l_s^2}{L^2}\, {\cal H}^2 \nn\\
\hspace{-0.4in} && \phantom{\supset \!\!\!\! \int \d^4 x \sqrt{|g_4|}\ \frac{M_p^2}{2}\ \Bigg[} + \frac{1}{(4\pi)^3}\ g_s^2 \, \frac{l_s^4}{L^4}\ l_s^2 {\cal H}^2 \ {\cal H}^2  + \frac{1}{(4 \pi)^3}\ g_s^2 \, \frac{l_s^2}{L^2}\ l_s^4 {\cal H}^4 \ {\cal H}^2 \Bigg] \ . \nn
\eea
Minimal requirements to justify approximations of a perturbative regime are that $l_s <L$ and $g_s <1$. This makes the last two terms negligible compared to ${\cal R}_4 \sim {\cal H}^2$, similarly to the previous 4d contributions.

The first term, due to the square of the 6d Riemann tensor, is not necessarily negligible. It corresponds to a cosmological constant term (or more generally would contribute to the potential of moduli that are here fixed). This is the first instance of a constant term that we will simply set aside; we will encounter more of them.\\

We are left in \eqref{Scurvestimate} with two terms, which provide an interesting correction in $\frac{l_s^2}{L^2}$ to the model-dependent axion kinetic terms. With $(\del_0 \vartheta)^2 = - (\del \vartheta)^2$, we get 
\beq
{\cal S}_{{\rm curv}} \supset  -\int \d^4 x \sqrt{|g_4|}\ \ \frac{l_s^2}{L^2}\ \frac{1}{2}\, \sum_i (\del \vartheta^i )^2 \ .
\eeq
In other words, the kinetic terms get the following correction
\beq
(\del \vartheta^i )^2 \ \longrightarrow \  \left(1 + \frac{l_s^2}{L^2} \right)  (\del \vartheta^i )^2 \ ,\qquad f_i \ \longrightarrow \  f_i' = \sqrt{1 + \frac{l_s^2}{L^2} } \ f_i \ .\label{correctionfi}
\eeq
While a common hierarchy to ensure a perturbative expansion would be $l_s \ll L$, we will however see in Section \ref{sec:CY} that some 6d lengths will here be close to $l_s$. This is due to the volume upper bound ${\cal V} \leq 25$ \eqref{boundVol} and the phenomenological wish to have internal anisotropies. As a consequence, we will only have here $l_s \lesssim L$. This situation places us at the boundary of control on the string perturbative expansion and the validity of the 4d theory, which require neglecting higher order corrections in e.g.~${\alpha'}^2$. As an example, for ${\cal V} = 25 = \left(\frac{L}{l_s}\right)^6$, we get $\frac{l_s^2}{L^2} \simeq 0.4 \sim \frac13 $. That value is a useful illustration: it agrees with a perturbative expansion, while providing a non-trivial correction in \eqref{correctionfi}. For the benchmark used below, ${\cal V}=5.9$, the same isotropic estimate instead gives
\beq
\epsilon_{\rm curv}\equiv\frac{l_s^2}{L^2}\sim{\cal V}^{-1/3}\simeq0.6\,.
\eeq
We use this benchmark value to estimate the impact of the correction on 4d axion physics.

Let us schematically recall the origin of this correction
\bea
&& \hspace{-0.3in} \tr |{\cal R}_{ie+}|^2 \ \longrightarrow \ {\cal R}_{mnpq} \, H_0{}^{mp}H_0{}^{nq} \ \longrightarrow \ \del_0 \vartheta^i \del_0 \vartheta^j \ \int {\rm vol}_6\ {\cal R}_{mnpq} \, \beta_i^{mp} \beta_j^{nq}  \\
&& \hspace{-0.3in} \d B_2 \w *_{10}\, {\rm CS}(\omega_+)|_{4,6d} \ \longrightarrow \  (H_{0mn} \, \d x^0 \d y^{mn}) \w  *_{10}\, ( H^a{}_{b0}\, \d x^0 \d \omega^b{}_{a} ) \ \longrightarrow \ \del_0 \vartheta^i \del_0 \vartheta^j\, \int \beta_i{}^a{}_b\ \beta_j \w  *_6\d \omega^b{}_{a} \nn
\eea
While we have given in \eqref{Scurvestimate} the correction with a positive sign, we cannot be sure at this stage of the actual sign of these 6d contributions. A slightly more accurate statement is then to allow for either sign in \eqref{correctionfi}, and have the corrected decay constant $f_i \rightarrow f_i'$ within the range
\begin{empheq}[box=\widefbox]{equation}
\sqrt{1-\epsilon_{\rm curv}}\, f_i \ \leq \ f_i' \ \leq \ \sqrt{1+\epsilon_{\rm curv}}\, f_i    \label{ficorr}
\end{empheq}
When considering eigenmodes of the mass matrix, we will investigate the effect of these corrections on the 4d physical axions. The physical decay constants and masses will be modified: we discuss this in Section \ref{sec:curveffectf}.\\

Having evaluated the curvature terms of ${\cal S}_{{\rm curv}}$, we finally turn to another action piece, set aside in Section \ref{sec:10d4d}, that is ${\cal S}_{\bar{{\cal F}}}$ \eqref{SF}. This 6d gauge contribution enters the scalar potential for moduli that we consider as stabilized or frozen here (e.g.~the gauge bundle moduli). This term is thus simply a constant here, it appears as a cosmological constant. We can rewrite it as follows
\bea
\hspace{-0.4in} {\cal S}_{\bar{{\cal F}}} = - \frac{1}{2 \kappa_{10}^2\, g_s^2}\, \frac{\alpha'}{4} \int \d^{10}x \sqrt{|g_{10}|} \  \tr |\bar{{\cal F}}|^2  &= & - \int \! \d^{4}x \sqrt{|g_{4}|} \, \frac{1}{8\pi \alpha_{{\rm YM}}}\, \frac{ \int_X \d^{6}x \sqrt{|g_{6}|} \ \tr |\bar{{\cal F}}|^2}{{\cal V} (2\pi l_s)^6} \nn\\
&= & - M_p^2 \int \! \d^{4}x \sqrt{|g_{4}|} \ \frac{l_s^2}{8}\ \frac{ \int_X \d^{6}x \sqrt{|g_{6}|} \ \tr |\bar{{\cal F}}|^2}{{\cal V} (2\pi l_s)^6} \ .
\eea
If, as above, we estimate $ \tr |\bar{{\cal F}}|^2 \sim L^{-4}$, one gets for this term $\frac{l_s^2}{L^2} \, \frac{1}{L^{2}}$. We cannot assess the size of this term, similarly to the contraction of 6d Riemann tensors discussed previously. We get here a second example of a constant term that we will set aside.\\

To summarize our findings, the curvature terms may impact the 4d physics by changing the decay constants $f_i$ as proposed in \eqref{ficorr}: we will investigate the quantitative impact on axion mass eigenstates in Section \ref{sec:curveffectf}. In addition, we identified two cosmological constant-like terms that we discard for now; we will however discuss them again when considering the final 4d scalar potential in Section \ref{sec:sum4d}\footnote{A related cancellation follows from the heterotic Bianchi identity, which relates the gauge and curvature four-form sources. Their harmonic components are not independent: after projection onto the zero modes, the total source must vanish in cohomology.
This can lead to (partial or complete) cancellation of the corresponding 4d contributions for certain gauge bundles, see~\cite{Lukas:1997fg}. This does not imply a pointwise cancellation of the local gauge and curvature densities.}. In the following, our starting point will then be the 4d theory \eqref{S4df3} without ${\cal S}_{{\rm curv}}$ and ${\cal S}_{\bar{{\cal F}}}$: it describes the physics of 4d axions in visible and hidden sectors. Only their scalar potential is missing, and we now turn to it.

\subsection{Scalar potential and mass spectrum}\label{sec:pot}

The 4d effective theory obtained from 10d heterotic string on a Calabi-Yau 3-fold $X$ can be formulated in an ${\cal N}=1$ language. The effective action is determined by the following K\"{a}hler potential $K$ and superpotential $W$ \cite{deCarlos:1992kox, Gukov:2003cy, Gaillard:2007jr, Serone:2007sv, Anderson:2010mh, Dundee:2010sb, Parameswaran:2010ec}: 
\begin{equation}
\label{KandW}
    \begin{aligned}
    & K=-\ln(S+\bar S)-\ln{\cal V}(\mathtt{T}^i,\bar{\mathtt{T}}^i)-\ln\left(-{\rm i} \int_{X}\Omega(z_a)\wedge \bar\Omega(\bar z_{\bar a})\right)+\cdots\\
    & W= \int_{X} H\wedge\Omega(z_a) +\sum_i A_i\, e^{-2\pi\, \mathtt{T}^i}+ \sum_i B_i\, e^{-a_i\, G_i(S,\mathtt{T}^j,z_b)}\quad,
    \end{aligned}
\end{equation}
where $S= s- {\rm i} \frac{a}{2\pi}$ denotes the axio-dilaton defined in \eqref{moduli}, $\mathtt{T}^i$ and $z_a$ stand respectively for the K\"ahler moduli and the complex structure and gauge bundle ones, ${\cal V}$ is the 6d volume to be specified below, and $\Omega$ is the holomorphic (3,0)-form. The ellipsis represents perturbative $\alpha'$, string loop and non-perturbative corrections to the K\"ahler potential~\cite{Shenker:1990uf,Candelas:1990rm,Barreiro:1997rp,Becker:2002nn,Anguelova:2010ed}. The first term in $W$ arises from the 3-form flux, the second denotes the worldsheet-instanton contributions, which arise from Euclidean fundamental strings wrapping isolated holomorphic 2-cycles of the Calabi–Yau manifold. The third term in $W$ instead represents non-perturbative contributions from hidden-sector gaugino condensation, whose gauge kinetic function receives threshold corrections depending on the compactification moduli.

We focus on the branch of perturbative heterotic compactifications in which the hidden $E_8$ gauge group is completely broken to Abelian factors by the gauge bundle and discrete Wilson lines, such that no hidden non-Abelian gauge sector survives at low energies and gaugino condensation is therefore absent. This restriction is motivated by the phenomenology of axion quintessence: hidden-sector gaugino condensation generates an additional non-perturbative superpotential whose axion dependence generically produces masses far above the scale required for late-time cosmology~\cite{Leedom:2025mlr}. Eliminating this contribution therefore isolates the worldsheet instantons as the dominant source of the axion potential. In this branch the gaugino condensation term is absent and, after integrating out the complex-structure and bundle fields, that we assume stabilized and heavy (see section \ref{sec:stab}), the superpotential reduces to
\begin{equation}
\label{superpot}
    W=W_0+\sum_i A_i\, e^{-2\pi\, \mathtt{T}^i}\,,
\end{equation}
where $W_0$ denotes the effective constant contribution to the superpotential after integration. Depending on the compactification, this may arise from the $H$-flux or from the stabilization mechanism itself. 
Since the Euclidean worldsheet action is proportional to the area of the wrapped curve, the corresponding contribution to the superpotential is exponentially suppressed in terms of ${\tt T}^i={\tt v}^i- {\rm i} \frac{{\tt b}^i}{2\pi}$. Those are the complexified K\"ahler fields, ${\tt v}^i$ are the geometric K\"ahler moduli, and ${\tt b}^i$ are the corresponding axions in the geometrical basis, before diagonalization of the K\"ahler metric. Note that we use here a slightly different notation than in \eqref{moduli}: there the fields were defined in an orthogonal basis for the $\beta_i$, that is, with a diagonal real K\"ahler metric. The two sets of fields are thus related by a change of basis. Finally, the prefactors $A_i$ are Pfaffian one-loop determinants which can depend on the complex-structure and gauge-bundle moduli~\cite{BuchbinderDonagiOvrut2002,BeasleyWitten2003,Curio2010}; they are treated as constants in the effective theory after these fields have been integrated out. In the following, we restrict ourselves to a minimal instanton structure in which each contribution depends on a single K\"ahler modulus. This assumption is not required by heterotic compactifications in general, since worldsheet instantons can wrap effective curves whose volumes involve linear combinations of K\"ahler moduli. It nevertheless provides a simple setting in which the hierarchy of axion masses can be directly related to the hierarchy of individual cycle volumes. More general instanton charge vectors, potentially leading to axion alignment, will be briefly discussed later.

Similarly, after integration of the additional fields, their expectation values contribute only an overall constant to the K\"ahler potential \eqref{KandW}. The effective K\"ahler potential for the K\"ahler moduli reads 
\begin{equation}
\label{Kahler_pot_Kahler}
    K=K_0-\ln(\mathcal{V})\,.
\end{equation}

In the following, we will use this material to derive the 4d scalar potential, but also the kinetic terms to eventually obtain and discuss the mass spectrum. Prior to this, we first say a word on possible mechanisms to realize the partial moduli stabilization mentioned above.

\subsubsection{Foreword on moduli stabilization}\label{sec:stab}

A complete analysis of moduli stabilization in perturbative heterotic compactifications is beyond the scope of this work and has recently been reviewed in detail in~\cite{Leedom:2025mlr}. Here we summarize the ingredients required for the effective theory developed below and specify the branch of the heterotic landscape in which our analysis applies.
Unlike type IIB compactifications, perturbative heterotic theories possess only the NS 3-form $H$-flux, resulting in a much more restricted landscape of flux vacua. In particular, recent analyses of heterotic flux vacua have shown that obtaining a sufficiently small flux superpotential $W_0$ is highly constrained, with supersymmetric solutions exhibiting important obstructions both at vanishing $W_0$ and in the large-complex-structure regime~\cite{Buchbinder:2026}. Consequently, moduli stabilization relies on a combination of fluxes, gauge bundle effects, worldsheet instantons, threshold corrections and perturbative and non-perturbative corrections to the K\"ahler potential. Since these ingredients are highly model dependent, our goal is not to construct a complete stabilization mechanism, but rather to identify the conditions under which a light axion effective field theory may consistently emerge.

The complex-structure and bundle moduli are assumed to be stabilized at scales well above those relevant for the K\"ahler sector. Several mechanisms capable of achieving this are known. In particular, heterotic compactifications admit stabilization through the interplay between gauge bundle holomorphy and anomalous U(1) D-terms, where the requirement of preserving bundle holomorphy constrains the complex-structure moduli. For anomalous U(1) symmetries the D-term takes the schematic form
\begin{equation}
    D=\frac{c}{S+\bar S}-q_C |C|^2
\end{equation}
where $c$ describes the FI term coefficient and $C$ are a representation of the gauge bundle moduli.
Depending on the compactification and bundle choice, this mechanism can stabilize part or all of the complex-structure sector in supersymmetric Minkowski vacua~\cite{Anderson:2010mh,Anderson:2011cza,Anderson:2011ty,Dundee:2010sb}. Our analysis requires only that these fields acquire masses sufficiently larger than those of the K\"ahler moduli so that they may be integrated out consistently.

In the absence of hidden-sector gaugino condensation, the dilaton ${\rm Re}[S]$ must instead be stabilized through corrections to its K\"ahler potential. Perturbative corrections and Shenker-type non-perturbative effects have been investigated as possible ingredients in dilaton stabilization and supersymmetry breaking in heterotic compactifications~\cite{Shenker:1990uf,Gaillard:2007jr,Leedom:2022zdm}. More recently, the interplay between moduli-dependent threshold corrections, non-perturbative dynamics, and vacuum structure has been explored in heterotic effective theories with automorphic symmetries~\cite{Leedom:2026nby}. Throughout this work we assume that the dilaton is fixed by such effects at a scale above the K\"ahler sector.

Under these assumptions the heavy complex-structure, bundle and dilaton fields can be consistently integrated out. The resulting low-energy theory contains only the K\"ahler moduli ${\tt T}^i$, namely the saxions ${\tt v}^i$ and their axionic partners ${\tt b}^i$, as well as the model-independent axion $a$ related to ${\rm Im}[S]$. In the following, we explicitly determine the axionic mass spectrum and investigate the hierarchy required for a single-field description. The stabilization of the saxionic sector, including its possible backreaction on the light axion potential, is not derived within the present toy model and constitutes an additional consistency requirement.
Our aim is therefore to identify the conditions under which such an axionic effective theory can arise.

\subsubsection{Non-perturbative scalar potential}
We now derive the scalar potential governing the dynamics of the K\"ahler moduli and their associated axions. In the four-dimensional $\mathcal{N}=1$ supergravity description, the scalar potential is entirely specified by the K\"ahler potential and the superpotential. We assume the superpotential to take the form \eqref{superpot}.
The K\"ahler potential in heterotic takes the form \eqref{Kahler_pot_Kahler}.
The 4d F-term scalar potential reads 
\begin{equation}
\label{Ftpot}
    V_F= e^K \left(K^{i\bar{j}} D_iWD_{\bar j}\bar W-3|W|^2\right)\,,
\end{equation}
where we denote
\begin{equation}
    D_iW=\partial_iW+(\partial_iK)W\,,\qquad K_i=\partial_{{\tt T}^i}K=-\frac{\partial_{{\tt T}^i}\mathcal{V}}{\cal V}\,,\qquad K_{i\bar j}=\partial_i\partial_{\bar{j}} K \ .
\end{equation}
To make the axion dependence explicit, we define
\begin{equation}
	X_i\equiv A_i e^{-2\pi\, \mathtt T^i}
	=A_i e^{-2\pi\, \mathtt v^i}e^{\mathrm i\,\mathtt b^i},
	\qquad
	W=W_0+\sum_iX_i \ ,
\end{equation}
where the superpotential was given in \eqref{superpot}. The K\"ahler-covariant derivative then becomes
\begin{equation}
	D_mW=	K_mW_0+\sum_i\left(K_m-2\pi\delta_{mi}\right)X_i.
\end{equation}
Substituting this expression into \eqref{Ftpot} and using the no-scale identity valid for K\"ahler moduli
\begin{equation}
	K^{i\bar j}K_iK_{\bar j}=3,
\end{equation}
the contribution proportional to $|W_0|^2$ cancels. Unlike in type IIB, this cancellation applies to the K\"ahler-moduli sector only after the remaining moduli have been integrated out consistently. 

As a result, we are left in $V_F$ schematically with two types of terms: those proportional to $W_0 X_i$ and to $X_i X_j$. We will assume in the following that the $\mathtt{v^i}$ are stabilized (meaning that ${\cal V}$ is as well, making $K$ in \eqref{Kahler_pot_Kahler} constant): this allows us to compare the values of $|W_0 X_i|$ and $|X_i X_j|$. The terms quadratic in the instanton factors are exponentially suppressed relative to the leading instanton terms, as 
\begin{equation}
	e^{-2\pi \mathtt{v}^i}\ll |W_0| \ ,
\end{equation}
such that we will neglect them. We are then left with the mixed term between $W_0$ and a single instanton: this gives
\begin{equation}
	V_F^{(1)}	=	2e^K	\sum_i	\operatorname{Re}	\Big[	\overline W_0X_i\left(K^{m\bar n}	\left(	K_m-2\pi\delta_{mi}	\right)	K_{\bar n}	-3	\right)\Big].
\end{equation}
Using the no-scale identities, the factor in parentheses becomes
\begin{equation}
	K^{m\bar n}	\left(	K_m-2\pi\delta_{mi}	\right)	K_{\bar n}	-3	=
		K^{m\bar n}K_mK_{\bar n}	-2\pi K^{i\bar n}K_{\bar n}	-3=	3+4\pi \mathtt{v}^i-3=	4\pi \mathtt{v}^i \ ,
\end{equation}
thanks to the relation \eqref{volume} between $\mathcal{V}$ and the $\mathtt{v}^i$. The leading instanton contribution is therefore
\begin{equation}
	V_F^{(1)}	=	8\pi e^K	\sum_i	\mathtt{v}^i	\operatorname{Re} 	\left(	\overline W_0\, A_i\, e^{-2\pi\, \mathtt{v}^i} e^{{\rm i}\, \mathtt{b}^i} \right)\ .
\end{equation}

Writing the relative phase as
\begin{equation}
	\overline W_0\, A_i	=	|W_0\, A_i|\, e^{{\rm i}\, \delta_i}\ ,
\end{equation}
and setting it to be $\pi$ for later convenience, without loss of generality, one obtains 
\begin{equation}
	V_F^{(1)}	=	-\sum_i	8\pi e^K {\tt v}^i \, |W_0A_i|\, e^{-2\pi\, \mathtt{v}^i}	\cos \mathtt{b}^i \ .
\end{equation}
At leading order in the world-sheet instanton expansion, the axion-dependent
part thus reduces to
\begin{equation}
	V_{F,{\rm ax}}	\simeq	-\sum_i	\Lambda_{{\rm WS},i}^4\, \cos \mathtt{b}^i,	\qquad	\Lambda_{{\rm WS},i}^4	=	8\pi e^K	\mathtt{v}^i\, |W_0A_i|\, e^{-2\pi\, \mathtt{v}^i}\ . \label{L4}
\end{equation}

The derivation above assumes a minimal instanton structure, in which each contribution depends on a single K\"ahler modulus. More generally, worldsheet instantons can generate superpotential terms of the form
\begin{equation}
W_{\rm WS}=\sum_A A_A\exp\left(-2\pi q_{Ai}T^i\right),
\end{equation}
where $q_{Ai}$ denote the corresponding instanton charge vectors.

Repeating the leading-order expansion of the F-term scalar potential gives
\begin{equation}
V_{F,\mathrm{ax}}\simeq-\sum_A\Lambda_A^4\cos(q_{Ai}b^i+\delta_A),
\end{equation}
where, under the same assumptions used above,
\begin{equation}
\Lambda_A^4=8\pi e^K(q_{Ai}v^i)|W_0A_A|
\exp(-2\pi q_{Ai}v^i).
\end{equation}

The diagonal ansatz considered previously corresponds to $q_{Ai}=\delta_{Ai}$. In the following, we first investigate this minimal case and its geometric requirements, before considering a simple non-diagonal charge configuration leading to axion alignment.

The axions also couple to gauge sectors through Chern-Simons interactions. After confinement, non-perturbative gauge dynamics generate additional contributions to the potential. The full axion potential may therefore be written schematically as
\beq
V=V_F+V_{QCD}\,.
\eeq
Both worldsheet instantons and strongly coupled gauge sectors generate periodic axion potentials. These contributions can be written uniformly as 
\begin{empheq}[box=\widefbox]{equation}
	V = -\Lambda_{\rm QCD}^4 \cos\!\left( C_I^{{\rm QCD}} \,\theta^I \right) 
	- \sum_\alpha^{\phantom{\alpha}} \Lambda_{{\rm WS},\alpha}^4 \cos\!\left( C^{{\rm WS},\alpha}_I\, \theta^I \right) \equiv -\sum_s\Lambda^4_s\cos\!\left(C_I^{s}\,\theta^I\right) \label{potential}
\end{empheq}
where $\theta^I= a, {\tt b}^1, ..., {\tt b}^{h^{1,1}}$, $C_I^{{\rm QCD}}$ and $C^{{\rm WS},\alpha}_I$ are the charge vectors for QCD and for each 
world-sheet instanton respectively. The QCD charge vector can be read from the visible sector in \eqref{S4df2} as being $C_I^{{\rm QCD}}= (1,n_i)$ for $i=1,...,h^{1,1}$, with $n_i$ defined in \eqref{eq:ni}; with a change of basis one can equivalently introduce $\mathtt{n}_i$ such that $n_i b^i =\mathtt{n}_i \mathtt{b}^i$, allowing us to use either set of fields. Regarding the world-sheet instantons, we consider the situation where each of them couples to a single geometric axion direction, so the corresponding charge vectors take the simple form $C^{{\rm WS},\alpha}_I = \delta_I^{\alpha}$, as can be seen in \eqref{L4}.

\subsubsection{Canonical normalization, masses, decay constants and couplings}\label{sec:cannormmass}

 Having obtained the scalar potential for the axions, we now turn to their kinetic terms. They are given by the K\"ahler metric which is a priori non-diagonal. They read
\begin{equation}
    \frac{M_p^2}{2}\, G_{IJ} \,\del_{\mu}\theta^I \del^{\mu} \theta^J
= \frac{M_p^2}{4\pi^2} \left( K_{S\bar{S}}\, \partial_\mu a\partial^\mu a+
K_{i\bar j}\,
\partial_\mu {\tt b}^i
\partial^\mu {\tt b}^j \right)= \frac12 f_a^2 (\del a)^2 + \frac12 f_i^2 (\del b^i)^2  \,. \label{kinterms4d}
\end{equation}
From $K \supset -\ln(S+\bar S)$ 
one obtains
\begin{equation}
K_{S\bar S}=\frac{1}{4s^2},
\qquad
G_{aa}=\frac{1}{4\pi^2} 2K_{S\bar S}=\frac{1}{4\pi^2} \frac{1}{2s^2}\,,
\end{equation}
and one can verify the matching with $f_a^2$. At tree level the model-independent axion does not kinetically mix with the K\"ahler axions. Thus the full axion metric is block diagonal
\begin{equation}
G
= \frac{1}{4\pi^2}
\begin{pmatrix}
\dfrac{1}{2s^2} & 0 \\
0 & 2 K_{i\bar j}
\end{pmatrix}\,, \label{metricaxion}
\end{equation}
and one needs to diagonalize only $K_{i\bar j}$. Doing so will give the eigenvalues of $K_{i\bar j}$, which should be equal to $\gamma_i^2/(4  {\cal V})$, defined in Section \ref{sec:pertdimred}. This diagonalization corresponds to the change of basis between ${\tt b}^i$ and $b^i$. The precise $K_{i\bar j}$ depends on the Calabi-Yau $X$ chosen: we will thus come back to it in Section \ref{sec:CY}.

To identify the physical axion mass spectrum one must pass to canonically normalized fields. We follow the procedure used in the Mathematica notebook {\tt Lightaxionk3fibredCY}. Rather than diagonalizing and rescaling in separate steps, we employ a Cholesky decomposition $G = L^T L$, which simultaneously canonically normalizes the axion metric while preserving its positive-definite structure. We define $\hat{\vartheta}=L\,\theta$, where $L$ is the Cholesky factor of the axion metric; the canonically normalized fields are $\vartheta=M_p \, \hat{\vartheta}$. The physical squared masses, $\{ m_A^2\}$, are then the eigenvalues of the rotated Hessian
\begin{equation}
	\widetilde{\mathcal{H}}_{\!ess} =\frac{1}{M_p^2} (L^{-T}) \, \mathcal{H}_{\!ess} \, (L^{-1}),\quad\quad(\mathcal{H}_{\!ess})_{IJ} = \left. \frac{\partial^2 V}{\partial \theta^I \partial \theta^J} \right|_{\theta=0}\,.
\end{equation}
These eigenvalues are obtained by diagonalizing $\widetilde{\mathcal{H}}_{\!ess}$ introducing an orthogonal matrix $O$, 
\beq
(m_A^2 )=O\,\widetilde{\mathcal{H}}_{\!ess}\,O^T \ .
\eeq
The physical mass eigenstates are defined as 
\beq 
\phi =M_p O \hat{\vartheta}=  M_p O L \theta
\eeq
The inverse relation 
\beq
\theta=\frac{1}{M_p}L^{-1}O^T\phi
\eeq
is used to rewrite the potential and gauge couplings in terms of the physical axions $\phi^A$. Consider a charge vector $C_I$, appearing in the scalar potential or in the Chern-Simons coupling, the corresponding axion combination becomes 
\beq
C_I^s\theta^I= \frac{1}{M_p}C^sL^{-1}O^T\phi=\sum_A\alpha_{A}^s\frac{\phi^A}{M_p}
\eeq
where $\alpha_{A}^s=\left(C^sL^{-1}O^T\right)_A$. 
The coupling of the mass eigenstate $\phi^A$ to that sector $s$ is therefore determined by $\alpha_{A}^s$, and the corresponding mass-basis decay constant is
\begin{equation}
	f_{A}^s=\frac{M_p}{|\alpha_{A}^s|}\,.
\end{equation}
Consequently, the hierarchy of instanton actions alone does not determine the physical axion spectrum. Canonical normalization mixes the geometric axions, and only the eigenvalues of the canonically normalized Hessian determine the physical masses. In particular, a geometric axion lifted by a very small instanton scale may have substantial overlap with directions lifted by larger world-sheet instantons or by QCD, producing a much heavier mass eigenstate than a naive inspection of the potential would suggest.

Having transformed to the physical basis, all couplings may be rewritten from \eqref{S4df3} in terms of the physical axions $\phi^A$ towards
\begin{equation}
\mathcal L
\supset
\sum_A
\frac{\lambda_{A}^s}{4f_{A}^s}
\phi^A\big({\cal F}_{\mu\nu}^{\mathfrak{a}} \tilde{{\cal F}}^{\mu\nu}_{\mathfrak{a}}\big) \,,
\end{equation}
where the dimensionless couplings \(\lambda_{A}^s\) are obtained by reading off the coefficients of the physical axions after performing the transformation.

\subsection{Summary: 4d model}\label{sec:sum4d}

In Section \ref{sec:pertdimred}, we derived a 4d theory of $h_X^{1,1}+1$ axions $\theta^I$ by dimensional reduction of 10d perturbative heterotic string on a 6d Calabi-Yau manifold $X$. The corresponding action is the one in \eqref{S4df3} without the two terms ${\cal S}_{{\rm curv}},\, {\cal S}_{\bar{{\cal F}}} $. Those two were discussed in Section \ref{sec:curv}, and we will come back to their impact. To that 4d action, one should add a non-perturbative scalar potential derived in Section \ref{sec:pot}, given in \eqref{potential}. In the resulting multifield 4d theory, one can compute the mass spectrum, proceeding as explained in Section \ref{sec:cannormmass}. In this way, one also determines the mass eigenstates, which are nothing but the physical axions. Those are linear combinations $\phi^A$ of the initial $\theta^I$. One can then rewrite the whole 4d theory in terms of the $\phi^A$ instead, determining in particular their decay constant $f_A$ and couplings $\lambda_A$. The numerical values depend on the details of the Calabi-Yau chosen. In Section \ref{sec:CY}, we will select one $X$ and give explicitly the 4d model of interest. Prior to this, there is one remaining step to eventually reach a 4d effective theory for a single, light field, as we now explain.

Having determined the full axion spectrum and couplings, we can now consider the regime in which one physical mass eigenstate is much lighter than all other axions. 
After canonical normalization and diagonalization of the Hessian, a configuration of interest contains a physical mass eigenstate $\phi$ satisfying 
\beq
m_\phi\ll m_A\,,\quad \text{with } A=2,\ldots, h^{1,1} + 1\,,
\eeq
where $m_A$ denotes the masses of the remaining axions. Such a hierarchy requires a compactification in which one instanton scale is much lighter than the others, along a geometric state whose kinetic mixing is negligible. If such a configuration can be realized for a suitable Calabi-Yau, then at energies relevant for late-time cosmology, the heavy fields may be integrated out, leaving an effective theory for a single light axion $\phi$ coupled to the visible and the dark sector as described by the following 4d model
\begin{empheq}[box=\widefbox]{align}
{\cal S}_{4d} &= \int \d^{4}x \sqrt{|g_{4}|} \left( \frac{M_p^2}{2} {\cal R}_{4} -\frac{1}{4}\left( {\cal F}_{\mu\nu}^{\mathfrak{a}} {\cal F}^{\mu\nu}_{\mathfrak{a}}\right)_v -\frac{1}{4}\left( {\cal F}_{\mu\nu}^{\mathfrak{a}} {\cal F}^{\mu\nu}_{\mathfrak{a}}\right)_h  \right) \nn\\
& + \int \d^{4}x \sqrt{|g_{4}|} \, \Bigg(\! -  \frac12\, (\del \phi)^2   -V_{\rm eff}(\phi) \Bigg) \label{S4dfinal}\\
&+ \int \d^{4}x \sqrt{|g_{4}|}\, \Bigg(\frac{\lambda_{v}}{4 f} \, \phi \, \big({\cal F}_{\mu\nu}^{\mathfrak{a}} \tilde{{\cal F}}^{\mu\nu}_{\mathfrak{a}}\big)_v +\frac{\lambda_{h}}{4 f} \, \phi \, \big({\cal F}_{\mu\nu}^{\mathfrak{a}} \tilde{{\cal F}}^{\mu\nu}_{\mathfrak{a}}\big)_h \Bigg) \nn
\end{empheq}
The effective scalar potential is given by $V_{\rm eff}= \Lambda^4\cos\left(\frac{\phi}{f}\right)+V_c$, where $V_c$ is a constant and $\Lambda^4$ is the scale of the lightest world-sheet instanton. The change of sign with respect to \eqref{potential} is obtained by a shift in $\phi$; the minimum is now at $\phi= \pi\, f$, and the axion mass obeys the well-known relation $m_{\phi}^2\, f^2 = \Lambda^4$.

In sections \ref{sec:curv} and \ref{sec:pot}, we have encountered various constant contributions to the scalar potential; these also include the terms in the potential \eqref{potential} for each stabilized heavy axion. All these constant contributions were not necessarily found negligible (see e.g.~Section \ref{sec:curv}). Their precise value depends on the details of moduli stabilization, which goes beyond the scope of this paper. To illustrate the possible cosmological solutions obtained in the present setting, we will consider in Section \ref{sec:cosmo} two simple cases: $V_c=0$ or $V_c=\Lambda^4$, that is
\bea
\text{Hilltop-like:}&&\quad V_{{\rm Hill}} = \Lambda^4 \,\cos \left(\frac{\phi}{f} \right) \ , \label{VHill0}\\
\text{Axionic:}&&\quad V_{{\rm Ax}} = \Lambda^4 \, \left( 1 +\cos \left(\frac{\phi}{f} \right) \right) \ . \label{VAx0}
\eea

The validity of this single field effective description relies on several assumptions, that we recapitulate here. First, all non-axionic moduli, including the saxionic partners of the K\"ahler axions, are assumed to be stabilized at masses sufficiently above the cosmological scales of interest $m_{\rm heavy }\gg H_0$. Under this condition, their fluctuations are negligible during the late-time cosmological evolution, and integrating them out generates only higher dimensional operators, suppressed by powers of $m_{\rm heavy}^{-1}$. Second, the remaining axionic directions must be sufficiently massive to be integrated out. Unlike the saxionic sector, this hierarchy can be investigated explicitly within the multifield axion effective theory, and will be verified for the illustrative benchmark considered below.  Third, the stabilization mechanism is assumed to fix the expectation values of the dilaton, complex-structure and bundle moduli, allowing the gauge couplings and the parameters of the axionic effective theory to be treated as approximately constant. Finally, the constant vacuum-energy contribution is treated as an independent parameter, as discussed above. The analysis therefore focuses solely on the (light) axion-dependent part of the potential.

Getting the adequate hierarchy among axions depends on the Calabi-Yau choice; we now focus on this matter. In turn, this will fix all constants in the 4d multifield axion theory made up of the action \eqref{S4df3} and the potential \eqref{potential}, and eventually determine the effective single-field model \eqref{S4dfinal}.

\section{Calabi-Yau selection }\label{sec:CY}

In this section, we investigate the geometric conditions under which heterotic compactifications may give rise to an effective theory containing a single ultralight axion presented in Section \ref{sec:sum4d}. We first explore explicit toric Calabi–Yau hypersurfaces and assess the compatibility of a large instanton action with a bounded overall volume and geometric control in Section \ref{sec:toric}. Motivated by the difficulties encountered in this search, we subsequently consider an illustrative K3-fibred volume form in Section \ref{sec:K3}, which allows us to construct and analyze a minimal axion effective theory. The latter provides a benchmark for the cosmological analysis, rather than a fully specified Calabi–Yau compactification with an explicitly constructed gauge bundle.
Together with an adequate gauge bundle, we will compute in Sections \ref{sec:4dmodelCY} and \ref{sec:curveffectf} the resulting quantities appearing eventually in the 4d single field model. This illustrative model will then be used in Section \ref{sec:cosmo} for cosmology.

Our aim is to identify the simplest compactification geometry capable of producing a single ultralight axion through the exponential suppression of a worldsheet-instanton contribution. Within our diagonal instanton ansatz, this motivates considering an anisotropic compactification with one sufficiently large two-cycle volume, while the remaining cycles are kept comparatively small and under perturbative control. We deliberately focus on a single large cycle, rather than more general configurations involving several instanton charge vectors or axion alignment, in order to isolate the geometric requirements of the minimal mechanism.

As discussed in \cite{Leedom:2025mlr}, such anisotropic configurations face non-trivial constraints. In particular, the overall compactification volume must remain bounded, while all effective curves should remain sufficiently large for the perturbative description to be reliable. We therefore impose the following geometric criteria:
\begin{enumerate}
\item
\label{crit:big_v}The existence of an effective curve class with sufficiently large volume, along which the instanton scale is exponentially suppressed, as in \eqref{L4}. In the toric scan below, generators of the toric Mori cone are used as a finite proxy for such classes.
\item \label{crit:small_mix}Sufficiently small off-diagonal K\"ahler metric components involving the  anisotropic direction, such that kinetic mixing does not spoil the axion mass hierarchy.
\item \label{crit:small_V}A bounded overall volume $\mathcal{V}\lesssim 25$, as required by the constraint \eqref{boundVol}.
\item \label{crit:K}The K\"ahler moduli lie strictly inside the K\"ahler cone. Equivalently, all effective curve volumes are positive, ensuring that the compactification geometry remains non-degenerate and that no cycle collapses to zero size.
\item \label{crit:vg1}The volumes of all generators of the toric Mori cone and all prime toric divisors are at least one in string units, such that $\alpha'$-corrections can remain under perturbative control.
\end{enumerate}
The five criteria listed above severely restrict the class of viable Calabi–Yau geometries. Prior to searching for them, let us detail a few useful notations. We expand the K\"ahler form $J$ in a basis $\beta_i$ of $H^{1,1}(X)$ as 
\begin{equation}
    J={\tt v}_i\, \beta_i\,.
\end{equation}
The Calabi-Yau volume is given by
\begin{equation}
\label{volume}
    \mathcal{V}
    =
    \frac16 k_{ijk}\, {\tt v}_i {\tt v}_j {\tt v}_k \, ,
\end{equation}
where
\begin{equation}
    k_{ijk}
    =
    \int_X \beta_i \wedge \beta_j \wedge \beta_k
\end{equation}
are the triple intersection numbers. The four-cycle volumes are then defined as
\begin{equation}
    \tau_i
    =
    \frac{\partial \mathcal{V}}{\partial {\tt v}_i}
    =
    \frac12 k_{ijk}\, {\tt v}_j {\tt v}_k \, .
\end{equation}
The Mori-cone is the cone of effective curve classes $C_a$, whose volumes are given by 
\begin{equation}
\label{Mori_cone_vol}
    {\rm Vol}(C_a)=\int_{C_a}J={\tt v}_i\int_{C_a}\beta_i\,.
\end{equation}
As explained below \eqref{superpot}, we use here ${\tt v}^i$ instead of $v^i$ to denote the real part of K\"ahler moduli, corresponding to 2-cycle volumes in string units as in \eqref{betavol}. Their definition is analogous and the two only differ by a change of basis. Indeed, the $v^i$ use a basis of harmonic 2-forms $\beta_i$ that are orthogonal, providing a diagonal K\"ahler metric; this is not assumed for ${\tt v}^i$. Technically, we should then use in this section a different notation for (a priori non-orthogonal) $\beta_i$ and the corresponding numbers $n_i$, to distinguish notations from those of Section \ref{sec:10d4d}. For simplicity we do not, as there should not be any ambiguity. Finally, in this section, we also lower the indices ${\tt v}^i \rightarrow {\tt v}_i$ for convenience.

\subsection{Exploratory scan of toric hypersurfaces}\label{sec:toric}

\subsubsection{Strategy and results}

In isotropic Calabi–Yau geometries, parametrically large two-cycle volumes imply a large overall volume. In Swiss-cheese geometries, keeping the total volume finite while increasing an individual two-cycle volume drives an effective curve towards the boundary of the K\"ahler cone~\cite{Leedom:2025mlr}. These asymptotic obstructions motivate a search for finite-volume configurations satisfying the geometric-control criteria
\begin{equation}
\max_a {\rm Vol}(C_a)\geq30 \ .
\end{equation}
This is a finite, basis-independent proxy for a large instanton-supporting effective curve; it does not enumerate all effective curve classes or determine their worldsheet-instanton prefactors.

To address this question, we performed an exploratory numerical scan using \texttt{CYTools}~\cite{Demirtas:2022hqf} over favorable toric hypersurface Calabi-Yau threefolds with
\begin{equation}
h^{1,1}=2,3,4.
\end{equation}
We considered at most 1000 favorable reflexive polytopes for each value of $h^{1,1}$. For each polytope, we generated up to 20 random fine, regular, star triangulations using the fast \texttt{CYTools} sampler, removed phases certified by \texttt{CYTools} to be trivially equivalent, and sampled K\"ahler-cone directions using Dirichlet parameters $0.3$, $1$ and $4$. For each parameter we drew 200 directions per inequivalent phase and seed. Along each direction we used the homogeneous scaling of curve, divisor and Calabi-Yau volumes with the radial K\"ahler coordinate to determine analytically whether the weak conditions
\begin{equation}
\max_a {\rm Vol}(C_a)\ge30 \ ,
\qquad
\mathcal V<25 \ ,
\label{bound_vol}
\end{equation}
could simultaneously be satisfied.

This first stage constitutes only a coarse geometric preselection. On each weakly selected direction, rather than in an independent second sample, we then imposed
\begin{equation}
\min_a {\rm Vol}(C_a)\geq1\,,\qquad
\min_A {\rm Vol}(D_A)\geq1\,,
\label{full_geometric_control}
\end{equation}
where $C_a$ run over the generators returned by the toric Mori cone and $D_A$ over all prime toric divisors. Thus the weak and controlled counts form a single sequential cut flow. A surviving configuration would subsequently require evaluation of the K\"ahler metric, canonical normalization and the physical axion spectrum, including criterion \ref{crit:small_mix}.

The complete scan was repeated with the five fixed seeds listed in Table \ref{tab:scan_summary}. The seed changes both the sampled triangulations and the K\"ahler-cone directions. The code records one summary file per seed and an aggregate table; only fully controlled candidates are exported individually.
Across the five seeds, we sampled 4,305,000 K\"ahler-cone directions as shown in Table~\ref{tab:scan_summary}. Of these, 45,354 passed the weak preselection, but none satisfied the full geometric-control conditions~\eqref{full_geometric_control}. Five direction evaluations were discarded due to non-positive volumes returned by floating-point arithmetic. No candidate therefore remained for the subsequent analysis of the canonically normalized axion spectrum.

Although the scan does not constitute a no-go theorem, the absence of controlled candidates indicates a strong tension between a single large effective curve, bounded compactification volume and geometric control in the sampled toric hypersurfaces. This result does not constrain more general instanton configurations, which we investigate below.

\begin{table}[t]
\centering
\small
\begin{tabular}{@{}rrrrrr@{}}
\toprule
\textbf{Seed} & \textbf{Triang.} & \textbf{Phases} & \textbf{Directions} & \textbf{Weak (fraction)} & \textbf{Controlled} \\
\midrule
20260713 & 1925 & 1435 & 861,000 & 9159 (1.1\%) & 0 \\
20260714 & 1923 & 1435 & 861,000 & 8922 (1.1\%) & 0 \\
20260715 & 1931 & 1435 & 861,000 & 8940 (1.1\%) & 0 \\
20260716 & 1939 & 1435 & 861,000 & 9103 (1.1\%) & 0 \\
20260717 & 1925 & 1435 & 861,000 & 9230 (1.1\%) & 0 \\
\midrule
\textbf{Total} & --- & --- & 4,305,000 & 45,354 (1.1\%) & 0 \\
\bottomrule
\end{tabular}
\caption{Five-seed summary of the sequential exploratory scan. Every seed covers 1279 favorable polytopes. The weak preselection requires $\max_a{\rm Vol}(C_a)\ge30$ and $\mathcal V<25$; the controlled column then applies $\min_a{\rm Vol}(C_a)\ge1$ and $\min_A{\rm Vol}(D_A)\ge1$ to those same directions. The reproducible implementation of the toric scan is provided in the ancillary files \textit{cytools\_reproducible\_scan.py} and \textit{Scan\_CYtools\_large\_v\_small\_V.ipynb}. }
\label{tab:scan_summary}
\end{table}

\subsubsection{The closest candidate}\label{sec:toriccandidate}

Among all geometries satisfying the weak preselection, the configuration coming closest to fulfilling the complete set of geometric requirements is the resolved degree-$12$ hypersurface
\begin{equation}
X_{12}\subset \mathbb P^4_{1,1,2,2,6}
\end{equation}
with
\begin{equation}
(h^{1,1},h^{2,1})=(2,128).
\end{equation}
In the divisor basis returned by \texttt{CYTools}, the only non-vanishing triple intersection numbers are
\begin{equation}
k_{122} =k_{212}=k_{221}=2\ ,
\qquad
k_{222}=-8 \ ,
\end{equation}
so that the Calabi-Yau volume is
\begin{equation}
\mathcal{V}=\frac{1}{6} k_{i j k}\, {\tt v}_i {\tt v}_j {\tt v}_k={\tt v}_1 {\tt v}_2^2-\frac{4}{3} {\tt v}_2^3={\tt v}_2^2\left({\tt v}_1-\frac{4}{3} {\tt v}_2\right)\,.
\end{equation}
The basis 4-cycle volumes are 
\begin{equation}
\tau_1=\frac{\partial \mathcal{V}}{\partial \,{\tt v}_1}={\tt v}_2^2\ ,\qquad \tau_2=\frac{\partial \mathcal{V}}{\partial \,{\tt v}_2}=2 {\tt v}_1 {\tt v}_2-4 {\tt v}_2^2 \ ,
\end{equation}
while a generating set for the toric Mori cone returned in the same integral basis is
\begin{equation}
r_1=(0,1)\,,\qquad r_2=(1,0)\,,\qquad r_3=(1,-2)\,.
\label{eq:h11twoMoriRays}
\end{equation}
Their volumes, given by \eqref{Mori_cone_vol}, are
\begin{equation}
\operatorname{Vol}\left(C_a\right)=\left({\tt v}_2, {\tt v}_1, {\tt v}_1-2 {\tt v}_2\right)\,.
\end{equation}

The best point identified by the scan is
\begin{equation}
({\tt v}_1,{\tt v}_2)\simeq(30.1,\,1.0)\,,\qquad\mathcal V\simeq25.0\,,
\end{equation}
for which
\begin{equation}
\operatorname{Vol}(C_a)\simeq(1.0,\,30.1,\,28.2)\,,
\end{equation}
and
\begin{equation}
\tau_{\min}=\tau_1\simeq0.9\,.
\end{equation}
Thus, even the closest finite-volume candidate lies outside the stretched K\"ahler cone according to our control criteria.

The origin of this obstruction can be understood: since
\begin{equation}
\frac{\partial\mathcal V}{\partial\, {\tt v}_1}={\tt v}_2^2>0\,,
\end{equation}
the Calabi-Yau volume increases monotonically with the large cycle throughout the physical K\"ahler cone. Consequently, for fixed upper bound
\begin{equation}
\mathcal V<25,
\end{equation}
increasing the large cycle necessarily decreases the largest attainable value of the small cycle. Imposing ${\tt v}_1\ge30$ and $\mathcal V\le25$ gives
\begin{equation}
30\,{\tt v}_2^2-\frac43{\tt v}_2^3\le25\,,
\end{equation}
whose maximal physical solution is
\begin{equation}
{\tt v}_2^{\rm max}\simeq1.0\,.
\end{equation}
Therefore no point in this Calabi-Yau phase can simultaneously satisfy
\begin{equation}
{\tt v}_1\ge30\ ,\qquad\mathcal V<25\ ,\qquad\operatorname{Vol}(C_a)\ge1 \ ,
\end{equation}
for all Mori-cone generators. This provides a finite-volume realization of the same geometric obstruction identified asymptotically in~\cite{Leedom:2025mlr}. While the previous argument established that effective curves collapse in the limit of parametrically large cycle volumes, the present numerical analysis shows that even moderate hierarchies, such as those required by our phenomenological construction, remain incompatible with geometric control in the closest candidate identified by the scan.

Although this geometry has no point satisfying simultaneously the single-large-cycle requirement and the control criteria of the minimal construction, it is instructive to examine whether its intersection structure can nevertheless support the kinematics of a more general instanton configuration.
As an illustrative choice, consider two worldsheet-instanton charge vectors \(q_H=(3,1)\) and \(q_L=(4,1)\). Both lie in the toric Mori cone \eqref{eq:h11twoMoriRays}. The former fixes the axionic combination \(3b^1+b^2\), leaving the primitive direction \(d=(1,-3)\), along which \(q_H\cdot d=0\) and \(q_L\cdot d=1\). At the point \(({\tt v}_1,{\tt v}_2)=(10.8,1.5)\), we obtain
\begin{equation}
\mathcal V\simeq18.7\,,\qquad
\operatorname{Vol}(C_a)\simeq(1.5,10.8,7.9)\,,\qquad
\min_A\operatorname{Vol}(D_A)\simeq2.2\,.
\end{equation}
Thus this aligned point, unlike the point optimized for the single-large-cycle preselection, lies inside the stretched region used in the scan. The instanton actions are \(S_H\simeq212.0\) and \(S_L\simeq279.6\), and the effective decay constant is \(f_{\mathrm{eff}}\simeq0.4M_{\mathrm P}\), using the heterotic K\"ahler metric derived from \(K=-\ln\mathcal V\). This explicit intersection structure is therefore kinematically compatible with the hierarchy of instanton actions and the axionic normalization required by the illustrative quintessence scenario discussed below.

This observation should not be interpreted as a microscopic realization of the proposed superpotential. Indeed,
\begin{equation}
\det\!\begin{pmatrix}3&1\\4&1\end{pmatrix}=-1\,,
\qquad q_L-q_H=(1,0)=r_2\,.
\end{equation}
The selected charges therefore form an integral lattice basis, and the light direction is not protected by the charge lattice itself. Each of the lower-action classes in \eqref{eq:h11twoMoriRays} has a non-vanishing projection along \(d\). At the aligned point their actions and projections are
\begin{equation}
\begin{array}{c|ccc}
Q &(0,1)&(1,0)&(1,-2)\\ \hline
2\pi Q\cdot {\tt v}&9.2&67.7&49.4\\
Q\cdot d&-3&1&7
\end{array}\,.
\label{eq:h11twoDangerousClasses}
\end{equation}
The genus-zero Gopakumar--Vafa invariants were computed using \texttt{CYTools} 1.4.8~\cite{Demirtas:2022hqf} with its \texttt{cygv} 0.2.2 backend~\cite{RiosTascon:cygv}, based on the algorithm of \cite{Demirtas:ComputationalMirrorSymmetry}. For these classes they are non-zero,
\begin{equation}
n^0_{(0,1)}=2496\,,\qquad
n^0_{(1,0)}=1\,941\,264\,,\qquad
n^0_{(1,-2)}=2\,.
\end{equation}
The fixed triangulation, cone data and invariant calculation are reproduced by the ancillary script \path{cytools_alignment_candidate.py}.
Consequently, these classes cannot be excluded on geometric grounds alone and would generically generate contributions much larger than the one associated with \(q_L\). A viable heterotic realization requires their contributions to vanish through fermionic zero modes or cancellations, while retaining non-zero contributions in the selected classes. This is a condition on isolated rational curves and on the Pfaffians determined by the gauge bundle and complex-structure moduli~\cite{BuchbinderDonagiOvrut2002,BeasleyWitten2003,Curio2010}; a non-zero Gopakumar--Vafa invariant by itself does not establish a non-zero heterotic superpotential. Establishing the required selection therefore lies beyond a purely toric scan.

We also performed a finite search for aligned instanton charge configurations on favorable toric Calabi-Yau hypersurfaces, covering 243 polytopes with \(h^{1,1}=3\) and 36 with \(h^{1,1}=2\), with one fine, regular, star triangulation per polytope and a total of 27,900 sampled K\"ahler directions. The search identified integral charge systems satisfying the stated geometric-control conditions and yielding phenomenologically relevant effective decay constants and selected instanton actions. After retaining one candidate per successful geometry, 55 candidates were subjected to a complete genus-zero invariant audit of the toric-cone lattice classes through the selected light-instanton action. Every retained candidate exhibited lower-action classes with nonzero projection on the proposed light direction and nonvanishing genus-zero Gopakumar-Vafa invariants. Thus, the search supports the geometric compatibility of the desired axion kinematics, but does not establish the required instanton selection, which remains dependent on an explicit gauge-bundle construction. Reproducibility files and detailed candidate data are provided in the ancillary archive \path{alignment_search_20261021_results.zip}.

The two-modulus example shows that aligned instanton charges can yield the required axion kinematics within the geometric-control region, although the suppression of competing instantons remains unresolved. Its K3-fibred structure motivates considering a three-modulus K3-fibred effective ansatz, with an additional blow-up direction, to investigate the axion spectrum and gauge-sector constraints.

\subsection{K3-fibred geometry}\label{sec:K3}

\subsubsection{A well-motivated example}

We consider a three-modulus K3-fibred volume ansatz with an additional diagonal blow-up mode, motivated by the LVS constructions of~\cite{Cicoli:2008va}. The anisotropic structure allows us to investigate the axion kinetic matrix and the hierarchy of instanton actions while keeping the total volume bounded. We treat this geometry as an effective ansatz, rather than a fully specified Calabi–Yau compactification.
\begin{equation}
\label{eq:volk3fibre}
    \mathcal{V} = \xi_1 {\tt v}_1 {\tt v}_2^2+\xi_2 {\tt v}_2^3+\xi_3 {\tt v}_3^3 \, ,
\end{equation}
where ${\tt v}_3$ is associated to a blow-up 2-cycle. The large worldsheet-instanton suppression relevant for the light axion is obtained by taking ${\tt v}_1\gg1$ while keeping the total volume bounded. This is achieved by scaling appropriately ${\tt v}_2$ and ${\tt v}_3$. The axionic kinetic matrix is determined by \eqref{metricaxion}
\begin{equation}
    G_{ij}=\frac{1}{4\pi^2}2K_{ij}=\frac{1}{8\pi^2}\frac{\partial^2 K}{\partial {\tt v}^i\partial {\tt v}^j}.
\end{equation}
The kinetic matrix reads
\begin{equation}
\begin{aligned}
     G_{11}&=\frac{1}{8\pi^2}\frac{\xi_1^2 {\tt v}_2^4}{\left(\xi_1 {\tt v}_1 {\tt v}_2^2+\xi_2 {\tt v}_2^3+\xi_3 {\tt v}_3^3\right){}^2}\,,\\
G_{22}&=\frac{1}{8\pi^2}\frac{{\tt v}_2^2 \left(2 {\tt v}_1^2 \xi_1^2+4 {\tt v}_1 {\tt v}_2 \xi_1 \xi_2+3 {\tt v}_2^2 \xi_2^2\right)-2 {\tt v}_3^3 \xi_3 ({\tt v}_1\xi_1+3 {\tt v}_2 \xi_2)}{\left({\tt v}_1 {\tt v}_2^2 \xi_1+{\tt v}_2^3 \xi_2+{\tt v}_3^3 \xi_3\right)^2}\,,\\
 G_{12}&= \frac{1}{8\pi^2} \frac{\xi_1 {\tt v}_2\left(\xi_2 {\tt v}_2^3-2 \xi_3 {\tt v}_3^3\right)}{\left(\xi_1 {\tt v}_1 {\tt v}_2^2+\xi_2 {\tt v}_2^3+\xi_3 {\tt v}_3^3\right){}^2}\,.
\end{aligned}
\end{equation}
In the regime in which the blow-up contribution is subleading, the (${\tt v}_1, {\tt v}_2$) sector controls the leading structure of the K\"ahler metric.
For ${\tt v}_1\gg {\tt v}_2\gtrsim {\tt v}_3 $, the dominant contribution is $\mathcal{V}\simeq \xi_1 {\tt v}_1 {\tt v}_2^2$, and the components reduce to
\begin{equation}
	G_{11}\sim \frac{1}{8\pi^2}\frac{1}{{\tt v}_1^2}\,,\qquad G_{22}\sim \frac{1}{4\pi^2 }\frac{1}{{\tt v}_2^2}\,,\quad G_{12}\sim\frac{1}{8\pi^2}\frac{\xi_2}{ \xi_1{\tt v}_1^2}  \,.\label{Gapprox}
\end{equation}
A convenient measure of the physical kinetic mixing is the dimensionless ratio $\epsilon_{\rm kin}\equiv \frac{G_{12}}{\sqrt{G_{11}G_{22}}}$, which at first order reads
\begin{equation}
    \epsilon_{\rm kin}\equiv \frac{G_{12}}{\sqrt{G_{11}G_{22}}}\sim \frac{{\tt v}_2\xi_2}{\sqrt{2}{\tt v}_1\xi_1}\ll1\,,
\end{equation}

The kinetic matrix at leading order is therefore diagonal in the ${\tt v}_1,\,{\tt v}_2$ directions, such that the axion  associated with ${\tt v}_1$ remains approximately aligned with the kinetic eigenstate. Including the blow-up term generates additional off-diagonal entries, but in the regime where the blow-up cycle is small, ${\tt v}_3 \ll {\tt v}_1$, these are suppressed by the relative size of the blow-up contribution.
This increases the likelihood that the light axion survives after canonical normalization. 

We consider the three-modulus K3-fibred volume form motivated by the LVS constructions of \cite{Cicoli:2008va}. We stress that the coordinates ${\tt v}_i$ are adapted to the fibred structure and need not correspond to an integral basis of $H^{1,1}(X,\mathbb{Z})$. Consequently, the coefficients
\begin{equation}
k_{122}=k_{212}=k_{221}=2\xi_1\,,\qquad k_{222}=6\xi_2\,,\qquad
k_{333}=6\xi_3\,.
\end{equation}
may be fractional and should not be interpreted as the triple intersection numbers in an integral divisor basis. Rather, the volume form is used as an effective geometric ansatz capturing the relevant features of a K3 fibration with a diagonal blow-up mode.
The first two four-cycle volumes follow directly as
\begin{equation}
\tau_1=\frac{\partial\mathcal{V}}{\partial {\tt v}_1}=\xi_1 {\tt v}_2^2 \, ,
\quad
\tau_2=\frac{\partial\mathcal{V}}{\partial {\tt v}_2}={\tt v}_2\left(2\xi_1{\tt v}_1+3\xi_2{\tt v}_2\right)\,.
\end{equation}
For the blow up direction, we chose ${\tt v}_3>0$ to denote its size. Since the direction enters the volume with negative orientation, $\xi_3<0$, the positive volume of the corresponding divisor is
\begin{equation}
    \tau_3=-\frac{\partial\mathcal{V}}{\partial {\tt v}_3}=
-3\xi_3{\tt v}_3^2 >0\,.
\end{equation}
For positive ${\tt v}_i$, positivity of these four-cycle volumes requires $\xi_1>0$, $2\xi_1{\tt v}_1+3\xi_2{\tt v}_2>0$, and $\xi_3<0$. Thus $\xi_2>0$ is a sufficient simplifying choice used below, but is not a general consequence of divisor positivity; the blow-up sign reflects the convention that ${\tt v}_3$ parametrizes the size of the blow-up divisor.

Expressing the volume in terms of the four-cycle moduli yields the familiar Swiss-cheese form
\begin{equation}
\mathcal{V}=\alpha\left[\sqrt{\tau_1}\left(\tau_2-\beta\,\tau_1\right)-\gamma\,\tau_3^{3/2}\right],
\end{equation}
where
\begin{equation}
\alpha=\frac{1}{2\sqrt{\xi_1}}\ ,\qquad\xi_1=\frac{1}{4\alpha^2}\ ,\qquad\xi_2=\frac{\beta}{4\alpha^2}\ ,\qquad\xi_3=-\frac{1}{3^{3}(\alpha\gamma)^2}\ .
\end{equation}
For concreteness, we adopt the normalization \cite{Cicoli:2008va}
\begin{equation}
\alpha=1\ ,\qquad\beta=\frac12\ ,\qquad\gamma=1\ .
\end{equation}
For this choice, the Calabi-Yau volume expressed in terms of the 2-cycle moduli becomes
\begin{equation}
\mathcal{V}=\left(\frac14{\tt v}_1+\frac18{\tt v}_2\right){\tt v}_2^2-\frac{1}{3^{3}}{\tt v}_3^3\,.
\end{equation}

Plugging in the values for the $\xi$s in Eq.~\eqref{Gapprox}, in the limit where ${\tt v}_1\gg {\tt v}_2,{\tt v}_3$, we find $\epsilon_{\rm kin}\sim \frac{{\tt v}_2}{2\sqrt{2} {\tt v}_1}$,
showing the absence of kinetic mixing, thus satisfying criterion \ref{crit:small_mix}. The remaining criteria are not automatically guaranteed and will be verified for a suitable point in the K\"ahler moduli space in Section \ref{sec:4dmodelCY}.

\subsubsection{Gauge bundle and topological coefficients}\label{sec:gauge}

Having identified a K3-fibred volume ansatz exhibiting the desired geometrical features, we now turn to the gauge bundle, whose topological data also enter the 4d effective theory.
The second Chern classes are defined as\footnote{\label{foot:trRR}The characteristic forms constructed from two smooth connections differ by an exact Chern-Weil transgression term. Their integrals against a closed $\beta_i$ on compact $X$ without boundary therefore agree, as reviewed in Appendix \ref{ap:curv}. This cohomological statement does not imply equality of the corresponding local curvature densities.}
\begin{equation}
c_2(V)= \frac{1}{8\pi^2} \tr\left( \bar{{\cal F}} \w \bar{{\cal F}} \right)\,,\quad c_2(T_X)= \frac{1}{8\pi^2} \tr\left( \bar{{\cal R}} \w \bar{{\cal R}} \right)\,.
\end{equation}
The anomaly cancellation condition~\cite{Green:1984sg}
\begin{equation}
    c_2(V_1) + c_2(V_2) = c_2(T_X)
\end{equation}
must then be satisfied in the absence of NS5-branes: this corresponds to the tadpole cancellation \eqref{tadpole} (see Footnote \ref{foot:trRR}). Contrary to the standard embedding~\cite{Witten:1985bz}, where $V_1=T_X$ and $V_2$ is trivial, this condition does not uniquely determine the second Chern classes of the two gauge bundles. The four-dimensional gauge group is instead given by the commutants of the respective bundle structure groups inside the two $E_8$ factors. An adequate choice of these groups and bundles is assumed, in particular in the hidden sector, to avoid unwanted non-perturbative contributions to the scalar potential.

In this case, the topological coefficients $n_i$ \eqref{eq:ni} read
\begin{equation}
    n_i=\int_X \beta_i\wedge \frac14\left(c_2(V_1)-c_2(V_2)\right)\,.
\end{equation}
The difference of the second Chern classes can be expanded in the dual basis of harmonic $(2,2)$-forms,
\begin{equation}
    c_2(V_1)-c_2(V_2) = d^i\,\hat D_i\,,
\end{equation}
where the basis $\hat D_i\in H^{2,2}(X)$ satisfies
\begin{equation}
    \int_X \beta_i \wedge \hat D_j = \delta_{ij}
    \qquad \Rightarrow \qquad
    n_i=\frac14\int_X\beta_i\wedge
    \left(c_2(V_1)-c_2(V_2)\right)=\frac14 d^i\,.
\end{equation}
Therefore, once the Calabi-Yau topology, the divisor basis and the gauge bundles are specified, the coefficients $n_i$ are completely determined. Compared to the standard embedding, the non-standard embedding allows for additional freedom in these coefficients, while their values remain subject to the anomaly cancellation condition and to the existence of consistent gauge bundles.\footnote{An alternative would have been the so-called {\it symmetric embedding}, where $c_2(V_1)=c_2(V_2)=\frac12 c_2(T_X)$. This implies $n_i=0$, which sets the corresponding model-dependent axion couplings $\lambda_{i,v/h}$ to zero and simplifies drastically the scalar potential \eqref{potential}, with $C_I^{\rm QCD}=(1,0,\ldots,0)$. While such a simpler setting is appealing, it would not provide the mixing between the model-independent and model-dependent axions considered in this work. We therefore focus on a more general non-standard embedding with non-vanishing $n_i$.}

We work with the standard $h^{1,1}=3$ LVS K3-fibred toy geometry, which is specified only through its intersection structure and is not associated with a fully explicit Calabi-Yau construction. Consequently, neither the individual second Chern classes of the gauge bundles nor their difference are uniquely determined.

In the minimal scenario, a large two-cycle volume suppresses the corresponding axion decay constant, while small kinetic mixing limits the enhancement from the other axionic directions. Consequently, obtaining a sizeable effective decay constant for the lightest mass eigenstate requires a sufficiently large topological coefficient \(n_1\). For the benchmark considered here, requiring \(f_{\rm eff}\simeq0.4M_p\) after canonical normalization and diagonalization of the full mass matrix leads to \(n_1=77\). This value is fixed by phenomenology; however, the one-loop gauge kinetic functions must remain positive in both gauge sectors, imposing 
\begin{equation}
    |n_i {\tt v}^i|<s\,.
\end{equation}
For the moduli values considered here, satisfying this bound would require a substantial cancellation from the remaining topological contributions. Such a cancellation would involve comparatively large coefficients of opposite sign, determined by the differences of the second Chern classes of the visible and hidden gauge bundles. Such topological data cannot be freely prescribed, since the two gauge bundles must simultaneously satisfy the integrated Bianchi identity, anomaly cancellation and the conditions required for supersymmetric bundle stability. 

Although these conditions do not provide a universal upper bound on the individual coefficients $n_i$, realizing such a large difference while maintaining a consistent gauge-bundle configuration would require a non-trivial microscopic construction. Moreover, satisfying the gauge-kinetic positivity bound would rely on a substantial cancellation between independently determined topological contributions.

\subsubsection{Aligned worldsheet instanton}

We now consider worldsheet-instanton contributions involving non-trivial charge vectors, with the aim of avoiding the large topological coefficients required in the minimal benchmark. Two contributions with different charges can stabilize one axionic combination while leaving an approximately flat direction, subsequently lifted by the more suppressed instanton~\cite{Kim:2004rp}\footnote{Related multi-instanton mechanisms have recently been realized explicitly in type IIB compactifications: poly-instanton effects can generate a double-exponentially suppressed scale suitable for axion hilltop quintessence~\cite{Caraffi:2026wzk}.}. 

We consider an effective worldsheet-instanton configuration with mixed charge vectors in the coordinates adapted to the K3 fibration. Since these coordinates need not define an integral divisor basis, the corresponding charges need not be integral in this parametrisation. In an explicit compactification, they must arise from integral effective curve classes under the appropriate change of basis. Here, we take the effective charge vectors
\begin{equation}
q_1=(4,1,0),\qquad q_2=(3,1,0),
\end{equation}
as an illustrative benchmark, leaving their microscopic realization to a global construction.

These correspond to the superpotential contributions
\[
W_{\rm WS}
=A_1 e^{-2\pi(4T^1+T^2)}
+A_2 e^{-2\pi(3T^1+T^2)}
+A_3 e^{-2\pi T^3}\,.
\]
We choose the K\"ahler moduli and topological coefficients as
\[
\begin{aligned}
{\tt v}_1&\simeq10.8\,,&
{\tt v}_2&\simeq1.5\,,&
{\tt v}_3&\simeq1.8\,,\\
n_1&=\frac14\,,&
n_2&=\frac14\,,&
n_3&=\frac14\,,
\end{aligned}
\]
giving an overall compactification volume
\[
\mathcal V\simeq5.8\,.
\]
Displayed decimal values are rounded upward to one decimal place for readability (in the mantissa for scientific notation); the calculations use the unrounded inputs.
The coefficients $n_i$ are understood here as effective coefficients in the adapted K\"ahler coordinates. Their relation to the quantized topological data is determined by the same change of basis used for the intersection form and the instanton charges. In the integral basis used in \eqref{eq:ni}, $n_i=d^i/4$, so the choice above corresponds to unit components $d^i$ when the two bases coincide. An explicit realization must still construct stable bundles with the required second Chern classes.
We have verified that varying these coefficients within a range of order-one values does not significantly affect the ultralight axion mass or its effective decay constant. This is expected, since the corresponding hierarchy is primarily determined by the K\"ahler metric and the aligned worldsheet-instanton contributions, rather than by the precise values of the $n_i$. The latter mainly affect the axion couplings to the gauge sectors and the one-loop corrected gauge kinetic functions.
We choose \(s=21.5\), so that \(g_s=\sqrt{{\cal V}/s}\simeq0.6\). The threshold contribution is \(n_i{\tt v}^i\simeq3.6\), and \eqref{physicalgaugecouplings} gives
\[
\begin{aligned}
\alpha_v^{-1}=s+n_i{\tt v}^i&=25.0>0\,,\\
\alpha_h^{-1}=s-n_i{\tt v}^i&\simeq18.0>0\,.
\end{aligned}
\]
Both gauge kinetic functions are therefore positive, and the threshold-corrected visible coupling matches the phenomenological target \(\alpha_v\simeq1/25\). The relative threshold correction is \(n_i{\tt v}^i/s\simeq0.2\). The finite value of $s$ is an assumed stabilized input: the stabilization sector is not constructed in this effective benchmark and must not lift the aligned ultralight axion.

The corresponding instanton actions are
\[
\begin{aligned}
S_1&=2\pi(4{\tt v}_1+{\tt v}_2)\simeq279.3\,,\\
S_2&=2\pi(3{\tt v}_1+{\tt v}_2)\simeq211.7\,.
\end{aligned}
\]
The second instanton contribution is less suppressed and therefore dominantly stabilizes the combination \(3b^1+b^2\). The corresponding orthogonal direction in axion-coordinate space is proportional to \(d=(1,-3,0)\), satisfying \(q_2\cdot d=0\). Along this direction, the first instanton has a non-vanishing projection, \(q_1\cdot d=1\), and generates the much smaller potential responsible for the ultralight axion. The hierarchy of instanton actions therefore produces an approximate flat direction, which is lifted by the more suppressed instanton contribution. Its physical mass and effective decay constant are determined after canonical normalization, while the fundamental axion periodicities depend on the underlying integral charge lattice.

The assumed instanton hierarchy requires that additional worldsheet-instanton contributions do not significantly lift the light axionic direction. In the aligned two-axion sector, an additional charge vector $Q=(p,q,r)$ has projection $Q\cdot d=p-3q$ along the approximate flat direction $d=(1,-3,0)$. Instantons satisfying $p=3q$ therefore do not directly lift this direction, whereas those with non-vanishing projection must be sufficiently suppressed.

More generally, after canonical normalization and mass diagonalization, the additional contributions must satisfy
\begin{equation}
\Lambda_Q^4
\left|Q\cdot u_{\rm light}\right|^2
\ll m_{\rm light}^2,
\end{equation}
where $u_{\rm light}$ denotes the canonically normalized light direction expressed in the original axion coordinates.

As in the two-modulus example, the absence of dominant contributions along the light direction requires a suitable selection of worldsheet instantons, controlled by bundle-dependent zero modes or cancellations~\cite{BuchbinderDonagiOvrut2002,BeasleyWitten2003,Curio2010}. Whether this selection can be realized for the present charge vectors remains an open question.

\subsection{4d model}\label{sec:4dmodelCY}
We now compute the physical axion spectrum and couplings for the aligned K3-fibred benchmark introduced in Section \ref{sec:K3}. The 2-cycle volumes can be fixed to
\begin{equation}
    {\tt v}_1\simeq10.8\,,\quad {\tt v}_2\simeq 1.5\,,\quad {\tt v}_3\simeq 1.8\,,
\end{equation}
giving the overall volume to be $\mathcal{V}\simeq 5.8<25$. We take $n_i=1/4$, $s=21.5$, and the charge vectors $q_H=(3,1,0)$, $q_L=(4,1,0)$, $q_3=(0,0,1)$. In the resulting 4d multifield theory describing four axions, this information is sufficient to determine the axion kinetic terms and the Chern-Simons coupling ($\lambda_a, \lambda_i$): see \eqref{S4df3} and \eqref{kinterms4d}.

Determining the scalar potential \eqref{potential} requires further specifying the amplitudes: we use the tree-level input $\alpha_{\rm YM}=1/s\simeq1/21.5$, corresponding after the threshold correction to $\alpha_v=1/25$, take $\Lambda_{{\rm QCD}}= 8 \times 10^{-20}M_{p}$, and for the instantons we take in \eqref{L4} $A_i \simeq 1$,  $e^K = \frac{1}{\mathcal{V}}$ and $W_0=10^{-1}$\footnote{Such a value for $W_0$ is motivated by scenarios involving fractional Chern-Simons contributions \cite{Cicoli:2013rwa}. Recent investigations of heterotic flux vacua have also shown that obtaining small \(W_0\) is highly constrained, particularly in the large complex-structure regime, although moderately sub-unitary values can arise away from this limit~\cite{Buchbinder:2026}.}. Since \(m_\phi^2\propto W_0\), order-one variations of this parameter do not qualitatively affect the axion mass hierarchy. The 4d multifield theory summarized in Section \ref{sec:sum4d} is then fully fixed as a gauge-coupling-matched EFT benchmark.

We can now proceed as described in Section \ref{sec:cannormmass} to compute the mass spectrum and express the theory in terms of the corresponding physical axions. We also extract this way the corresponding decay constants and Chern-Simons couplings. We find
  
\begin{equation}\label{4fieldvalues}
\begin{aligned}
& m_{A}^2\simeq M_p^2 \begin{pmatrix}
        9.0 \times 10^{-3} \\
        1.5 \times 10^{-72}\\
        1.3 \times 10^{-86}\\
        7.2 \times 10^{-120}
    \end{pmatrix} \,,
    \quad  f_A\simeq M_p \begin{pmatrix}
        2.8\times 10^{-2}\\
        5.2 \times 10^{-3}\\
        3.4 \times 10^{-3}\\
        3.5\times 10^{-1}
    \end{pmatrix}\,,\\[0.15in]
& \lambda_{A,v}\simeq \begin{pmatrix}
        2.7 \times 10^{-3}\\
        6.4 \times 10^{-3}\\
        4.4 \times 10^{-18}\\
        -1.5\times 10^{-50}
    \end{pmatrix}\,,
    \quad  \lambda_{A,h}\simeq \begin{pmatrix}
        -3.8\times 10^{-3}\\
        8.6 \times 10^{-3}\\
        1.5 \times 10^{-3}\\
        -8.9\times 10^{-3}
    \end{pmatrix} \,.
\end{aligned}
\end{equation}
The heaviest state is mainly associated with the third worldsheet-instanton contribution, while the next-to-heaviest state corresponds to the QCD axion. The remaining two states are much lighter, with the lightest one providing the quintessence candidate. The mass-basis Chern-Simons coefficients use the common physical gauge normalization $1/[8\pi(s\pm n_i v^i)]$ after projection onto each eigenstate. For the decay constant, we get $f_{\rm QCD} \simeq 5.2 \cdot 10^{-3}\, M_p \simeq 1.4 \cdot 10^{-2}\, l_s^{-1}$. This value is slightly above the parametric lower bound $f_{\rm QCD}\gtrsim10^{-2}l_s^{-1}$ proposed in~\cite{Reece:2025zva}. Given the approximate nature of this bound, the two scales are comparable, and no significant tension arises at this level.

The lightest axion obeys
\begin{equation}
m_4\simeq2.7\times10^{-60}\, M_p\ ,\qquad f_4\simeq3.5\times 10^{-1}\, M_p \ .
\end{equation}
This realizes the situation described in Section \ref{sec:sum4d}: one eventually obtains a single field 4d model describing the physics of this lightest field, whose action is given in \eqref{S4dfinal}.
We have also investigated the stability of the effective single-field description under the relaxation of the heavier axionic directions. Using the full multifield axion potential, we determine the shifts of the heavy-field minima induced by a displacement of the lightest axion. Within the linearized approximation, the resulting correction to the effective potential satisfies \(|\Delta V|/\Lambda_{\rm eff}^4\sim10^{-30}\) over the field range considered. This demonstrates that the backreaction of the heavier axions is negligible in this approximation and provides a quantitative consistency check of the effective single-field description.

The light axion coupling with the visible-sector is negligibly small,
\begin{equation}
\lambda_{4,v}\simeq-1.5\times10^{-50},
\end{equation}
whereas its hidden-sector coupling remains finite,
\begin{equation}
\lambda_{4,h}\simeq-8.9\times10^{-3}.
\end{equation}
The corresponding effective coupling scales are therefore of order
\begin{equation}
f_{4,v}^{\rm eff}=\frac{f_4}{|\lambda_{4,v}|}\sim\mathcal{O}(10^{49})\, M_p\ ,\qquad f_{4,h}^{\rm eff}=\frac{f_4}{|\lambda_{4,h}|}\sim\mathcal{O}(10^1)\, M_p\ .
\end{equation}
The lightest axion is thus effectively decoupled from the visible gauge sector while retaining a Planck-suppressed coupling to the hidden Abelian sector, as expected~\cite{Agrawal:2024ejr, Leedom:2025mlr}. A non-negligible coupling to the visible sector would induce an additional mass contribution at the QCD scale, which would overwhelm the ultralight mass. 

The resulting single axion field 4d model is given by
\begin{empheq}[box=\widefbox]{align}
{\cal S}_{4d} &= \int \d^{4}x \sqrt{|g_{4}|} \left( \frac{M_p^2}{2} {\cal R}_{4} -\frac{1}{4}\left( {\cal F}_{\mu\nu}^{\mathfrak{a}} {\cal F}^{\mu\nu}_{\mathfrak{a}}\right)_v -\frac{1}{4}\left( {\cal F}_{\mu\nu}^{\mathfrak{a}} {\cal F}^{\mu\nu}_{\mathfrak{a}}\right)_h  \right) \nn\\
& + \int \d^{4}x \sqrt{|g_{4}|} \, \Bigg(\! -  \frac12\, (\del \phi)^2   -V_{\rm eff}(\phi)  +\frac{\lambda_{h}}{4 f} \, \phi \, \big({\cal F}_{\mu\nu}^{\mathfrak{a}} \tilde{{\cal F}}^{\mu\nu}_{\mathfrak{a}}\big)_h \Bigg)  \label{S4dfinalcosmo}
\end{empheq}
where for the effective scalar potential $V_{\rm eff}$, we will consider the two options \eqref{VHill0} and \eqref{VAx0}, that we recall here for convenience
\bea
\text{Hilltop-like:}&&\quad V_{{\rm Hill}} = \Lambda^4 \,\cos \left(\frac{\phi}{f} \right) \ , \label{VHill}\\
\text{Axionic:}&&\quad V_{{\rm Ax}} = \Lambda^4 \, \left( 1 +\cos \left(\frac{\phi}{f} \right) \right) \ . \label{VAx}
\eea
We will use the above values for the Hilltop-like case \eqref{VHill}: we read
\begin{equation}
{\rm Hill}:\quad \Lambda^4 =m^2 f^2 \simeq 8.8 \times 10^{-121}\, M_p^4\ ,\quad f\simeq 3.5 \times 10^{-1}\, M_p\ , \quad |\lambda_h|\simeq 8.9\times 10^{-3}\ . \label{Eq: Pot_values}
\end{equation}
For the Axionic case \eqref{VAx}, we slightly change the instanton charges to $q_1=(4,1,0)$, $q_2=(5,1,0)$ and the displayed cycle volumes to $({\tt v}_1,{\tt v}_2,{\tt v}_3)\simeq(8.6,1.6,1.9)$, while keeping the remaining inputs fixed. Proceeding as above, we obtain the following parameter values in the 4d model
\begin{equation}
{\rm Ax}:\quad \Lambda^4 =m^2 f^2 \simeq 4.5 \times 10^{-121}\, M_p^4\ ,\quad f\simeq 4.3 \times 10^{-1}\, M_p\ , \quad |\lambda_h|\simeq 1.3 \times 10^{-2}\ .
\label{Eq: Pot_values Axionic}
\end{equation}
At this second point, ${\cal V}\simeq5.6$, $g_s\simeq0.6$ and $n_i{\tt v}^i\simeq3.1$, so the same dilaton input gives $\alpha_v^{-1}\simeq24.6$ and $\alpha_h^{-1}\simeq18.5$. It therefore remains close to the target visible coupling while satisfying positivity in both sectors.
This minor adjustment will be justified by phenomenological requirements. We are now ready to use this explicit 4d model for cosmology. Prior to this, we return to the possible effect of the curvature terms discussed in Section \ref{sec:curv} on the above parameters.

\subsection{Effect of the curvature terms}\label{sec:curveffectf}

In Appendix \ref{ap:curv} and Section \ref{sec:curv}, we discussed curvature terms contributing to the 10d action and eventually to the 4d multifield theory. We concluded that these terms may impact the decay constants $f_i$, which get modified, $f_i\rightarrow f_i'$, within the estimated range \eqref{ficorr}. For the benchmark ${\cal V}=5.9$, we have $\epsilon_{\rm curv}={\cal V}^{-1/3}\simeq0.6$, and thus
\beq
\sqrt{1-\epsilon_{\rm curv}}\,f_i\ \leq\ f_i'\ \leq\ \sqrt{1+\epsilon_{\rm curv}}\,f_i\,.
\label{ficorr2}
\eeq
Here we evaluate quantitatively the effect of each extremal change in $f_i$,
\bea
\text{Lowered $f_i$:}&&\quad f_i'=\sqrt{1-\epsilon_{\rm curv}}\,f_i\,, \label{lowfi}\\
\text{Increased $f_i$:}&&\quad f_i'=\sqrt{1+\epsilon_{\rm curv}}\,f_i\,.\label{incrfi}
\eea

These corrections can be recast as modifications of the corresponding part of the non-diagonal K\"ahler metric, as can be seen in \eqref{kinterms4d} and \eqref{metricaxion}, namely $G_{ij}\rightarrow(1\pm\epsilon_{\rm curv})G_{ij}$. We recall that the physical axions, corresponding to mass eigenstates, are linear combinations of the initial axions. All physical masses and decay constants thus get affected by the corrections \eqref{lowfi} and \eqref{incrfi}. Proceeding as described in Section \ref{sec:cannormmass}, we compute the resulting new masses and decay constants: we detail our results in Table \ref{tab:curvature_comparisonLowfi} for \eqref{lowfi} and Table \ref{tab:curvature_comparisonIncrfi} for \eqref{incrfi}. The common rescaling remains an estimate of the possible size of the effect, rather than a microscopic determination of its tensor structure.

\begin{table}[ht!]
\centering
\renewcommand{\arraystretch}{1.2}
\begin{tabular}{c c c c c c c c c c c c}
\toprule
Axion &
Dominant &
$m^2_{\rm old}$ &
$m^2_{\rm new}$ &
$\dfrac{m^2_{\rm new}}{m^2_{\rm old}}$ &
$f_{\rm old}$ &
$f_{\rm new}$ &
$\dfrac{f_{\rm new}}{f_{\rm old}}$ \\
\midrule

$A_1$ &
WS3 &
$9.0\times10^{-3}$ &
$2.0\times10^{-2}$ &
$2.3$ &
$2.8\times10^{-2}$ &
$1.9\times10^{-2}$ &
$0.7$
\\

$A_2$ &
QCD &
$1.5\times10^{-72}$ &
$1.5\times10^{-72}$ &
$1.1$ &
$5.2\times10^{-3}$ &
$5.1\times10^{-3}$ &
$1.0$
\\

$A_3$ &
WS12 &
$1.3\times10^{-86}$ &
$2.8\times10^{-86}$ &
$2.3$ &
$3.4\times10^{-3}$ &
$2.3\times10^{-3}$ &
$0.7$
\\

$\phi$ &
WS12 &
$7.2\times10^{-120}$ &
$1.6\times10^{-119}$ &
$2.3$ &
$3.5\times10^{-1}$ &
$2.3\times10^{-1}$ &
$0.7$
\\

\bottomrule
\end{tabular}
\caption{Comparison of the axion mass spectrum and decay constants before and after the illustrative volume-dependent kinetic-metric deformation \eqref{lowfi}. Quantities are given in Planck units.}\label{tab:curvature_comparisonLowfi}
\end{table}

\begin{table}[ht!]
\centering
\renewcommand{\arraystretch}{1.2}
\begin{tabular}{c c c c c c c c c c c c}
\toprule
Axion &
Dominant &
$m^2_{\rm old}$ &
$m^2_{\rm new}$ &
$\dfrac{m^2_{\rm new}}{m^2_{\rm old}}$ &
$f_{\rm old}$ &
$f_{\rm new}$ &
$\dfrac{f_{\rm new}}{f_{\rm old}}$ \\
\midrule

$A_1$ &
WS3 &
$9.0\times10^{-3}$ &
$5.8\times10^{-3}$ &
$0.6$ &
$2.8\times10^{-2}$ &
$3.5\times10^{-2}$ &
$1.2$
\\

$A_2$ &
QCD &
$1.5\times10^{-72}$ &
$1.5\times10^{-72}$ &
$1.0$ &
$5.2\times10^{-3}$ &
$5.2\times10^{-3}$ &
$1.1$
\\

$A_3$ &
WS12 &
$1.3\times10^{-86}$ &
$8.1\times10^{-87}$ &
$0.7$ &
$3.4\times10^{-3}$ &
$4.3\times10^{-3}$ &
$1.3$
\\

$\phi$ &
WS12 &
$7.2\times10^{-120}$ &
$4.7\times10^{-120}$ &
$0.7$ &
$3.5\times10^{-1}$ &
$4.3\times10^{-1}$ &
$1.2$
\\

\bottomrule
\end{tabular}
\caption{Comparison of the axion mass spectrum and decay constants before and after the illustrative volume-dependent kinetic-metric deformation \eqref{incrfi}. Quantities are given in Planck units.}\label{tab:curvature_comparisonIncrfi}
\end{table}

Within the illustrative kinetic-metric deformations considered here, the strong axion mass hierarchy is preserved, and the effective decay constant of the ultralight mode remains sub-Planckian. This is a sensitivity result for the specified uniform deformations, not a general robustness statement for arbitrary higher-derivative or off-diagonal corrections. A microscopic evaluation would require the internal Calabi-Yau metric, as discussed in the Outlook in Section \ref{sec:ccl}. For the cosmological analysis we therefore retain the uncorrected parameter values of Section \ref{sec:4dmodelCY}.

\section{Cosmology: quintessence solutions}\label{sec:cosmo}

In Sections \ref{sec:10d4d} and \ref{sec:CY}, we derived the general axion effective theory motivated by perturbative heterotic compactifications and specified an illustrative benchmark using a K3-fibred volume ansatz. Under the stated assumptions concerning moduli stabilization and nonperturbative contributions, the lightest axionic direction is described by the effective single-field action \eqref{S4dfinalcosmo}, that we rewrite here for convenience
\bea
{\cal S}_{4d} &&= \int \d^{4}x \sqrt{|g_{4}|} \left( \frac{M_p^2}{2} {\cal R}_{4} -  \frac12\, (\del \phi)^2   -V(\phi) \right)\label{S4dcosmo}\\
&& + \int \d^{4}x \sqrt{|g_{4}|} \, \Bigg(\!  -\frac{1}{4}\left( {\cal F}_{\mu\nu}^{\mathfrak{a}} {\cal F}^{\mu\nu}_{\mathfrak{a}}\right)_v -\frac{1}{4}\left( {\cal F}_{\mu\nu}^{\mathfrak{a}} {\cal F}^{\mu\nu}_{\mathfrak{a}}\right)_h  +\frac{\lambda_{h}}{4 f} \, \phi \, \big({\cal F}_{\mu\nu}^{\mathfrak{a}} \tilde{{\cal F}}^{\mu\nu}_{\mathfrak{a}}\big)_h \Bigg) \ ,\nn
\eea
with two scalar potentials to be considered
\bea
\text{Hilltop-like:}&&\quad V_{{\rm Hill}} = \Lambda_s^4 \,\cos \left(\frac{\phi}{f} \right) \ , \label{VHillcosmo}\\
\text{Axionic:}&&\quad V_{{\rm Ax}} = \Lambda_s^4 \, \left( 1 +\cos \left(\frac{\phi}{f} \right) \right) \ , \label{VAxcosmo}
\eea
and the parameter values slightly adjusted for each of them
\bea
{\rm Hill}:\quad && \Lambda_s^4 \simeq 8.8 \times 10^{-121}\, M_p^4\ ,\quad f\simeq 3.5 \times 10^{-1}\, M_p\ , \quad |\lambda_h|\simeq 8.9\times 10^{-3}\ , \label{parvaluesHill}\\
{\rm Ax}:\quad && \Lambda_s^4 \simeq 4.5 \times 10^{-121}\, M_p^4\ ,\quad f\simeq 4.3 \times 10^{-1}\, M_p\ , \quad |\lambda_h|\simeq 1.3\times 10^{-2}\ . \label{parvaluesAx}
\eea
The potential scale $\Lambda_s$ fixes the axion mass through the relation $\Lambda_s^4 =m^2 f^2$. In this section, we use this model to describe dark energy in our late universe, in the form of quintessence.

To that end, one should add to the action \eqref{S4dcosmo} an extra term capturing (non-relativistic) matter: while heterotic string possesses all necessary ingredients to derive it, doing this explicitly goes far beyond the scope of this work. We simply assume here a minimally coupled (i.e.~only gravitationally) matter sector. Turning to radiation, we recall from Section \ref{sec:4dmodelCY} the extremely weak axion coupling to visible radiation in our model, hence its absence in \eqref{S4dcosmo}. Only the coupling to dark radiation (hidden sector) might play a role: as a consequence, we will at first neglect that coupling in Sections \ref{sec:form} and \ref{sec:sol}, and come back to its possible effects in Section \ref{sec:radcoupling}. 

Therefore, in our 4d model, we are only left with the first line in \eqref{S4dcosmo}, together with matter. Such a setting is among the simplest quintessence models: see e.g.~a review in \cite[Sec.3.2]{Andriot:2026lac}. As we will recall in Section \ref{sec:form}, the two potentials of interest, \eqref{VHillcosmo} and \eqref{VAxcosmo}, fall in the category of thawing quintessence models. From those, realistic cosmological solutions describing our late universe can be found, as reviewed in \cite{Andriot:2024sif}. We will provide such solutions in Section \ref{sec:sol} and compare them to the latest observational data.

\subsection{Thawing quintessence models: formalism}\label{sec:form}

We consider the single field quintessence model given by the first line of \eqref{S4dcosmo} together with matter (and possibly radiation). The standard formalism to study background cosmological solutions from such a model is reviewed in \cite[App.B]{Andriot:2026lac}. We briefly recall the main points here; we do not discuss perturbations. We then provide a first description of a realistic solution, as the one we will be looking for.\\

We focus on (background) solutions for which the metric is of the type Friedmann-Lema\^itre-Robertson-Walker (FLRW), with scale factor $a_s(t)$, and where the scalar field only depends on time, $\phi(t)$ and $\del_t \phi \equiv \dot{\phi}$. The Hubble parameter ${\cal H}=\dot{a}_s/a_s$ is positive, as one considers solutions for an expanding universe. For simplicity, we do not consider any spatial curvature, $k=0$. In addition, the formalism considers each universe constituent to be described as a perfect fluid $n$ with energy density $\rho_n$, pressure $p_n$ and equation of state parameter $w_n=p_n/\rho_n$. These definitions allow us to introduce the energy density parameters $\Omega_{n} = \rho_n/(3 M_p^2 {\cal H}^2)$, which indicate the proportion of each constituent at a given time in the universe as can be seen through \eqref{F1}. Furthermore, for the scalar field, which will play the role of dark energy, one has
\beq
\rho_{\phi} = \frac12 \dot{\phi}^2 + V \ ,\quad w_{\phi} = \frac{\frac12 \dot{\phi}^2 - V }{\frac12 \dot{\phi}^2 + V } \ .
\eeq

We turn to the equations of motion of the quintessence model, to find an adequate cosmological solution. The Einstein equations boil down to the two Friedmann equations, and one has in addition the scalar field equation of motion. These three equations are however related: it is sufficient to solve two of them, together with one specific condition at a given time \cite[Sec.2.2]{Andriot:2024sif}. The time of interest will here be today: $t=t_0$. We denote quantities today with an index ${}_0$. We fix without loss of generality $a_s(t_0)=1$, and take the following fiducial values \cite{Planck:2018vyg}
\beq
\Omega_{r0} = 0.0001 \ ,\ \Omega_{m0}=0.3149 \ ,\ \Omega_{\phi0}=0.6850 \ ,\quad {\rm and} \ {\cal H}_0^2 = 3.45 \cdot 10^{-121} M_p^2\ . \label{Omega0}
\eeq
The values of the energy density parameters today obey the first Friedmann equation
\beq
1 = \sum_n \Omega_n \ .\label{F1}
\eeq
This condition is sufficient to reduce the problem to solving the other two equations, namely the second Friedmann equation and the scalar field equation of motion, which can be formulated in different fashions. We use here as a variable the number of e-folds, $N\equiv \ln a_s$. We then write the two equations as follows \cite[(2.20)]{Andriot:2024sif}
\bea
&& 2 \tilde{{\cal H}} \partial_N \tilde{{\cal H}} + 4 \Omega_{r0}\, e^{-4N} + 3 \Omega_{m0}\, e^{-3N} + \tilde{{\cal H}}^2 \left(\frac{\partial_N\phi}{M_p} \right)^2= 0 \ , \label{F2} \\[0.1in]
&& \tilde{{\cal H}}^2 \partial_N^2 \phi + \tilde{{\cal H}} \partial_N \tilde{{\cal H}}\, \partial_N\phi + 3 \tilde{{\cal H}}^2 \partial_N \phi + \del_{\phi} \tilde{V} = 0 \ , \label{Ep}
\eea
where $\tilde{V}\equiv V/{\cal H}_0^2$, $\tilde{{\cal H}}\equiv {\cal H}/{\cal H}_0$, for a given potential $V$. A solution to this system is the set of functions $(\tilde{{\cal H}}(N), \phi(N))$ together with specified initial conditions. They are chosen at $t_0$, i.e.~at $N=0$. In particular, $\tilde{{\cal H}}(N=0)=1$. We are left to specify $\phi(0),\del_N \phi (0)$: those can be traded for $\Omega_{\phi0}, w_{\phi0}$ using the relations
\beq
\frac{\del_N \phi (0)}{M_p} = \sqrt{3 \Omega_{\phi0}(1+w_{\phi0})} \ ,\quad \frac{\tilde{V}(\phi(0))}{M_p^2} = \frac32 \Omega_{\phi0}(1 - w_{\phi0}) \ ,
\label{initial speed}
\eeq
where we choose a positive field speed today. One still has to invert the potential, and we do so for the two potentials of interest \eqref{VHillcosmo} and \eqref{VAxcosmo}. With $\tilde{\Lambda}_s^4=\Lambda_s^4/{\cal H}_0^2$, one gets
\bea
\text{Hilltop-like:}&&\quad \phi (0) = f \arccos\left(\frac{M_p^2}{\tilde{\Lambda}_s^4}\frac{3}{2}\, \Omega_{\phi0}(1 - w_{\phi0}) \right)   \ , \\
\text{Axionic:}&&\quad \phi (0) = f \arccos\left(\frac{M_p^2}{\tilde{\Lambda}_s^4}\frac{3}{2}\, \Omega_{\phi0}(1 - w_{\phi0}) -1 \right) \ .
\label{initial position}
\eea

In summary, to find a quintessence solution, one only needs to solve \eqref{F2} and \eqref{Ep}, with the only remaining initial condition to specify: the value of $w_{\phi0}$. The scalar potential also needs to be given: it can be found for our model in \eqref{VHillcosmo} and \eqref{VAxcosmo}, with parameter values \eqref{parvaluesHill} and \eqref{parvaluesAx}.\\

Among all solutions that can be found this way, only some may qualify as realistic. Those are the ones which display the expected successive domination phases, namely radiation domination, followed by matter domination, and finally dark energy dominating today (the latter being ensured by the values \eqref{Omega0}). Examples will be provided in Figures \ref{fig:Axionic_Omegas} and \ref{fig:Hilltop_Omegas}. As was observed in \cite{Andriot:2024jsh, Andriot:2024sif} for various thawing models, obtaining radiation and matter domination phases in the past, of adequate durations, can only be achieved for highly tuned values of $w_{\phi0}$. In the following, we will then pick such values of $w_{\phi0}$, completely fixing the quintessence solution.

While such solutions will have the appropriate evolution of their $\Omega_n$, displaying the desired successive domination phases, the evolution of $w_{\phi}$ will be fixed without any further freedom. It is however this quantity which needs to be compared to the latest observational data. The defining property of a thawing model is that $w_{\phi} \simeq -1$ during matter domination, and increases in the late universe, which is precisely the behaviour obtained here. Qualitatively, this trend is consistent with the observations, which display a growing equation of state parameter for dark energy. We will be left to compare the evolution of $w_{\phi}$ quantitatively to observations.

\subsection{Quintessence solutions and comparison to observations}\label{sec:sol}

\begin{figure}[htb!]
    \centering
        \begin{subfigure}[b]{0.45\textwidth}
        \centering
        \includegraphics[width=\textwidth]{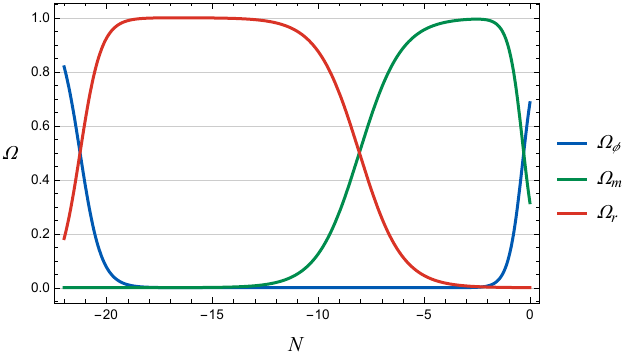}
        \caption{}
        \label{fig:Axionic_Omegas}
    \end{subfigure}
    \hfill
    \begin{subfigure}[b]{0.45\textwidth}
        \centering
        \includegraphics[width=\textwidth]{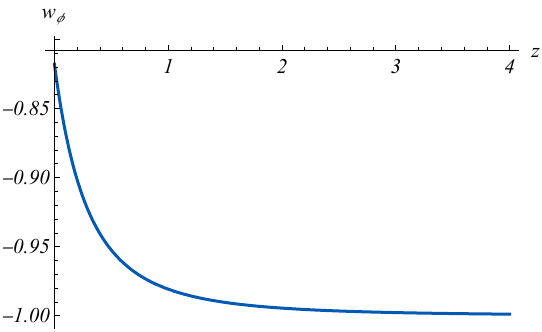}
        \caption{}
        \label{fig:Axionic_wphi}
    \end{subfigure}
    
    \vspace{0.35cm} 

    \begin{subfigure}[b]{0.45\textwidth}
        \centering
        \includegraphics[width=\textwidth]{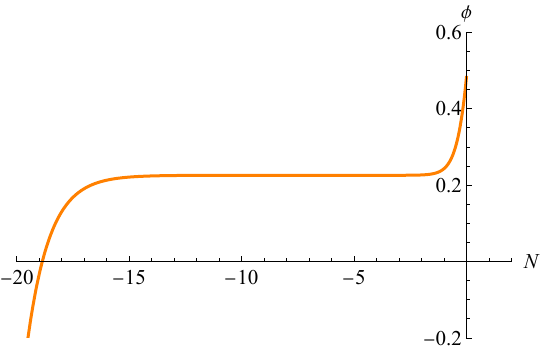}
        \caption{}
        \label{fig:Axionic_Phi(N)}
    \end{subfigure}
    \qquad \qquad \qquad
    \begin{subfigure}[b]{0.3\textwidth}
        \centering
        \includegraphics[width=\textwidth]{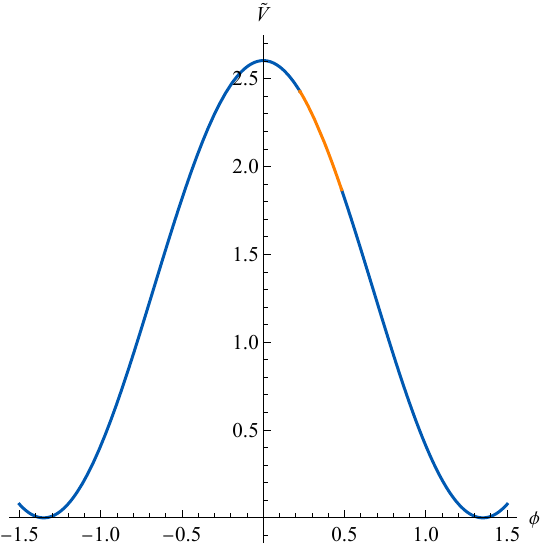}
        \caption{}
        \label{fig:Axionic_phi_on_pot}
    \end{subfigure}
\caption{Cosmological solution obtained as described in Section \ref{sec:form} for the Axionic potential \eqref{VAxcosmo} with parameter values \eqref{parvaluesAx}, for $w_{\phi0}= -0.818$. The evolution of $\Omega_n(N)$, $w_{\phi}(z)$ and $\phi(N)$ are displayed on Figure \ref{fig:Axionic_Omegas}, \ref{fig:Axionic_wphi} and \ref{fig:Axionic_Phi(N)}; we recall the relation between the number of e-folds $N$ and the redshift $z$: $N=-\ln (1+z)$. The field rolling down the potential, here $\tilde{V}$, from its frozen value $ \phi_\text{frozen}=0.225\,M_p$ until its value today, is shown on Figure \ref{fig:Axionic_phi_on_pot}.}\label{fig:Axionic}
\end{figure}

\begin{figure}[htb!]
    \centering
    \begin{subfigure}[b]{0.45\textwidth}
        \centering
        \includegraphics[width=\textwidth]{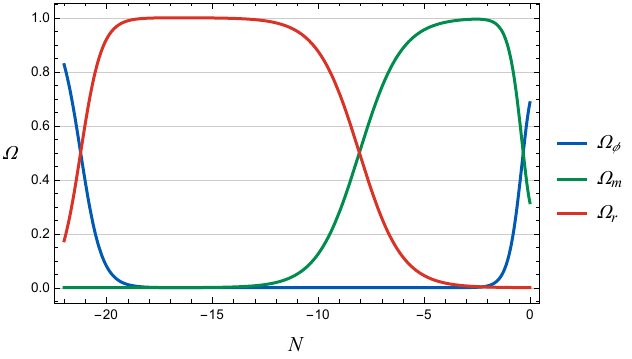}
        \caption{}
        \label{fig:Hilltop_Omegas}
    \end{subfigure}
    \hfill
    \begin{subfigure}[b]{0.45\textwidth}
        \centering
        \includegraphics[width=\textwidth]{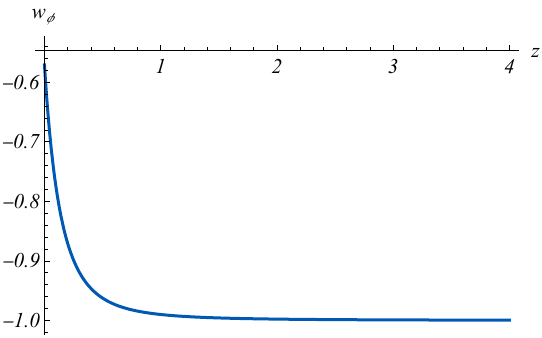}
        \caption{}
        \label{fig:Hilltop_wphi}
    \end{subfigure}
    
    \vspace{0.35cm} 

    \begin{subfigure}[b]{0.45\textwidth}
        \centering
        \includegraphics[width=\textwidth]{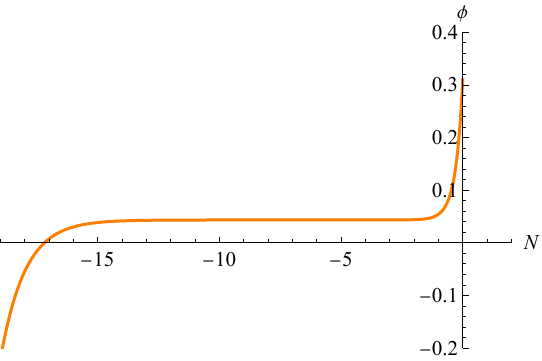}
        \caption{}
        \label{fig:Hilltop_Phi(N)}
    \end{subfigure}
    \qquad \qquad \qquad
    \begin{subfigure}[b]{0.3\textwidth}
        \centering
        \includegraphics[width=\textwidth]{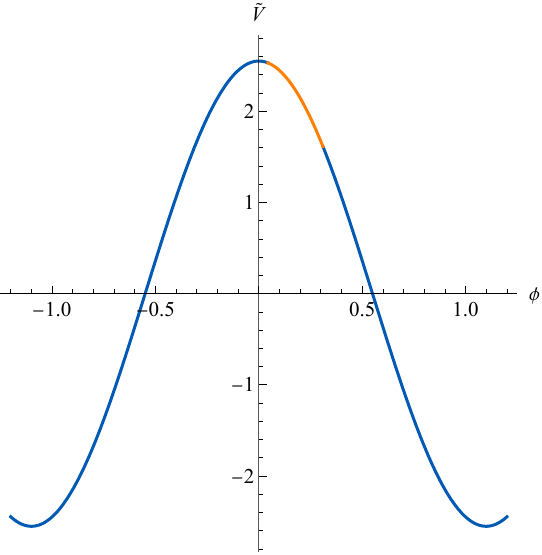}
        \caption{}
        \label{fig:Hilltop_phi_on_pot}
    \end{subfigure}
\caption{Cosmological solution obtained as described in Section \ref{sec:form} for the Hilltop-like potential \eqref{VHillcosmo} with parameter values \eqref{parvaluesHill}, for $w_{\phi0}= -0.571$. The evolution of $\Omega_n(N)$, $w_{\phi}(z)$ and $\phi(N)$ are displayed on Figure \ref{fig:Hilltop_Omegas}, \ref{fig:Hilltop_wphi} and \ref{fig:Hilltop_Phi(N)}; we recall the relation between the number of e-folds $N$ and the redshift $z$: $N=-\ln (1+z)$. The field rolling down the potential, here $\tilde{V}$, from its frozen value $ \phi_\text{frozen}=4.3 \times 10^{-2}\,M_p$ until its value today, is shown on Figure \ref{fig:Hilltop_phi_on_pot}.}\label{fig:Hilltop}
\end{figure}

We consider the single axion field quintessence model, given by the first line of \eqref{S4dcosmo}, together with matter and radiation. The two scalar potentials to consider were given in \eqref{VHillcosmo} and \eqref{VAxcosmo}, with parameter values \eqref{parvaluesHill} and \eqref{parvaluesAx}. We proceed as described in Section \ref{sec:form} to find realistic cosmological solutions, making use of the Mathematica notebook {\tt Figures}. For each potential, we find an adequate value of $w_{\phi0}$, giving us the solution of interest. We display the resulting solutions in Figures \ref{fig:Axionic} and \ref{fig:Hilltop}. One can verify in Figures \ref{fig:Axionic_Omegas} and \ref{fig:Hilltop_Omegas} that the $\Omega_n$ display the successive domination phases, while Figures \ref{fig:Axionic_wphi} and \ref{fig:Hilltop_wphi} exhibit a growing $w_{\phi}$ in the recent universe, as expected. For both solutions, we also show the evolution of the scalar field in Figures \ref{fig:Axionic_Phi(N)} and \ref{fig:Hilltop_Phi(N)}: from an initial kination phase, it gets frozen during radiation and matter domination, and finally rolls down the potential in the recent universe. This late-time rolling is displayed on the potential itself, in Figures \ref{fig:Axionic_phi_on_pot} and \ref{fig:Hilltop_phi_on_pot}. Since the solutions were integrated backward from today, the early kination branch has not been established as a physical prediction: its numerical stability and its effect on the expansion rate during BBN require a forward-integration and $\Omega_\phi(T)$ analysis.\\

We turn to the comparison of these background cosmological solutions to the latest observations. In recent years, hints of dynamical dark energy were given by DES \cite{DES:2024jxu} and DESI \cite{DESI:2024mwx, DESI:2025zgx}. The recent recalibration of SNe Ia  datasets \cite{DES:2025sig, Hoyt:2026fve} has brought the discrepancy between models of dynamical dark energy and $\Lambda$CDM to $3.2 - 3.4 \,\sigma$, while the latest full-shape analysis combining DESI DR1 and DR2 \cite{DESI:2026lnd} gives it between $3.0 - 3.5 \,\sigma$, depending on the dataset. This situation motivates us to consider a scenario of dynamical dark energy and compare our models and solutions to such observations by DESI.

We start with the single-field model with axionic potential \eqref{VAxcosmo}. This potential was actually tested directly against the data in \cite{DESI:2025fii}. Depending on the SNe Ia data sets, the best fit was obtained for the following parameter values
\beq
\hspace{-0.1in}\frac{f}{M_p}:\quad \left|\begin{array}{l} 0.74^{+0.84}_{-0.34}\quad \text{(Pantheon+)} \\[0.1in]
0.51^{+1.67}_{-0.28} \quad \text{(Union3)} \\[0.1in]
0.81^{+2.90}_{-0.49} \quad \text{(DESY5)}\end{array}\right. \ , \qquad \ \frac{m}{M_p}:\quad \left|\begin{array}{l} 8.78^{+6.13}_{-3.84}\ \cdot10^{-61}\quad \text{(Pantheon+)} \\[0.1in]
 13.0^{+11.7}_{-6.48}\ \cdot10^{-61} \quad \text{(Union3)} \\[0.1in] 9.63^{+7.90}_{-4.80}\ \cdot 10^{-61} \quad \text{(DESY5)}\end{array}\right.
\eeq
On our side, we could adjust the string compactification to obtain the parameter values \eqref{parvaluesAx}, corresponding to
\beq
\frac{f}{M_p} = 0.5 \ ,\qquad \frac{m}{M_p} = 1.6 \times 10^{-60} \ .
\eeq
For a qualitative comparison, $f$ lies within the quoted marginalized intervals, while $m$ does so for the Union3 data set.

The Hilltop-like potential \eqref{VHillcosmo} was not tested in the same way against data. To offer a comparison, we then use $w_{\phi}$; its evolution in our solution was displayed in Figure \ref{fig:Hilltop_wphi}. We perform a linear projection of our model-generated $w_{\phi}(a_s)$ onto the CPL form provided by DESI
\beq
w_{\rm DE} = w_0 + w_a\, (1-a_s) \ .
\eeq
From our solution, we obtain the following projection coefficients
\beq
{\rm Hilltop-like}:\quad  w_0^{\rm proj}=-0.80\ ,\quad w_a^{\rm proj}=-0.29 \ ,
\eeq
while the values provided by DESI in the latest analysis \cite{DESI:2026lnd} are
\beq
w_0= -0.821 \pm 0.054\ , \quad w_a= -0.65 \pm 0.20 \ .
\eeq
The projected $w_0^{\rm proj}$ lies near the quoted central range, while $w_a^{\rm proj}$ shows a larger discrepancy, although it is less tightly constrained by observations. For completeness, the same projection for the axionic potential gives $w_0^{\rm proj}=-0.87$ and $w_a^{\rm proj}=-0.18$. Overall, these comparisons show qualitative proximity to some observationally preferred ranges.

Still, our solutions do not reproduce the phantom regime ($w_{\rm DE}<-1$) favored by the DESI fits. This is also reflected in the relatively small slope $|w_a|$ obtained in our solutions. This mismatch was clear from the start when considering our single-field models, since by definition there, $w_{\phi}\geq-1$; this is independent of the choice of potential and parameter values. As reviewed in \cite{Andriot:2025los}, a convenient mechanism to generate an effective phantom regime is to have a coupling of the quintessence field to matter. We did not consider such a coupling so far in this work. We will discuss this idea, and possible extensions of our setting, in Section \ref{sec:ccl}.\\

Let us make a few final comments on the parameter values. Even though the string compactification admits some freedom to adjust their precise value, it remains a non-trivial fact that the correct orders of magnitude can be obtained. Let us start with cosmological constraints. For either potential, we find that realistic solutions could be obtained for $f \gtrsim 0.1\, M_p$. In turn, axion decay constants are generally expected to be sub-Planckian, $f\leq 1 \, M_p$, at least when built in a quantum gravity context; see for example arguments related to the weak gravity conjecture (WGC) \cite[3.5.8]{Palti:2019pca}. Similarly, the scale parameter $\Lambda_s^4/(M_p^2 {\cal H}_0^2)$ is constrained. From \eqref{initial speed}, a canonical scalar with $w_{\phi0}\geq-1$ obeys
\beq
\frac{V(\phi(0))}{M_p^2{\cal H}_0^2}
=\frac32\Omega_{\phi0}(1-w_{\phi0})
\leq 3\Omega_{\phi0}\,.
\eeq
For a potential whose maximum is $c\Lambda_s^4$, existence of the specified present-day state requires
\beq
\frac{\Lambda_s^4}{M_p^2{\cal H}_0^2}
\geq \frac{3\Omega_{\phi0}(1-w_{\phi0})}{2c}\,,
\qquad c=1\ \text{(Hilltop-like)},\quad c=2\ \text{(Axionic)}\,.
\eeq
The values $3\Omega_{\phi0}=2.06$ and $3\Omega_{\phi0}/2=1.03$ are therefore sufficient potential capacities to include the limiting case $w_{\phi0}=-1$; they are not necessary lower bounds for every solution. Numerically, the solutions studied here favour $\Lambda_s^4/(M_p^2{\cal H}_0^2)$ of order unity to ${\cal O}(10)$, with the precise interval depending on $f$ and on the chosen initial condition.
It is a priori not obvious that a string compactification would provide such parameter values, but our setting turned out to allow for this. The most delicate one is $\Lambda_s$, since the relation to ${\cal H}_0$ \eqref{Omega0} makes it tied to the tiny value of the (would-be) cosmological constant. To reach such a small scale, we had to invoke non-perturbative effects, within an anisotropic 6d Calabi-Yau manifold. Getting an adequate value for $f$ also required picking specific 6d geometric features, namely, in the minimal scenario, the values of the second Chern class components and numbers $n_i$ are required to be large, creating tension with the positivity of the one-loop corrected gauge kinetic function. The aligned-instanton configuration provides an alternative, allowing the desired axion mass hierarchy and an effective decay constant with order-one topological coefficients. It is interesting to see how requirements on the cosmological parameters provide valuable guidelines to select adequate string compactifications.

\subsection{Axionic friction term and solutions}\label{sec:radcoupling}

We finally investigate whether the coupling of the light axion to the hidden gauge sector modifies significantly the cosmological evolution. A dynamical axion can transfer energy to the gauge fields, which in turn can act as an effective friction term in the axion equation of motion, and contribute to a dark radiation component~\cite{Anber:2009ua}.
Varying the action \eqref{S4dcosmo} with respect to the scalar field gives
\begin{equation}
	\ddot{\phi}+3\mathcal{H}\dot{\phi}	+\partial_\phi V	=\frac{\lambda_h}{4f}	\left(	{\cal F}_{\mu\nu}^{\mathfrak a}	\widetilde{\cal F}^{\mu\nu}_{\mathfrak a}	\right)_h .
\end{equation}
It is convenient to rewrite the gauge-field source as an effective friction term,
\begin{equation}
\ddot{\phi}+\left(3\mathcal{H}+\Upsilon\right)\dot{\phi}+\partial_\phi V=0,
\label{EU}
\end{equation}
where
\begin{equation}
\Upsilon\equiv-\frac{\lambda_h}{4f\dot{\phi}}\left({\cal F}_{\mu\nu}^{\mathfrak a}\widetilde{\cal F}^{\mu\nu}_{\mathfrak a}
	\right)_h .
\end{equation}
The quantity \(\Upsilon\) is in general time dependent, and should be understood as a convenient parametrization of the backreaction of the produced gauge fields on the homogeneous axion evolution.

For solutions of \eqref{EU}, the scalar energy density obeys~\cite{Anber:2009ua,Berera:1995ie}
\begin{equation}
\dot{\rho}_\phi+3\mathcal{H}(1+w_\phi)\rho_\phi=-\Upsilon\dot{\phi}^{\,2}.
\label{phicont}
\end{equation}
Assuming that the energy transferred to the hidden gauge sector behaves as radiation, conservation of the total energy-momentum tensor gives
\begin{equation}
\dot{\rho}_{\rm dr}+4\mathcal{H}\rho_{\rm dr}=\Upsilon\dot{\phi}^2.
\label{rhodrcont}
\end{equation}
This form is analogous to the interacting dark-radiation description considered in ~\cite{Poulin:2023lkg}. Notice that the axionic Chern-Simons interaction itself is metric independent and therefore does not directly contribute to the energy-momentum tensor. The produced gauge fields, however, carry an energy density, represented here by $\rho_{\rm dr}$, which does contribute to the Friedmann equations.

Therefore the set of differential equations \eqref{F2} and \eqref{Ep}  used to perform our simulation becomes 
\bea
&& 2 \tilde{{\cal H}} \partial_N \tilde{{\cal H}} + 4 \Omega_{r0}\, e^{-4N} + 3 \Omega_{m0}\, e^{-3N} + \tilde{{\cal H}}^2 \left(\frac{\partial_N\phi}{M_p}\right)^2\, + \frac{4}{3} \tilde{\rho}_{dr}= 0 \ , \label{F2 Gauge} \\[0.1in]
&& \tilde{{\cal H}}^2 \partial_N^2 \phi + \tilde{{\cal H}} \partial_N \tilde{{\cal H}}\, \partial_N\phi + 3 \tilde{{\cal H}}^2 \partial_N \phi + \del_{\phi} \tilde{V}\,+ \tilde{\Upsilon} \tilde{\cal H} \partial_N\phi= 0 \ , \label{Ep Gauge} \\[0.1in] 
&&  \partial_N \tilde{\rho}_{dr}\, +4 \tilde{\rho}_{dr}\, -\tilde{\Upsilon} \tilde{\cal H} \left(\frac{\partial_N\phi}{M_p}\right)^2= 0 \ , \label{rhodrcont in e fold}
\eea
where
\begin{equation}
\tilde{\rho}_{\rm dr}\equiv\frac{\rho_{\rm dr}}{M_p^2{\cal H}_0^2},\qquad\tilde{\Upsilon}\equiv\frac{\Upsilon}{{\cal H}_0}.
\end{equation}

Axion-induced gauge-field production has been extensively investigated in Abelian and non-Abelian sectors, particularly in inflationary and spectator-field scenarios; see, for example, \cite{Anber:2009ua,Dimastrogiovanni:2016fuu,Peloso:2016gqs,Dimastrogiovanni:2023juq}. 
Here we restrict attention to the Abelian hidden sector of our construction.
At late times, the cosmological evolution approaches a quasi-de Sitter regime. In this limit, gauge-field production can be estimated using the standard large-instability-parameter approximation~\cite{Anber:2009ua,Barnaby:2011qe}, provided that the instability parameter $\xi$, defined by
\begin{equation}
\xi\equiv\frac{\lambda_h\dot{\phi}}{2f\mathcal{H}}=\frac{\lambda_h}{2f}\partial_N\phi,
\end{equation}
varies sufficiently slowly, and $|\xi|\gg1$. Under these assumptions, one obtains
\begin{equation}
\frac{1}{4}\left({\cal F}_{\mu\nu}^{\mathfrak a}\widetilde{\cal F}^{\mu\nu}_{\mathfrak a}
	\right)_h=
\left|\left\langle{\bf E}\cdot{\bf B}\right\rangle\right|\simeq2.4\times10^{-4}\,\mathcal{H}^4\frac{e^{2\pi|\xi|}}{|\xi|^4}, 
\end{equation}
and correspondingly
\begin{equation}
\left|\Upsilon_{\rm qdS}\right|\simeq1.2\times10^{-4}\,\frac{\lambda_h^2{\cal H}^3}{f^2}\frac{e^{2\pi|\xi|}}{|\xi|^5}. 
\end{equation}

Evaluating $\xi$ on the unperturbed Hilltop trajectory, the instability parameter becomes very small in the late-time quasi-de Sitter regime.
The large-$|\xi|$ approximation is therefore not applicable to our solutions, and significant tachyonic amplification is not expected in this regime.
We instead examine the earlier kination epoch, during which the scalar kinetic energy dominates, and
\begin{equation}
    \rho_\phi\simeq\frac12\dot{\phi}^2
    \propto a_s^{-6},
    \qquad
    \mathcal{H}\propto a_s^{-3},
    \qquad
    \dot{\phi}\propto a_s^{-3}.
\end{equation}
Consequently, the instability parameter remains
approximately constant,
\begin{equation}
    \xi=\frac{\lambda_h\dot{\phi}}{2f\mathcal{H}}
    \simeq {\rm const.}
\end{equation}

For the Hilltop benchmark \eqref{VHill}, with
$|\lambda_h|\simeq8.9\times10^{-3}$ and
$f\simeq0.4M_p$, the unperturbed scalar trajectory gives
\begin{equation}
    |\xi|\simeq3.1\times10^{-2}.
\end{equation}
This estimate assumes that the scalar background
evolution is unchanged and should not be interpreted
as an independent numerical solution.

For an ideal kination background,
\begin{equation}
    a_s(t)\propto t^{1/3},
    \qquad
    a_s(\tau)\propto\tau^{1/2},
    \qquad
    a_s\mathcal{H}=\frac{1}{2\tau},
\end{equation}
where $\tau$ stands for the conformal time given by $d\tau=dt/a_s(t)$.
The gauge-field helicity modes satisfy the differential equation
\begin{equation}
    A_\pm''+
    \left(k^2\pm\frac{k\xi}{\tau}\right)A_\pm=0.
    \label{kination_modes}
\end{equation}
For $\xi>0$, the negative-helicity mode has
\begin{equation}
    A_-''+
    \left(k^2-\frac{k|\xi|}{\tau}\right)A_-=0.
\end{equation}

Its effective frequency becomes
negative for
\begin{equation}
    k\tau<|\xi|,
    \qquad
    \frac{k}{a_s\mathcal{H}}<2|\xi|.
\end{equation}
For the benchmark considered here, this corresponds to
\begin{equation}
    \frac{k}{a_s \mathcal{H}}\lesssim6.2\times10^{-2}.
\end{equation}

A formal WKB estimate within the tachyonic region
gives the amplification exponent of the gauge-field helicity modes
\begin{equation}
    S=\int_{(k\tau)_i}^{|\xi|}
    d(k\tau)\,\sqrt{\frac{|\xi|}{(k\tau)}-1},
    \qquad
    S\leq\frac{\pi|\xi|}{2}.
\end{equation}
The corresponding upper estimate is
\begin{equation}
    e^S\lesssim1.1.
\end{equation}
Thus, the coupling does not lead to an appreciable tachyonic
amplification in this approximation.

There remains, however, an important limitation. During kination,
\begin{equation}
	\frac{k}{a_s\mathcal{H}}\propto a_s^2,
\end{equation}
so the relevant modes move out of, rather than into, the tachyonic regime as the Universe expands. Since kination coincides with the earliest epoch included in our cosmological evolution, the modes relevant for the instability are already non-adiabatic at the beginning of our solution. Their initial amplitudes therefore cannot be fixed by imposing an adiabatic-vacuum condition within the kination regime itself.

The above analysis consequently determines the possible amplification of a specified initial gauge-field state, but it does not uniquely determine the absolute values of
$\rho_{\rm dr}$ and $\left\langle{\bf E}\cdot{\bf B}\right\rangle $, without further information about the gauge-field state prior to the onset of kination. A fully quantitative prediction would require either specifying this initial state or extending the cosmological evolution to an earlier epoch and matching the gauge modes across the transition.
Our preliminary numerical implementation indicates that gauge-field production may be negligible; however, we leave a fully backreacted analysis, including the determination of the initial gauge-field state, for future work.

\section{Summary and outlook}\label{sec:ccl}

In this work, we have computed the effective-theory ingredients needed to investigate axionic quintessence in perturbative heterotic compactifications, including the axion kinetic terms, gauge couplings and non-perturbative potential, combining these ingredients with geometric and cosmological requirements.

Sections \ref{sec:10d4d} and \ref{sec:pot} derive the axion effective action, its gauge couplings and non-perturbative potential under the stated stabilization assumptions. The curvature analysis in Appendix \ref{ap:curv} identifies local contributions whose microscopic evaluation requires the internal Calabi-Yau metric. The two uniform kinetic-metric deformations studied in Section \ref{sec:curveffectf} preserve the strong mass hierarchy and a sub-Planckian effective decay constant; this is a sensitivity test, not a calculation of the full correction matrix. The exploratory toric scan in Section \ref{sec:toric} finds no sampled configuration satisfying simultaneously the single-large-cycle requirement, the volume bound and the full geometric-control cuts, motivating more general instanton charges. Aligned worldsheet instantons relax these restrictions: the explicit two-modulus example supports the required axion kinematics within the stated geometric-control region, while the three-modulus K3-fibred benchmark yields an ultralight axion with a phenomenologically relevant decay constant, positive gauge kinetic functions and a visible coupling of $1/25$. Heavy-axion relaxation gives a negligible correction to the light potential within the approximation considered. The required instanton selection must still be established using explicit bundle data. In Section \ref{sec:cosmo}, we use that model for quintessence to describe dark energy of our late-time universe. We first discuss in Section \ref{sec:form} the formalism to be used for thawing quintessence models, to which our model belongs. We present the equations to solve, given a model and an FLRW background metric. We also discuss the way to select realistic cosmological solutions by choosing initial conditions. In Section \ref{sec:sol}, we use this method to obtain specific cosmological solutions with a dynamical dark energy, for each of the two potentials. We compare those to the latest observational data by DESI \cite{DESI:2025fii,  DESI:2026lnd}, obtaining a reasonably good agreement. We discuss how cosmological constraints provide bounds on the various parameter values, which in turn inform us on the adequate string compactifications. In Section \ref{sec:radcoupling}, we finally discuss the effect of the axionic coupling to dark radiation, left aside so far.

A complete heterotic realization remains open. In addition to establishing the required instanton selection, it must construct consistent gauge bundles realizing the assumed topological coefficients, verify that possible St\"uckelberg gaugings leave the light axion dynamical~\cite{Choi:2019xij}, and stabilize the dilaton and all remaining moduli without spoiling the hierarchy. The constant vacuum-energy contribution also remains an independent input. Subject to these conditions, the theory supports quintessence at the relevant scales. 

A closely related open question concerns the cosmological initial conditions of the light axion. In the present analysis, the complex-structure and gauge-bundle moduli are assumed to be stabilised at scales above the K\"ahler sector and are therefore integrated out, so that the worldsheet-instanton prefactors $A_i$ can be treated as constants. More generally, however, these prefactors are Pfaffians which can depend non-trivially on bundle moduli and may vanish on special loci in moduli space~\cite{BuchbinderDonagiOvrut2002,Curio2010}. This raises the possibility that the axion potential relevant during an earlier inflationary epoch differs substantially from the late-time potential considered here.
In particular, one may envision an additional worldsheet-instanton contribution of the form
$W_{\rm lock}=A_{\rm lock}(\Phi)\,e^{-2\pi T_{\rm lock}}$,
where $\Phi$ collectively denotes bundle moduli. If $A_{\rm lock}(\Phi)$ is non-vanishing during inflation, this contribution could temporarily lift the axionic direction that eventually plays the role of quintessence and generate an inflationary mass $m_\phi^{\rm inf}\gtrsim H_{\rm inf}$. Such an inflationary locking mechanism would suppress the stochastic fluctuations of an otherwise ultralight axion and dynamically localise it close to a definite field value. After inflation, the bundle moduli could relax towards a region of moduli space where the corresponding Pfaffian becomes strongly suppressed, or vanishes on an appropriate zero locus, so that this additional contribution ceases to be relevant. The much more suppressed worldsheet-instanton term responsible for the late-time quintessence potential would then remain as the dominant source of shift-symmetry breaking.
If the relative instanton phases and charge vectors are such that the field value selected by the inflationary contribution corresponds to, or lies sufficiently close to, the maximum of the late-time potential, this mechanism could provide a dynamical origin for the hilltop initial condition used in the cosmological solutions studied here. A complete realization would require analysing the inflationary dynamics and stabilisation of the bundle moduli, the structure of the relevant Pfaffian zero loci, and the resulting time-dependent axion mass matrix.

Let us now comment on two further directions to pursue. A first one is the coupling to matter. Having a quintessence field, it is natural to ask whether, and how, it couples to (dark) matter, and look for any phenomenological consequence (fifth force, variation of fundamental constants, etc.). Interestingly, when it comes to axions, couplings are specific; a recent work evaluating such coupling from string theory is \cite{Benabou:2026lcu}, highlighting a derivative coupling.

As reviewed recently in \cite{Andriot:2025los}, functional couplings of the quintessence field to matter can lead to an effective phantom dark energy, i.e.~one with $w_{\rm DE}<-1$ (see \cite{Chanda:2026dmx} for an interesting recent alternative, that may connect to Section \ref{sec:radcoupling}). Such a mechanism would be useful to reproduce the recent parametric or non-parametric reconstructions of $w_{\rm DE}$ by DESI from observational data \cite{DESI:2025fii}, which exhibit a phantom behaviour. While a direct coupling of the quintessence field to (dark) matter could exist, another interesting path is to use two axions, as recently revisited in \cite{deSouza:2026qyf} (see also \cite{Garcia:2026rnj}). There, one oscillating axion mimics (dark) matter, while a second one acts as the quintessence field, the two being possibly coupled. It would be interesting to realize any of those coupled scenarios from string theory; the framework developed in this work should help in this respect.

Secondly, it could be worth revisiting the choice of the compactification manifold, and the resulting evaluation of curvature terms. To start with, one could focus on Riemann-flat manifolds, which are essentially built as a torus quotiented by a discrete group. Those were recently investigated in \cite{Bento:2025wuu, DallAgata:2025jii, ValeixoBento:2026jdx, Aparici:2026wdm}. It would be interesting to study whether some of these manifolds could be adequate for our purposes. In particular, the Riemann-flatness property would eliminate the local curvature corrections considered here. It is however less clear how the tadpole cancellation and the gauge bundle requirements would be satisfied. The simplest example could be to consider a flat $T^2 \times T^2 \times T^2$, obtained as an orbifold or a discrete quotient of $T^6$. This would allow more explicit calculations of the quantities involved. A drawback is the more straightforward volume formula, which may offer less freedom in getting a large 2-cycle while satisfying the volume upper bound \eqref{bound_vol}. 

If one sticks to a 6d Calabi-Yau manifold, it could be interesting to have a proper evaluation of the curvature terms instead of the estimate considered here in Sections \ref{sec:curv} and \ref{sec:curveffectf}. To that end, one could use the recent developments that determine (at least locally) the Calabi-Yau metric numerically, as e.g.~in  \cite{Anderson:2023viv, Hendi:2024yin, Constantin:2026bky, Constantin:2026rgm, Lust:2026mys}; we also note some recent analytical progress in \cite{Bykov:2026ytj}. The numerical methods may allow us to determine more precisely the various correction terms involved, with possibly interesting physical effects on e.g.~the decay constants. An interesting direction would be to investigate whether the aligned instanton configurations studied here can be realized in explicit heterotic compactifications. Our results show how late-time cosmological requirements provide non-trivial guidance in selecting heterotic compactifications.

\vspace{0.2in}

\vspace{0.4in}

\subsection*{Acknowledgements}

We warmly thank C.~Delaunay, C.~Eloy, A.~Lukas, M.~Reig and A.~Westphal for helpful exchanges during the completion of this work. MP's work is supported by the NWA through the ORC programme Emergence at All Scales. LV's work was supported by the French Agence Nationale de la Recherche (ANR) under grant ANR-25-CE57-1445-02 (project DarES). MP thanks CPTGA for support and LAPTh for hospitality during her visit.

\newpage

\begin{appendix}

\section{Computation of curvature terms}\label{ap:curv}

In Section \ref{sec:pertdimred}, while considering the dimensional reduction of 10d heterotic string to 4d, we identified various curvature terms, related to the Riemann tensor ${\cal R}^M_{+\, NPQ}$, gathered in the action ${\cal S}_{{\rm curv}}$ \eqref{Scurv}. In the literature, these terms are often discarded, as one focuses on 4d axion physics. In \cite{Svrcek:2006yi}, the 4d background is considered to be asymptotically flat, making it plausible that 4d contributions to such terms vanish; we will see here that 6d terms may still play a role. The possibility of getting contributions due to gravitational instantons is also mentioned in \cite{Svrcek:2006yi}, but not investigated. More generally, these curvature terms are discussed in e.g.~\cite{Becker:2009df, Anguelova:2010ed, Marques:2015vua, Eloy:2020dko, Ortin:2020xdm}, sometimes with an evaluation on certain backgrounds.

In this appendix, we explicitly compute the curvature terms in ${\cal S}_{{\rm curv}}$ (as well as those in the $H$-flux Bianchi identity \eqref{BI}) on a 4d cosmological background and a 6d Calabi-Yau manifold. Some of these curvature terms appear at order $\alpha'$ or at 1-loop, which are the same orders as the gauge field contributions used in axion physics. So an evaluation of these curvature terms seems necessary.

The explicit expressions below are obtained under the simplifying restrictions stated in the derivation. For a generic Calabi-Yau metric, harmonicity of the forms $\beta_i$ does not imply covariant constancy or determine their pointwise contractions. The local internal contractions appearing below should therefore be regarded as the starting point for the parametric estimate in Section \ref{sec:curv}, rather than as a universal microscopic correction matrix.

We will start with a computation of ${\cal R}_{+\, MNPQ}$, relevant to several contributions. We will also indicate several restrictions, often corresponding to ansatz specifications, that will allow us to simplify the computation. Then, we will compute each term appearing in ${\cal S}_{{\rm curv}}$ \eqref{Scurv} (and the Bianchi identity \eqref{BI}), schematically denoted respectively by $\tr |{\cal R}_{ie+}|^2$, $\tr({\cal R}_+ \w {\cal R}_+)$, $\tr {\cal R}_{10+}^4 $ and $\d B_2 \w *_{10}\, {\rm CS}(\omega_+)|_{4,6d}$. We finally gather all calculations to give an expression of ${\cal S}_{{\rm curv}}$ in \eqref{Scurvcomputedap}. An estimate of this final result is discussed in Section \ref{sec:curv}, together with its effect on 4d axion physics.\\

\begin{itemize}
    \item ${\cal R}_{+\, MNPQ}$ and restrictions
\end{itemize}

We first recall from \cite[(A.29)]{Andriot:2011iw} that
\beq
{\cal R}_{+\, MNPQ} = {\cal R}_{MNPQ} +\frac12 H_{SM[P}H^S{}_{Q]N} + \nabla_{[P} H_{Q] MN} \ .
\eeq
To compute these components, we introduce a first set of restrictions corresponding to the compactification ansatz described in Section \ref{sec:pertdimred}. To start with, the spacetime is a direct product of its 4d and 6d subspaces, giving a 10d metric of the form
\beq
\d s_{10}^2 = g_{\mu\nu}(x) \d x^{\mu} \d x^{\nu} + g_{mn}(y) \d y^m \d y^n \ ,\label{10dmetric}
\eeq
with $\mu,\nu$ 4d spacetime indices and $m,n$ 6d internal ones. In that case, the 10d Riemann tensor (with one lowered index) ${\cal R}_{MNPQ}$ has no mixed component between 4d and 6d (see e.g.~\cite{Andriot:2017oaz}). The same holds for the 10d Christoffel symbols; the 10d covariant derivative then becomes directly either the 4d or the 6d one. Second, the $H$-flux at order $(\alpha')^0$ is restricted to have two types of components: a pure 4d one and a mixed one, as in \eqref{Hflux}. The mixed one will come from $\d B_2$ and have one 4d index and two 6d ones.\footnote{In a cosmological background, allowing for $B_2$ with one 4d index would make that index to be along $\d t$, while the 4d dependence would also be in $t$. Eventually, this would give no $H$-flux with two 4d indices. For a Calabi-Yau compactification, one may also invoke the absence of 1-cycle to develop $B_2$, forbidding as well such an $H$-flux component.} In other words, we can write
\beq
H = \frac{1}{3!} H_{\mu\nu\rho} \d x^{\mu\nu\rho} + \frac12 H_{\mu mn} \d x^{\mu} \w \d y^{mn} + {\cal O}(\alpha') \ ,
\eeq
using abbreviations for wedge products of 1-forms. The component $ H_{\mu\nu\rho} $ will come from $H|_{4d}$ at order $(\alpha')^0$, which only depends on 4d coordinates. This implies
\beq
\nabla_{P} H_{Q\mu\nu}  = \nabla_{\pi} H_{\rho\mu\nu}\, \delta_P^{\pi} \delta_Q^{\rho} \ .
\eeq
From this ansatz, we deduce the following 10d components
\begin{subequations}\label{Rcompo}
\begin{align}
\left.{\cal R}_{+\, MNPQ}\right|_{MN=\mu\nu} =&\ \left( {\cal R}_{\mu\nu\pi\rho}   + \frac12 H_{\sigma\mu[\pi}H^{\sigma}{}_{\rho]\nu} + \nabla_{[\pi} H_{\rho] \mu\nu} \right) \delta_P^{\pi} \delta_Q^{\rho} \\
&\ \!+\, \frac12 H_{s\mu[p}H^{s}{}_{q]\nu} \, \delta_P^{p} \delta_Q^{q} +{\cal O}(\alpha') \nn \\
\left.{\cal R}_{+\, MNPQ}\right|_{MN=\mu n} =&\ \ \ \ \, \frac14 \left( H_{\sigma\mu \pi}H^{\sigma}{}_{qn} - H_{s\mu q}H^{s}{}_{\pi n} + 2 \nabla_{\pi} H_{q \mu n}  \right) \delta_P^{\pi} \delta_Q^{q} \\
&\ \!+\, \frac14 \left( H_{s\mu p}H^{s}{}_{\rho n} - H_{\sigma\mu \rho}H^{\sigma}{}_{pn} - 2 \nabla_{\rho} H_{p \mu n}  \right) \delta_P^{p} \delta_Q^{\rho} \nn\\
&\ \!+\, \nabla_{[p} H_{q] \mu n}\, \delta_P^{p} \delta_Q^{q}  +{\cal O}(\alpha') \nn\\
\left.{\cal R}_{+\, MNPQ}\right|_{MN=mn} =& \ \left( {\cal R}_{mnpq}   + \frac12 H_{\sigma m[p}H^{\sigma}{}_{q]n}  \right) \delta_P^{p} \delta_Q^{q} \\
&\ \!+\, \left( \frac12 H_{s m[\pi}H^{s}{}_{\rho]n} + \nabla_{[\pi} H_{\rho] mn} \right)\, \delta_P^{\pi} \delta_Q^{\rho} \nn\\
& \ \!+\, \frac12 \nabla_{p} H_{\rho mn} \, \delta_P^{p} \delta_Q^{\rho} - \frac12 \nabla_{q} H_{\pi mn} \, \delta_P^{\pi} \delta_Q^{q} + {\cal O}(\alpha')\nn
\end{align}
\end{subequations}\\

We now provide a second set of restrictions, specifying more the ansatz, to simplify further the above expressions. Anticipating on the use of an FLRW 4d metric, we consider the latter to be block diagonal with specific coordinate dependence, as follows
\beq
\d s_4^2 = g_{00}(t)\, \d t^2 + g_{ij}(t,x)\, \d x^i \d x^j \ .
\eeq
The restricted dependence of $g_{00}$ gives $\Gamma^0_{i0}=0$, which will be useful. We turn to the indices and dependence on 4d coordinates of the $H$-flux components and the axions. By restricting $a$ to depend on time only, one gets from \eqref{daH4} that $*_4 H|_{4d} \propto \d t$ (at order $(\alpha')^0$), where the proportionality factor may only depend on time. Given the block diagonal 4d metric, we deduce that $H|_{4d}$ has no leg along $\d t$, i.e.~that it is proportional to the 3d space volume-form, with again a proportionality factor depending at most on time. Spatially covariant derivatives of $H|_{4d}$ contain expansion-dependent connection terms. For the slowly evolving background considered here, we neglect these terms, which are negligible for late-time dynamics, and keep only temporal derivatives.
This implies that $\nabla_i H_{\mu\nu\rho} = 0$. Similarly, restricting $B_2$, i.e.~$b^i$, to depend only on time among the 4d coordinates, we obtain that $H_{\mu mn}$, coming from $\d B_2$, is only non-zero for $\mu=0$, and has the same 4d dependence, that is only on time. Using that $\Gamma^0_{i0}=0$, one deduces $\nabla_i H_{\mu mn}=0$. We summarize our second set of restrictions discussed so far, or ansatz specifications, as follows (at order $(\alpha')^0$)
\beq
H_{0\nu\rho} =0  \ ,\ H_{\mu mn} = \delta_{\mu}^0\, H_{0 mn} \ , \quad \nabla_{\sigma} H_{\mu\nu\rho} =\delta_{\sigma}^0\, \nabla_{0} H_{\mu\nu\rho} \ , \ \nabla_{\sigma} H_{\mu mn} =\delta_{\sigma}^0 \delta_{\mu}^0\, \nabla_{0} H_{0 mn} \ .
\eeq
This leads to the following simplifications
\beq
H_{s\mu[p}H^{s}{}_{q]\nu}=0 \ ,\ H_{\sigma\mu \pi}H^{\sigma}{}_{qn} = 0 \ ,\ H_{s m[\pi}H^{s}{}_{\rho]n} = 0 \ ,\  \nabla_{[\pi} H_{\rho] mn} = 0 \ .
\eeq

We will consider one further restriction, that we first motivate. We recall that $B_2 \propto \sum_i b^i(t) \beta_i$ with a constant proportionality coefficient, and the $\beta_i$ are harmonic 2-forms. One deduces that $\d B_2\propto \sum_i \beta_i \w \d b^i(t) $, meaning that $H_{\mu mn}$ is proportional to $\beta_{i\, mn}$ which carries the whole 6d coordinate dependence. This implies for the 6d covariant derivative that $\nabla_p H_{\mu mn}$ is given by the combination of the $\nabla_p \beta_{i\, mn} $. For the purpose of obtaining an explicit estimate, we impose the additional simplifying assumption
\beq
{\rm Assumption:}\quad \nabla_p \beta_{i\, mn} = 0 \ .
\eeq
This does not follow from harmonicity and is not satisfied by generic harmonic $(1,1)$-forms on a Calabi-Yau threefold. The expressions derived from it below are therefore conditional on this simplifying ansatz.
This implies the following restriction
\beq
\nabla_p H_{\mu mn} = 0 \ .
\eeq
This whole second set of restrictions leads us to simplify the components \eqref{Rcompo} towards
\begin{subequations}\label{Rcomposimpl}
\begin{align}
\left.{\cal R}_{+\, MNPQ}\right|_{MN=\mu\nu} \, =&\ \left( {\cal R}_{\mu\nu\pi\rho}   + \frac12 H_{\sigma\mu[\pi}H^{\sigma}{}_{\rho]\nu} + \nabla_{[\pi} H_{\rho] \mu\nu} \right) \delta_P^{\pi} \delta_Q^{\rho} +{\cal O}(\alpha')  \\
\left.{\cal R}_{+\, MNPQ}\right|_{MN=\mu n} \,=& \ \  \frac14 \left( H_{s\mu p}H^{s}{}_{\rho n} - 2 \nabla_{\rho} H_{p \mu n}  \right) \delta_P^{p} \delta_Q^{\rho} \\
\, & - \frac14 \left( H_{s\mu q}H^{s}{}_{\pi n} - 2 \nabla_{\pi} H_{q \mu n}  \right) \delta_P^{\pi} \delta_Q^{q} +{\cal O}(\alpha') \nn\\
\left.{\cal R}_{+\, MNPQ}\right|_{MN=mn} \, =&\  \left( {\cal R}_{mnpq}   + \frac12 H_{\sigma m[p}H^{\sigma}{}_{q]n}  \right) \delta_P^{p} \delta_Q^{q}  + {\cal O}(\alpha')
\end{align}
\end{subequations}
We will use those to compute the different curvature terms appearing in ${\cal S}_{{\rm curv}}$ and in the $H$-flux Bianchi identity.\\

Before doing so, let us provide a third set of restrictions, again motivated by our ansatz. First, we will consider that
\beq
H_{0mn}H_0{}^{mp} = H_{0mq}H_0{}^{mq}\, \delta_n^p \ ,\quad H_{0mq}\, \nabla_{0} H_0{}^m{}_p\ \d y^{pq}=0 \ . \label{HHortho}
\eeq
This is an additional pointwise algebraic assumption on the harmonic representatives, namely $\beta_{i\, np}\beta_{j}{}^{nm} = \delta_{ij}\delta_p^m\, \beta_{i\, nq}\beta_{j}{}^{nq}$. It is stronger than their integrated orthogonality and does not follow from the absence of 1-cycles.

In addition, we will consider the FLRW metric in 4d, taking the form 
\beq
{\rm FLRW}:\quad \d s_4^2 = - \d t^2 + a_s^2(t) \,\tilde{g}_{ij}(x) \d x^i \d x^j \ ,
\eeq
i.e.~with $g_{00}=-1$, and $a_s$ the scale factor. The non-zero Christoffel symbols and Riemann tensor components are given as follows (the dot denotes $\del_t = \del_0$)
\bea
&&\Gamma^i_{jk} = \tilde{\Gamma}^i_{jk} \ , \ \Gamma^0_{ij} = \dot{a}_s a_s  \, \tilde{g}_{ij} \ , \ \Gamma^i_{0j} = \Gamma^i_{j0} = \frac{\dot{a}_s}{a_s}\, \delta^i_j \ ,\\
&& {\cal R}^i{}_{jkl} = \tilde{{\cal R}}^i{}_{jkl} + \dot{a}_s^2 \left( \delta^i_k \, \tilde{g}_{jl} - \delta^i_l \, \tilde{g}_{jk} \right) \ ,\\
&& {\cal R}^i{}_{0j0} = - {\cal R}^i{}_{00j}  = - \frac{\ddot{a}_s}{a_s}\, \, \delta^i_j \ ,\ {\cal R}^0{}_{i0j} = - {\cal R}^0{}_{ij0} = \ddot{a}_s a_s \, \tilde{g}_{ij} \ ,
\eea
where all components in the last line are related by the symmetries of the Riemann tensor. We also introduce the Hubble parameter ${\cal H}$
\beq
\frac{\dot{a}_s}{a_s} = {\cal H}  \ , \quad \frac{\ddot{a}_s}{a_s} = \dot{{\cal H}} + {\cal H}^2 \ .
\eeq
We will consider for simplicity no 3d spatial curvature, meaning
\beq
k=0:\quad \tilde{{\cal R}}^i{}_{jkl}=0 \ ,\quad {\cal R}_4 = 6\left( {\cal H}^2 + \frac{\ddot{a}_s}{a_s} \right) \ .
\eeq
This ends the list of our restrictions, i.e.~ansatz specifications and assumptions. Using those, we now compute the various curvature terms.\\

\begin{itemize}
    \item $\tr |{\cal R}_{ie+}|^2$
\end{itemize}

We first compute the contraction of the Riemann tensor components defined in \eqref{trsquare}, which appears in ${\cal S}_{{\rm curv}}$. Using \eqref{Rcomposimpl}, we obtain
\bea
\hspace{-0.4in} 2\, \tr |{\cal R}_{ie+}|^2 &=& \left( {\cal R}_{\mu\nu\pi\rho}   + \frac12 H_{\sigma\mu[\pi}H^{\sigma}{}_{\rho]\nu} + \nabla_{[\pi} H_{\rho] \mu\nu} \right)^2 + \left( {\cal R}_{mnpq}   + \frac12 H_{\sigma m[p}H^{\sigma}{}_{q]n}  \right)^2\\
\hspace{-0.4in} && + \frac14 \left( H_{\mu ps}H_{\rho n}{}^s + 2 \nabla_{\rho} H_{\mu pn} \right)^2 + {\cal O}(\alpha') \ ,\nn
\eea
where here and in the following, {\sl the square stands for the contraction of free indices,} {\bf without numerical factor}. Using above specifications, these squares can be computed and simplified. In particular, the first relation in \eqref{HHortho} implies that $\left( H_{0 m[p}H_{0|q]n}  \right)^2=0$. Altogether, we obtain
\bea
2\, \tr |{\cal R}_{ie+}|^2 &=& \left( {\cal R}_{\mu\nu\pi\rho}  \right)^2 +\frac{1}{48} \left( H_{ijk} \right)^2  \left( H_{ijk} \right)^2 -\frac12 \left( \nabla_{0} H_{ijk} \right)^2 \label{trR21}\\
&&- {\cal R}_{ijkl}  H_{h}{}^{ik}H^{hjl} - 2 {\cal R}_{ij 0 k} \nabla_{0} H^{ijk} \nn\\
&&+ \left( {\cal R}_{mnpq} \right)^2 + {\cal R}_{mnpq} \, H_0{}^{mp}H_0{}^{nq} \nn\\
&& + \frac14 \left( H_{0mn} \right)^2 \left( H_{0mn} \right)^2 + \frac12 \left(  \nabla_{0} H_{0mn} \right)^2  + {\cal O}(\alpha') \ .\nn
\eea
Once again, the squares are on free indices, which do not include a $0$. For the latter, we have extracted $g_{00}$ and set $g_{00}=-1$ in \eqref{trR21}, anticipating on FLRW. Using the 4d Riemann tensor components obtained for FLRW, we further get
\bea
2\, \tr |{\cal R}_{ie+}|^2 &=& 12 \left({\cal H}^4 + \left( \dot{{\cal H}} + {\cal H}^2 \right)^2 \right)+ \left( {\cal R}_{mnpq} \right)^2 \\
&&- {\cal H}^2 \left( H_{ijk} \right)^2  + {\cal R}_{mnpq} \, H_0{}^{mp}H_0{}^{nq} \nn\\
&& +\frac{1}{48} \left( H_{ijk} \right)^2  \left( H_{ijk} \right)^2 -\frac12 \left( \nabla_{0} H_{ijk} \right)^2 \nn\\
&& + \frac14 \left( H_{0mn} \right)^2 \left( H_{0mn} \right)^2 + \frac12 \left(  \nabla_{0} H_{0mn} \right)^2  + {\cal O}(\alpha') \ .\nn
\eea

We are left to replace the $H$-flux components by their expressions in terms of axions. To that end, let us first recall here a few definitions and relations. We start with the model independent one $a$. We recall the relation $\d a = 2\pi^2 \alpha' M_p^2 *_4 H|_{4d}$ and $f_a^2= 2/(M_p^2 (4\pi^2 \alpha')^2)$. At order $(\alpha')^0$, we have (with $g_{00}=-1$)
\bea
&& H_{ijk} = - \frac{1}{2\pi^2 \alpha' M_p^2}\, \sqrt{|g_4|} \epsilon_{ijk}\ \del_0 a  \ , \quad \left( \nabla_{0} H_{ijk} \right)^2 = \frac{6}{(2\pi^2 \alpha' M_p^2)^2} \left(\del_0^2 a \right)^2 \ ,\\
&& {\rm vol}_4 \left( H_{ijk} \right)^2  = 6 H|_{4d} \w *_4 H|_{4d} = 6\,{\rm vol}_4 \, \frac{2}{M_p^2} \, f_a^2\, (\del_0 a)^2 \ .
\eea
We turn to the model dependent ones $b^i$. With $\beta_i = \frac12 \beta_{i\, mn} \d y^{mn}$ and the mixed $H$ coming from $\d B_2$, we have (with implicit sums on $i,j$)
\bea
&& H_{0mn} = 2\pi l_s^2\ \   \del_0 b^i \, \beta_{i\, mn} \ ,\\
&& \int_X {\rm vol}_6 \left(  \nabla_{0} H_{0mn} \right)^2 = (2\pi l_s^2)^2 \ \del_0^2 b^i\, \del_0^2 b^j \int_X \beta_i \w *_6 \beta_j =  \frac{(2\pi l_s)^6}{4\pi^2}\, \gamma_i^2 \left(\del_0^2 b^i \right)^2 \ .
\eea
We also recall that $f_i^2 = \gamma_i^2 M_p^2 / (8 \pi^2 {\cal V}) $, and the definition of the canonically normalized fields $\vartheta_a = f_a \, a$, $\vartheta^i = f_i\, b^i$ (without sum).

We now have all ingredients to compute the term in $\tr |{\cal R}_{ie+}|^2$ appearing in ${\cal S}_{{\rm curv}}$ \eqref{Scurv}. We obtain for this action term the following expression
\bea
{\cal S}_{{\rm curv}} \supset \!\!\!\! && \frac{1}{2 \kappa_{10}^2\, g_s^2}\, \frac{\alpha'}{4} \int \d^{10}x \sqrt{|g_{10}|} \,  \tr |{\cal R}_{ie+}|^2 \label{actiontrR2}\\
&& = \frac{M_p^2}{2} \, \frac{\alpha'}{4} \int {\rm vol}_4 \ \, \frac{1}{{\cal V} (2\pi l_s)^6}\int {\rm vol}_6\ \tr |{\cal R}_{ie+}|^2 \nn\\
&& = \int \d^4 x \sqrt{|g_4|}\ \Bigg[ \frac{M_p^2}{2}\ \frac32\, l_s^2 \, \big({\cal H}^4 + \big( \dot{{\cal H}} + {\cal H}^2 \big)^2 \big) + M_p^2\, {\cal O}({\alpha'}^2) \nn\\
&& \phantom{= \int \d^4 x \sqrt{|g_4|}\ \Bigg[} + \frac{M_p^2}{2} \ \frac18 \ \frac{1}{{\cal V} (2\pi l_s)^6}\int {\rm vol}_6\ \ l_s^2 \ {\cal R}_{mnpq} {\cal R}^{mnpq}  \nn\\
&& \phantom{= \int \d^4 x \sqrt{|g_4|}\ \Bigg[} + \frac{1}{2} \,  (\del_0 \vartheta_a)^2 \left(-\frac{3}{2}\, l_s^2\, {\cal H}^2  +\frac{3}{8}  \, \frac{l_s^2}{M_p^2} \, (\del_0 \vartheta_a)^2 \right) -\frac38\, l_s^2\, \left(\del_0^2 \vartheta_a \right)^2  \nn\\
&& \phantom{= \int \d^4 x \sqrt{|g_4|}\ \Bigg[} +\frac{1}{16} \,\frac{l_s^2}{M_p^2} \, \sum_i (\del_0 \vartheta^i)^4  + \frac{1}{16} \,  l_s^2\, \sum_i \left(\del_0^2 \vartheta^i \right)^2  \nn\\
&& \phantom{= \int \d^4 x \sqrt{|g_4|}\ \Bigg[} + \frac18\, \del_0 \vartheta^i \del_0 \vartheta^j \ l_s^2  \int {\rm vol}_6\ {\cal R}_{mnpq} \, \frac{ \beta_i^{mp} \beta_j^{nq} }{(2\pi l_s)^2\, \gamma_i \gamma_j} \Bigg] \ ,\nn
\eea
where to get an explicit expression (proportional to $(\del_0 \vartheta^i)^4$) when computing the term in $\left( H_{0mn} \right)^2 \left( H_{0mn} \right)^2$, we took for simplicity $\beta_{i\, mn} \beta_j^{mn}$ constant; one is otherwise left with the integral $\int {\rm vol}_6\ \beta_{i\, mn} \beta_j^{mn} \beta_{k\, pq} \beta_l^{pq}$. Eventually, that term will not provide a relevant contribution in any case.\\

\begin{itemize}
    \item $\tr({\cal R}_+ \w {\cal R}_+)$
\end{itemize}

We turn to the terms involving the 2-form ${\cal R}_+^M{}_N = \frac12 {\cal R}_+^M{}_{NPQ}\, \d x^P \w \d x^Q$, appearing in ${\cal S}_{{\rm curv}}$ as well as in the Bianchi identity. Using the various restrictions and specifications above, we first compute the components of this 2-form at order $(\alpha')^0$
\begin{subequations}\label{R2fcompo}
\begin{align}
{\cal R}_+^M{}_N|_{MN=\mu\nu} &= \frac12 {\cal R}^{\mu}{}_{\nu\pi\rho} \, \d x^{\pi\rho} +\delta^{\mu}_i \delta_{\nu}^j \left( \frac14 H^{i}{}_{hk}H^{h}{}_{jl} \, \d x^{kl} + \frac12 \nabla_{0} H^{i}{}_{jk}\, \d x^{0 k} \right) \nn\\
&= \ \ \delta^{\mu}_i \delta_{\nu}^j \ g_{jk} {\cal H}^2   \, \d x^{ik}
+ \delta^{\mu}_i \delta_{\nu}^0\ \frac{\ddot{a}_s}{a_s} \, \d x^{0i}
+ \delta^{\mu}_0 \delta_{\nu}^i\ \frac{\ddot{a}_s}{a_s} g_{ij} \, \d x^{0j} \\
& \ \ +\, \delta^{\mu}_i \delta_{\nu}^j \left( \frac14 H^{i}{}_{hk}H^{h}{}_{jl} \, \d x^{kl} + \frac12 \nabla_{0} H^{i}{}_{jk}\, \d x^{0 k} \right) \nn \\
{\cal R}_{+\, MN}|_{MN=\mu n} &= -\frac14 \delta_{\mu}^0 \left( H_{0 ps}H_{0 n}{}^{s} - 2 \nabla_{0} H_{0 n p}  \right)  \d x^{0} \w \d y^p \\
{\cal R}_+^M{}_N|_{MN=mn}&= \bar{{\cal R}}^m{}_{n} + \frac14 H_{0 p}{}^m H_{0qn} \d y^{pq}
\end{align}
\end{subequations}
with $\bar{{\cal R}}^m{}_{n}= \frac12 {\cal R}^m{}_{npq} \, \d y^{pq}$. We are left with the replacement of $H$-flux components in terms of axions, but this will only be done when calculating the next, related quantity. 

We can now compute $\tr({\cal R}_{10+} \w {\cal R}_{10+}) = - {\cal R}_+^A{}_B  \w {\cal R}_+^B{}_A $ at order $(\alpha')^0$, as follows
\bea
\tr({\cal R}_{10+} \w {\cal R}_{10+}) &=& - {\cal R}_+^{\mu}{}_{\nu}  \w {\cal R}_+^{\nu}{}_{\mu}  - 2 {\cal R}_+^{\mu}{}_n  \w {\cal R}_+^n{}_{\mu}  - {\cal R}_+^m{}_n  \w {\cal R}_+^n{}_m \\
&=& - {\cal R}_+^{i}{}_{j}  \w {\cal R}_+^{j}{}_{i}  - {\cal R}_+^m{}_n  \w {\cal R}_+^n{}_m \nn\\
&=& \left( {\cal H}^2   - \frac14 (H_{fgh})^2 \right) \, \nabla_{0} H_{ijk}\ \d x^{0ijk}   +\, \tr (\bar{{\cal R}}  \w  \bar{{\cal R}} )  + \frac14 H_{0r}{}^m H_{0 s}{}^n \, \bar{{\cal R}}_{mnpq} \, \d y^{pqrs}  \nn\\
&=& - 6\sqrt{2} \left( {\cal H}^2   - \frac{3}{M_p^2} \, (\del_0 \vartheta_a)^2  \right) \, \frac{1}{M_p} \ \del_0^2 \vartheta_a \ {\rm vol}_4   +\, \tr (\bar{{\cal R}}  \w  \bar{{\cal R}} ) \nn\\
&&+\, \pi^2 l_s^4 \ \del_0 b^i \del_0 b^j  \ \beta_{i\, r}{}^m \beta_{j\, s}{}^n \, \bar{{\cal R}}_{mnpq} \, \d y^{pqrs} \ , \nn
\eea
where the $b^i$ could be further expressed in terms of $\vartheta^i$; we will rewrite them in the final expression of ${\cal S}_{{\rm curv}}$.

Under the restrictions above, there is no mixed term, as anticipated in \eqref{trR+2}, especially for the $H$-flux Bianchi identity \eqref{BI}. We also read the 4d and 6d contributions
\bea
\tr({\cal R}_{+} \w {\cal R}_{+}) &=& - 6\sqrt{2} \left( {\cal H}^2   + \frac{3}{M_p^2} \, (\del \vartheta_a)^2  \right) \, \frac{1}{M_p} \ \del_0^2 \vartheta_a \ {\rm vol}_4 \ , \\
\tr(\bar{{\cal R}}_{+} \w \bar{{\cal R}}_{+}) &=& \tr (\bar{{\cal R}}  \w  \bar{{\cal R}} ) + \pi^2 l_s^4 \ \del_0 b^i \del_0 b^j  \ \beta_{i\, r}{}^m \beta_{j\, s}{}^n \, \bar{{\cal R}}_{mnpq} \, \d y^{pqrs} \ .\label{trRRap}
\eea
The 4d term $\tr({\cal R} \w {\cal R})$ had been evaluated in a cosmological context in \cite{Alexander:2004us}, on an FLRW background together with gravitational waves: it was found that the background contribution vanishes. We verify this background evaluation here: we get a non-zero result for $\tr({\cal R}_{+} \w {\cal R}_{+})$ only thanks to the $H$-flux contribution, here related to $\del_0 \vartheta_a$. We do not consider geometric fluctuations, in particular no gravitational wave, in this computation.

It is interesting to note that the 6d contribution $\tr(\bar{{\cal R}}_{+} \w \bar{{\cal R}}_{+}) $ is actually 4d-dependent, through its last term. As we will see, that term is however subdominant in such cosmological setting and may eventually be neglected: this is due to time derivatives being considered much smaller than 6d space derivatives in this context. This would allow us to trade $\tr(\bar{{\cal R}}_{+} \w \bar{{\cal R}}_{+}) $ for the purely geometric 6d quantity $\tr (\bar{{\cal R}}  \w  \bar{{\cal R}} )$.

We can now evaluate the contribution of these quantities to ${\cal S}_{{\rm curv}}$ \eqref{Scurv}, as follows
\bea
{\cal S}_{{\rm curv}} \supset \!\!\!\! && -\, \frac{1}{16 \pi^2} \int \tr({\cal R}_+ \w {\cal R}_+) \left(  a  +   b^i \ \, \frac{1}{12 \times 16 \pi^2}  \int_X \beta_i \w \tr (\bar{{\cal R}}_+ \w \bar{{\cal R}}_+) \right) \nn\\
&& =\, \frac{3}{2} \int {\rm vol}_4 \ l_s^2\ \vartheta_a \  \del_0^2 \vartheta_a \left( {\cal H}^2   - \frac{3}{M_p^2} \, (\del_0 \vartheta_a)^2  \right)  \label{SctrRR}\\
&&\ +\, \frac{\sqrt{2}}{2 (4\pi)^4} \int {\rm vol}_4 \ \frac{1}{M_p} \, \del_0^2 \vartheta_a \left( {\cal H}^2   - \frac{3}{M_p^2} \, (\del_0 \vartheta_a)^2  \right) \, b^i \ \int_X \beta_i \w \tr (\bar{{\cal R}}  \w  \bar{{\cal R}} ) \nn\\
&&\hspace{-0.6in} +\, \frac{2\sqrt{2}}{4^4} \int {\rm vol}_4 \ \frac{1}{M_p} \, \del_0^2 \vartheta_a \left( {\cal H}^2  - \frac{3}{M_p^2} \, (\del_0 \vartheta_a)^2  \right) \, b^i \ \del_0 b^j \del_0 b^k \  l_s^6\ \int_X \beta_i \w \, \frac{\beta_{j\, r}{}^m \beta_{k\, s}{}^n}{(2\pi l_s)^2} \, {\cal R}_{mnpq} \, \d y^{pqrs} \nn
\eea\\

\begin{itemize}
    \item $\tr {\cal R}_{10+}^4$
\end{itemize}

Another quantity entering ${\cal S}_{{\rm curv}}$ is $\tr {\cal R}_{10+}^4$. Up to a sign (unknown to us), it is given by the following contractions and wedge products
\beq
\tr {\cal R}_{10+}^4 = \mp {\cal R}^M_{+\, N} \w {\cal R}^N_{+\, P} \w {\cal R}^P_{+\, Q} \w {\cal R}^Q_{+\, M} \ .
\eeq
With the specifications and simplifying assumptions above, one obtains
\beq
\tr {\cal R}_{10+}^4 =0 \ .
\eeq
Indeed, this 8-form cannot be purely 6d nor 4d: it must then be a mixture. But the contraction of indices then forces to have mixed indices, as can be seen through the 2-form components \eqref{R2fcompo}, and this must happen twice at least, i.e.~${\cal R}^{\mu}_{+\, n}$ and ${\cal R}^{m}_{+\, \nu}$. However, the wedge product of those two 2-forms is proportional to $\d x^0 \w \d x^0$ which vanishes.\\

\begin{itemize}
\item $\d B_2 \w *_{10}\, {\rm CS}(\omega_+)|_{4,6d}$
\end{itemize}

To compute the last term in ${\cal S}_{{\rm curv}}$, we now focus on the Chern-Simons 3-form 
\beq
{\rm CS}(\omega_+) = - {\cal R}^A_{+\, B} \w \omega^B_{+\, A} + \frac13 \omega^A_{+\, B} \w \omega^B_{+\, C} \w \omega^C_{+\, A} \ .
\eeq
We are interested in certain mixed 4d and 6d components. Such a mixed 3-form must be exact, as we will verify explicitly: indeed, $\d {\rm CS}(\omega_+) = \tr({\cal R}_{10+} \w {\cal R}_{10+})$, which we recall has no mixed component. To compute the components of interest, we first express the spin connection 1-form coefficient as follows in terms of vielbeins
\beq
\omega^A_{+\, BN} = \omega^A{}_{BN} + \frac12 H^A{}_{BN} = e^A{}_M ( \del_N e^M{}_B + e^P{}_B \Gamma^M_{NP} ) + \frac12 H^A{}_{BN} \ .
\eeq
Given the 10d metric ansatz \eqref{10dmetric}, the 10d vielbeins are block diagonal with separate coordinate dependence. The same holds for the Christoffel symbols. Therefore, there is no geometric spin connection coefficient $\omega^A{}_{BN}$ with mixed 4d and 6d indices. However, the $H$-flux can provide such mixed components.

More precisely, we are interested in the mixed component ${\rm CS}(\omega_+)|_{0mn}$: this is because it appears in ${\cal S}_{{\rm curv}}$ contracted with $\d B_2$, which only admits such a component. Using above specifications, including the fact that $\omega^{0}_{\ \,0 0}=0$ for FLRW, we obtain at order $(\alpha')^0$
\bea
\frac12 {\rm CS}(\omega_+)|_{0mn} \d x^0 \w \d y^{mn} &=&  {\cal R}_{+\, 0 m} \w H^{m}{}_{0n} \d y^n  - \frac12 {\cal R}^m_{+\, n} \w H^n_{\ \, m0} \d x^0\\
&& - \frac18 H_0{}^{mn} H_{0mp} H_{0nq}\, \d x^0 \w \d y^{pq}  + \frac12 H^a{}_{b0}\, \d x^0 \w \omega^b_{\ \, c} \w \omega^c_{\ \, a} \nn\\
&=&  +\frac14 \left( H_{0 np}H_{0}{}^{nm} - 2 \nabla_{0} H_0{}^m{}_p  \right) H_{0mq}\, \d x^{0} \w \d y^{pq} \nn\\
&& - \frac12 H^a{}_{b0}\, \d x^0 \w  \left( \bar{{\cal R}}^b{}_{a}  - \omega^b_{\ \, c} \w \omega^c_{\ \, a} \right) \ .\nn
\eea
As expected this is a closed form: indeed, an exterior derivative on the first line would produce a $\d x^0$, making it vanish, while for the second line, we recall the relation $ \bar{{\cal R}}^a{}_{b} =  \d \omega^a_{\ \, b} + \omega^a_{\ \, c} \w \omega^c_{\ \, b}$. We can further simplify this expression using the restriction \eqref{HHortho}, and replacing the $H$-flux components: we obtain
\bea
\hspace{-0.2in} \frac12 {\rm CS}(\omega_+)|_{0mn} \d x^0 \w \d y^{mn} &=& \frac14  H_{0 mn}H_{0}{}^{mn} \, H_{0pq}\, \d x^{0} \w \d y^{pq} - \frac12 H^a{}_{b0}\, \d x^0 \w \d \omega^b{}_{a} \\
&=&  (2\pi l_s^2)^3\ \del_0 b^i \del_0 b^j \del_0 b^k\ \frac12 \beta_{i\, mn}\beta_{j}{}^{mn} \, \d x^{0} \w \beta_{k}  \nn\\
&&\!\! - \pi l_s^2\, \del_0 b^i\, \beta_i{}^a{}_b\, \d x^0 \w \d \omega^b{}_{a} \ .\nn
\eea
This allows us to compute the corresponding term in ${\cal S}_{{\rm curv}}$ \eqref{Scurv}
\bea
{\cal S}_{{\rm curv}} \supset \!\!\!\! &&  -\frac{1}{2 \kappa_{10}^2\, g_s^2}\, \frac{\alpha'}{4} \int \d B_2 \w *_{10}\, {\rm CS}(\omega_+)|_{4,6d}  \label{ScCS}\\
&& =  \int \d^4 x \sqrt{|g_4|}\ \frac{1}{2} \frac{l_s^2}{M_p^2}  \sum_i (\del_0 \vartheta^i)^4   \nn\\
&& - \int \d^4 x \sqrt{|g_4|}\ \frac{l_s^2}{8} \, \del_0 \vartheta^i \del_0 \vartheta^j\, \frac{1}{(2\pi l_s)^2} \int \frac{\beta_i{}^a{}_b\ \beta_j}{\gamma_i \gamma_j} \w  *_6\d \omega^b{}_{a} \ . \nn
\eea

Finally, let us point out that the 4d components of the Chern-Simons 3-form at order $(\alpha')^0$ necessarily boil down to contributions from 4d forms only, given the above ansatz
\beq
{\rm CS}(\omega_+)|_{4d} = - {\cal R}^{\alpha}_{+\, {\beta}} \w \omega^{\beta}_{+\, {\alpha}} + \frac13 \omega^{\alpha}_{+\, {\beta}} \w \omega^{\beta}_{+\, {\gamma}} \w \omega^{\gamma}_{+\, {\alpha}} \ .
\eeq
We also verify that each of these 4d contributions only depends on 4d coordinates. This proves that $H|_{4d}$ only depends on 4d coordinates at order $\alpha'$, as initially considered in the main text.

\begin{itemize}
    \item ${\cal S}_{{\rm curv}}$
\end{itemize}

We finally gather all previous results to provide the expression of ${\cal S}_{{\rm curv}}$ \eqref{Scurv} conditional on the assumptions above. We introduce conveniently $\hat{\vartheta}= \vartheta/M_p$. The resulting expression is the following
\bea
{\cal S}_{{\rm curv}} \!\!\!\! && = \int \d^4 x \sqrt{|g_4|}\ \frac{M_p^2}{2}\ \Bigg[ \frac32\, l_s^2 \, \big({\cal H}^4 + \big( \dot{{\cal H}} + {\cal H}^2 \big)^2 \big) + \ {\cal O}({\alpha'}^2) \label{Scurvcomputedap}\\
&& \phantom{= \int \d^4 x \sqrt{|g_4|}\ \frac{M_p^2}{2}\ \Bigg[} - \frac{3}{2} \, l_s^2\, {\cal H}^2 \,  (\del_0 \hat{\vartheta}_a)^2 + 3\, l_s^2\,  {\cal H}^2\ \hat{\vartheta}_a\, \del_0^2 \hat{\vartheta}_a  \nn\\
&& \phantom{= \int \d^4 x \sqrt{|g_4|}\ \frac{M_p^2}{2}\ \Bigg[}  + \frac{3}{8}\, l_s^2 \,  (\del_0 \hat{\vartheta}_a)^4  - \frac34\, l_s^2\, \left(\del_0^2 \hat{\vartheta}_a \right)^2  - 9\ l_s^2\ \hat{\vartheta}_a\, \del_0^2 \hat{\vartheta}_a \, (\del_0 \hat{\vartheta}_a)^2    \nn\\
&& \phantom{= \int \d^4 x \sqrt{|g_4|}\ \frac{M_p^2}{2}\ \Bigg[} + \frac{9}{8} \, l_s^2 \, \sum_i (\del_0 \hat{\vartheta}^i)^4  + \frac{1}{8} \,  l_s^2\, \sum_i \left(\del_0^2 \hat{\vartheta}^i \right)^2  \nn\\
&& \phantom{= \int \d^4 x \sqrt{|g_4|}\ \frac{M_p^2}{2}\ \Bigg[} + \frac18 \ \frac{1}{{\cal V} (2\pi l_s)^6}\int {\rm vol}_6\ \ l_s^2 \ {\cal R}_{mnpq} {\cal R}^{mnpq}  \nn\\
&& \phantom{= \int \d^4 x \sqrt{|g_4|}\ \frac{M_p^2}{2}\ \Bigg[} +\frac12\, l_s^2\ \del_0 \hat{\vartheta}^i \del_0 \hat{\vartheta}^j  \int {\rm vol}_6\ {\cal R}_{mnpq} \ \frac{ \beta_i^{mp} \beta_j^{nq} }{2\, (2\pi l_s)^2\, \gamma_i \gamma_j} \nn\\
&& \phantom{= \int \d^4 x \sqrt{|g_4|}\ \frac{M_p^2}{2}\ \Bigg[} + \frac{1}{2} \, l_s^2\ \del_0 \hat{\vartheta}^i \del_0 \hat{\vartheta}^j\, \int \frac{-\beta_i{}^a{}_b\ \beta_j}{2\, (2\pi l_s)^2\, \gamma_i \gamma_j} \w  *_6\d \omega^b{}_{a} \nn\\
&& \phantom{= \int \d^4 x \sqrt{|g_4|}\ \frac{M_p^2}{2}\ \Bigg[} + \frac{1}{(4\pi)^3}\ \del_0^2 \hat{\vartheta}_a \left( {\cal H}^2   - 3 \, (\del_0 \hat{\vartheta}_a)^2  \right) \, \frac{\sqrt{{\cal V}}}{M_p^2}\ \hat{\vartheta}^i \, \int \frac{\beta_i}{\gamma_i} \w \tr (\bar{{\cal R}}  \w  \bar{{\cal R}} ) \nn\\
&&\hspace{-0.5in} + \frac{1}{(4 \pi)^3}\, \del_0^2 \hat{\vartheta}_a \left( {\cal H}^2  - 3 \, (\del_0 \hat{\vartheta}_a)^2  \right) \, \hat{\vartheta}^i \, \del_0 \hat{\vartheta}^j \del_0 \hat{\vartheta}^k \ \frac{\sqrt{{\cal V}}}{M_p^2} \, {\cal V} (2\pi l_s)^6 \! \int \frac{\beta_i}{\gamma_i} \w \frac{\beta_{j\, r}{}^m \beta_{k\, s}{}^n}{2\, (2\pi l_s)^2\, \gamma_j \gamma_k} \, {\cal R}_{mnpq} \, \d y^{pqrs}  \Bigg] \nn
\eea
An estimate of this expression is presented in Section \ref{sec:curv}.

\end{appendix}

\newpage

\addcontentsline{toc}{section}{References}
\bibliographystyle{JHEP}
\bibliography{References}

\end{document}